# Electrochemical Electron Transfer: Key Concepts, Theories, and Parameterization via Atomistic Simulations

Mengke Zhang[1,2], Yanxia Chen[1,*], Marko M. Melander,[3,*] Jun Huang[2,4,*]

*[1]Hefei National Research Center for Physical Science at Microscale and Department of Chemical Physics, University of Science and Technology of China, 230026 Hefei, China*

*[2]Institute of Energy Technologies, IET-3: Theory and Computation of Energy Materials, Forschungszentrum Jülich GmbH, 52425 Jülich, Germany*

*[3]Department of Chemistry, Nanoscience Center, University of Jyväskylä, P.O. Box 35 (YN), FI-40014, Jyväskylä, Finland*

*[4]Theory of Electrocatalytic Interfaces, Faculty of Georesources and Materials Engineering, RWTH Aachen University, 52062 Aachen, Germany*

*Corresponding Authors: ju.huang@fz-juelich.de, marko.m.melander@jyu.fi, yachen@ustc.edu.cn

## Abstract

Electron transfer (ET) at electrochemical interfaces lies at the heart of numerous energy conversion and storage processes, yet its theoretical description and computation modeling remain dynamic areas of research. This review is aimed at elucidating key concepts and theories of ET kinetics, focusing on the coupling between classical solvent fluctuations and quantum electronic states of metallic electrodes and redox species. We begin with fundamental rate theories, reaction coordinates, and timescales relevant to electrochemical systems, and then systematically explore the regimes of weak, strong, and intermediate electronic coupling. Special attention is given to solvent dynamics and the structure of the electrical double layer (EDL), both of which critically impact ET kinetics. Atomistic simulations—particularly density functional theory (DFT) and molecular dynamics (MD) are highlighted as useful tools for assessing key assumptions such as linear response and determining key parameters such as solvent reorganization energy, electronic coupling strengths, and those describing nuclear dynamics. We conclude by outlining opportunities for advancing the field through multiscale, quantum-classical models that incorporate EDL effects, multiple reaction coordinates, solvent-controlled dynamics, and transitions between adiabatic and non-adiabatic regimes. This review aims to serve as both a conceptual guide and a practical resource for researchers seeking to integrate theory and simulation in the study of electrochemical ET across diverse systems.

# 1. Introduction

## 1.1. Preamble

Electron transfer (ET) is a fundamental process underlying a wide range of chemical, electrochemical, and biological phenomena and applications. Its fundamental concepts and theories have been established over several decades though seminal works[1–11] and now systematically presented in standard textbooks[12,13]. However, achieving a detailed understanding of ET kinetics at electrochemical interfaces (electrochemical ET) remains a daunting challenge due to several layers of complexities. First, the experimentally observable electric current is due to the reaction kinetics of all elementary ET steps which could form a complex reaction pathway. While simplified concepts such as potential determining step[14,15] and rate determining step[15,16] have been developed and widely used, they have been shown to be insufficient in many cases[17,18], and a full microkinetic model with all ET steps[19–26] is required.

Second, each elementary ET step occurs in a highly complex, heterogeneous nanoscale interfacial region, namely, the electrical double layer (EDL) where the local reaction environment can be drastically different from the bulk conditions. In this sense, a comprehensive understanding of electrochemical ET needs to integrate the theory of the ET itself and the EDL theory. EDL effects on electrochemical reactions have been extensively discussed in recent literature[19,20,24–35].

Third, each elementary ET step in the EDL is a complex process involving the tunnelling of electrons between the electrode surface and redox species, accompanied by reorganization of the nuclear configuration from that of the initial state to that of the final state. Such nuclear reorganization, in a simple yet common scenario, includes reorganization of both the redox species itself (inner-sphere reorganization), i.e., adjustments of bond lengths and angles within the redox species, as well as its solvent environment (outer-sphere or solvent reorganization), e.g., changes in solvent polarization in response to charge redistribution[12]. This scenario corresponds to outer-sphere ET, and represents the theoretical minimum of ET, as the chemical structure of the redox species is preserved. In more complex cases, bond formation or breaking may occur during the ET process, e.g., chemisorption of the redox species onto the electrode surface, as in ion-coupled ET, or intramolecular bond dissociation, as in bond-breaking ET; such processes are referred to as inner-sphere ET[36]. In general, descriptions of the ET process are based on the separation of fast electron tunnelling and slow nuclear reorganization, in which the nuclei first reorganize to a prerequisite configuration that permits instantaneous electron tunnelling. The electron tunneling can be viewed as electronic transitions between the manifold of electronic states of the electrode surface and the valence electronic states of the redox species, which are coupled with nuclear dynamics. Slower nuclear dynamics and stronger electronic coupling between the electrode surface and redox species both lead to stronger electron-nuclear coupling during the tunneling process (as discussed in Section 1.2.5). As the electron-nuclear coupling strength, which can be characterized by the Landau-Zener parameter, increases from weak to strong, the electron tunnelling probability increases, and the ET process crosses over from the non-adiabatic to the adiabatic limit.

Overall, ET theory at electrochemical interfaces necessitates theoretical treatments of nuclear reorganization, electron transitions between the electrode surface and redox species, and nuclear dynamics in both non-adiabatic and adiabatic reactions. This review aims to elucidate how these physical aspects contribute to ET kinetics and address the underlying concepts, theoretical basis, and computational parameterization of ET theories. This is achieved through detailed derivations of key equations that are frequently met in the literature while their derivations cannot be easily traced. By providing these details that have often been considered trivial in old papers, the present review will hopefully serve as an instrumental resource for undergraduate and graduate students who are not satisfied with just using the equations but want to know how they are obtained. To use these theoretical concepts in practice as explanatory and predictive methods, we discuss the application of theoretical computational methods, such as density functional theory (DFT) and molecular dynamics (MD) simulations, to simulate diabatic and adiabatic free energy surfaces (FESs) and quantities such as electronic coupling constants, reorganization energies, and solvent dynamics that influence ET rates. The examined quantities include nuclear reorganization energy, electronic coupling strength between the electrode surface and redox species, as well as those relevant to nuclear dynamics, such as the nuclear frequency, nuclear relaxation time, and friction. Important know-hows and existing challenges in computational parameterization of ET theories will be discussed.

In the reminder of the introduction section, we clarify in detail several central concepts and approximations used in ET theory, and outline the historical development of ET theories. In Section 2, we introduce the general reaction rate theory, reaction coordinates (RCs), and timescales in electrochemistry. In Section 3, we focus on the Marcus theory of ET kinetics with emphasis on nuclear reorganization, in which the inner sphere (i.e., the redox species) is treated as a set of harmonic vibrational modes and the outer sphere (i.e., the solvent) as a dielectric continuum, and a single effective nuclear coordinate is introduced to describe their reorganization. Sections 4 and 5 formulate ET rates under weak and strong electronic coupling between the electrode surface and redox species, primarily based on time-dependent perturbation theory and model Hamiltonian approach, respectively, while Section 6 addresses the intermediate coupling regime with explicit consideration of nuclear dynamics based on generalized Langevin equation and Landau-Zener theory. As two commonly encountered types of inner-sphere ET in electrocatalytic reactions, namely chemisorption and bond-breaking ET, these processes can be treated theoretically by incorporating additional nuclear coordinates, i.e., reaction coordinates, associated with bond formation or breaking, such as the distance between the electrode surface and redox species in chemisorption, or the distance between two fragments in bond-breaking ET. Corresponding theoretical considerations are presented in Section 5.3. The impact of the EDL on ET kinetics is discussed in Section 7, mainly within a recent semiclassical continuum model of the EDL[37–43]. This gradual progression underscores the importance of nuclear reorganization, the electronic structure of the electrode surface, electrocatalytic effects, adiabaticity, and nonergodicity.

Given the breath and great variety of electrochemical ET, this review is by no means comprehensive. Focusing on metallic electrodes, this review does not cover ET at semiconductors and other materials; interested readers are referred to the classical review by Gerischer[44] and more recent review by Santos and Schmickler[45]. Long-range electrochemical ET reactions and ET reactions at semiconductor electrodes, which are usually considered nonadiabatic and share many concepts with ET reactions at metallic electrodes, will be briefly discussed in Section 4.5. A more comprehensive review has recently been provided by Nazmutdinov and Ulstrup and their coworkers[46]. Neither do we discuss proton-coupled electron transfer (PCET), as its theoretical description depends on the relative timescales of proton motion in different regimes and treatment of nuclear quantum effects for the transferring proton; excellent reviews are available in the literature[47–49]. Furthermore, we do not extensively discuss the applications of ET theory but refer the readers to other excellent sources[50–53].

## 1.2. Concepts

### 1.2.1. Reaction plane/volume and work term

The ET rate inherently depends on the distance from the electrode surface, as both electronic coupling strength between the electrode surface and redox species[54,55], solvent environment, and local concentration of redox species vary with the distance. Given that the electronic coupling strength decays exponentially from the electrode surface with a characteristic length of only a few angstroms, the electron tunnelling probability rapidly becomes negligible beyond this distance such that appreciable tunnelling occurs only for redox species within a finite volume near the electrode surface, referred to as the reaction volume. The reversible work is required for the redox species in the bulk solution to pass through the diffuse layer of the EDL to reach the reaction volume, which has a characteristic thickness corresponding to the Debye length of typically several nanometers. This work term mainly involves overcoming the changes in the solvation free energy and electrostatic potential energy due to the presence of the interfacial electric field; hence it depends on the local reaction environment, and its effect on ET kinetics is discussed in Section 7.1. As discussed, as the redox species approaches the electrode surface, the electron tunnelling probability increases rapidly due to stronger electronic coupling. However, the local concentration of the redox species decreases sharply because of repulsive interactions with the electrode surface. As a result, ET faces two competing effects: shorter distances would increase the electron tunnelling probability and thus increase the intrinsic rate constant but simultaneously reduce the local reactant concentration and thereby suppress the overall ET rate by the law of mass action. Therefore, ET has a certain optimal distance where the tunnelling probability and reactant concentration have the maximal contribution to the ET rate. This optimal distance can be identified as the reaction plane. Herein, we consider the reduced species donating an electron to the electrode surface (oxidation reaction) and oxidized species accepting an electron from the electrode surface (reduction reaction), each at its respective reaction plane for simplicity. A more precise evaluation of the ET rate can be achieved by integrating over the reaction volume[56,57], which will be discussed in Section 7.

### 1.2.2. Two-state/level model for ET

As the electron donor or acceptor in electrochemical ET is semi-infinite surface, the electronic structure of the electrode surface is essential for determining the intrinsic ET kinetics. Unlike the discrete electronic levels of an atom or a molecular, the electronic states of an electrode have a continuum of electronic states, i.e., energy bands, due to the overlap of atomic orbitals in the solid. These energy bands exhibit a continuous energy distribution characterized by the electron density of states (DOS). For metallic electrodes, electrons are filled up to the Fermi level at absolute zero temperature. At finite temperatures, the electron occupancy probability transitions smoothly from unity to zero due to the temperature-dependency of the Fermi-Dirac distribution. This transition, dictated by the Fermi-Dirac distribution, spans an energy interval of several $k_{\mathrm{B}}T$, approximately $100\ \mathrm{meV}$ at room temperature.

Before treating all electronic states, it is beneficial to first consider the ET between a specific electronic state $k$ on the metal surface and the valence electronic state $a$ of the redox species (two-level system) to develop conceptual understanding. In this case, depending on whether the electron resides in the electronic state $a$ or $k$, we can immediately recognize two stable electronic states of the ET system: a reduced one, including the reduced species and its solvation structure, and an oxidized one, including the oxidized species, its solvation structure, and the electron in the electronic state $k$. It is expected that the reduced and oxidized states will reach their minimum free energies at their respective equilibrium nuclear configurations, which includes both the redox species and the solvent. The rate constant is then given as the number of transitions per unit time and every transition between the reduced and oxidized states, where each transition involves the quantum transition of the electron between the electronic states $k$ and $a$ along with changes in nuclear configuration.

### 1.2.3. Franck-Condon principle, diabatic free energy surface, and nuclear reorganization

Owing to the almost instantaneous nature of electron tunneling compared to the timescale of nuclear motion (see Section 2.5 for details), as a first approximation the nuclear configuration can be assumed to remain frozen during electron transitions between the different electronic states: this is the *Franck-Condon* principle. When the system is in the oxidized state at its equilibrium nuclear configuration, the corresponding reduced state at this configuration has significantly higher free energy as this nuclear configuration is far from the equilibrium for the reduced state. This means that there is a substantial "Franck-Condon barrier" or a vertical energy gap for the electron transition and the tunneling probability is low. In other words, the system is far from a nuclear configuration where the two states have the same energy and where the resonance between the states leads to high electron tunnelling probability. To remove this barrier and to achieve the resonance condition, the nuclei of the redox species and solvent need to gradually reorganize into a configuration away from the equilibrium in the oxidized state towards that of the reduced state. In this process, the oxidized state moves out of equilibrium, raising its free energy, while the reduced state approaches its equilibrium, lowering its free energy. At some non-equilibrium nuclear configurations resulting from the nuclear reorganization, the oxidized and reduced states have the same free energy, the resonance condition is satisfied, and electron tunnelling can take place at the maximal probability. The rearrangement of the nuclei from the equilibrium configuration of the reactant state (oxidized or reduced state) to these non-equilibrium configurations is referred to as nuclear reorganization.

The nuclear reorganization is driven by thermal fluctuations of the nuclei. For the ET system consisting of $N$ nuclei in both the redox species and the surrounding solvent, thermal fluctuations of the nuclear coordinates naturally lead to $3N$-dimensional free energy surfaces (FESs) for the oxidized and reduced states. Rather than using the high complex $3N$-dimensional FES, the $3N$ nuclear coordinates can be projected on an effective one-dimensional nuclear coordinate $\xi$ as the RC which captures the overall or collective nuclear reorganization driving the ET; the corresponding FESs along this nuclear coordinate are schematically plotted in Figure 1a. The two minima correspond to the oxidized (blue line) and reduced (red line) states at their respective equilibrium nuclear coordinates. Small thermal fluctuations about the minima lead to parabolic FESs, as shown in Figure 1a, which constitute the typical graphical representation of Marcus theory. The Marcus parabolic representation becomes insufficient for ET reactions accompanied by major structural reorganizations, such as ligand substitution process[58] and inner-sphere ET[59,60], for which more specific theoretical treatments are required. The inner-sphere ET reactions, such as chemisorption and bond-breaking ET, are specifically discussed in Section 5.3. The difference in free energy between the minimum of the product and that of the reactant represents the reaction free energy, also known as the thermodynamic driving force of the reaction. If there is no electronic coupling between the electronic states $k$ and $a$, corresponding to an infinite barrier or distance for electron tunneling, electron tunnelling cannot occur, even at the intersection of the two FESs where the Franck-Condon barrier is practically zero. In this case, the electronic states of the oxidized or reduced states remain unchanged as the nuclear configuration fluctuates. This contrasts with the adiabatic approximation (Born-Oppenheimer approximation), where the motions of electrons and nuclei are concerted and electrons adjust instantaneously to nuclear motion. For this reason, we refer to the FESs of the oxidized and reduced states as the *diabatic* FESs, and label them according to their electronic states as the diabatic state $k$ and the diabatic state $a$, respectively.

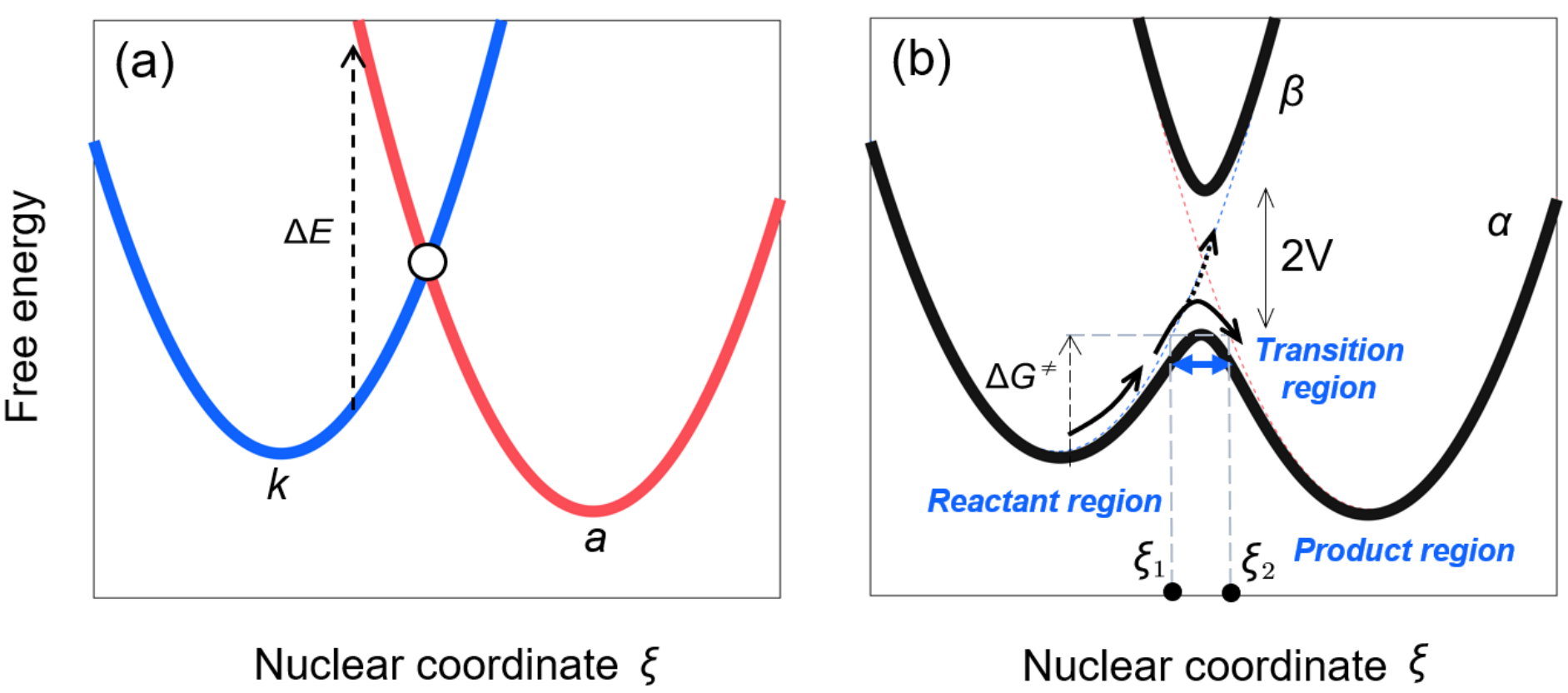

Figure 1. (a) Free energy surfaces (FESs) of the diabatic states $k$ and $a$ as functions of the effective nuclear coordinate $\xi$. The diabatic state $k$ corresponds to the oxidized state with the electron residing in the electronic state $k$ of the electrode surface, whereas the diabatic state $a$ corresponds to the reduced state with electron residing in the valence electronic state $a$ of the redox species. The energy gap between the diabatic states is denoted as $\Delta E$, as indicated. (b) The adiabatic ground state $\alpha$ and adiabatic excited state $\beta$, which are the construction and destruction of the diabatic states $k$ and $a$, respectively. The energy gap between the adiabatic states at the point where the diabatic FESs cross is the twice the coupling matrix element $V$ between the electronic states $k$ and $a$. The transition region, in which the electron transition actually occurs is denoted by $\Delta\xi$, with $\xi_1$ and $\xi_2$ representing its boundaries. In the reactant region ($\xi < \xi_1$) and product region ($\xi > \xi_2$), the adiabatic state $\alpha$ resembles the diabatic state $k$ and $a$, respectively. As indicated by the solid arrows, ET proceeds via thermal fluctuations of nuclear configuration into the transition region, followed by crossing of this region and subsequent relaxation to the equilibrium nuclear configuration of the product. The activation free energy $\Delta G^{\neq}$ is the free energy difference between the top of the adiabatic FES of state $\alpha$ and the minimum in the reactant region. Upon crossing the transition region, there is a finite probability for the system to undergo a transition to the excited state $\beta$, as indicated by the dotted arrow.

### 1.2.4. Adiabatic free energy surface

If sufficiently large electronic coupling exists between the electronic states $k$ and $a$, the corresponding diabatic states are split into an adiabatic excited state $\beta$ and an adiabatic ground state $\alpha$, which give rise to two adiabatic FESs, as shown in Figure 1b. The coupling strength between the electronic states $k$ and $a$ can be characterized by the coupling matrix element, $V$, with twice its value corresponding to the energy gap between the adiabatic states at the point where the diabatic FESs cross. The characteristic region of the resonance splitting, i.e., sufficient electronic coupling, is denoted as $\Delta\xi$, with $\xi_1$ and $\xi_2$ representing its boundaries. Beyond this region, the system remains in the adiabatic state $\alpha$ due to the large energy gap between the adiabatic states $\alpha$ and $\beta$. It is clear that the adiabatic state $\alpha$ resembles the diabatic state $k$ for $\xi < \xi_1$ and the diabatic state $a$ for $\xi > \xi_2$. Therefore, these two regions can be referred to as the reactant region and the product region, respectively. In the region within $\Delta\xi$, the lower adiabatic ground state $\alpha$ is a constructive superposition of the diabatic states $k$ and $a$, while the upper or excited adiabatic state $\beta$ results from the destructive superposition of the diabatic states. Adiabatic ET then proceeds within the $\Delta\xi$ on the adiabatic FES of the ground state $\alpha$ and $\Delta\xi$ is the region where the electron transition actually occurs, and it is thus referred to as the transition region herein. However, when the system crosses this region, the state $\alpha$ has a certain probability of being excited to the state $\beta$ due to the relatively small vertical energy gap compared to the reactant and product regions. Evidently, this probability increases as the coupling strength $V$ decreases. In other words, a stronger electronic coupling leads to a higher probability of electron transition along the adiabatic FES of the state $\alpha$.

### 1.2.5. Effective nuclear frequency, activation free energy, and electronic transmission coefficient

As shown in Figure 1b, for ET to occur, the nuclear configuration first fluctuates to a prerequisite configuration at the nuclear coordinate $\xi_1$, after which the electron transition proceeds as the system crosses the transition region $\Delta\xi$. Within this picture, the rate constant $k_{\mathrm{ET}}$, defined as the number of successful electron transitions per unit time, therefore depends on three key factors: (i) the effective nuclear frequency of thermal fluctuations ($\nu_{\mathrm{n}}$), which sets the number of attempts for the system to cross the transition region per unit time; (ii) the free energy barrier that the system must overcome ($\Delta G^{\neq}$), defined as the free energy difference between the top of the adiabatic ground state FES and the minimum in the reactant region. This barrier determines the probability of observing the prerequisite nuclear configuration through the Boltzmann factor ($e^{-\frac{\Delta G^{\neq}}{k_{\mathrm{B}}T}}$), where $k_{\mathrm{B}}$ is the Boltzmann constant and $T$ is the absolute temperature, assuming thermal equilibrium within the reactant region; and (iii) the probability of successful electron transition crossing the transition region, denoted by $\kappa_{\mathrm{el}}$—the electronic transmission coefficient or tunnelling factor[46,61]. Combined, these elementary considerations give the rate constant as

$$k_{\mathrm{ET}} = \nu_{\mathrm{n}}\kappa_{\mathrm{el}}e^{-\frac{\Delta G^{\neq}}{k_{\mathrm{B}}T}}, \tag{1}$$

which is the standard expression for elementary ET in a two-level model or homogeneous reactions[12,46,61].

### 1.2.6. Non-adiabatic, adiabatic, and dynamics-controlled ET

Based on the electronic coupling strength and nuclear dynamics, three limiting regimes of ET can be distinguished.

- Non-adiabatic ET at extremely weak electronic coupling. In this case, the transition region is narrowed down to the intersection of the diabatic FESs. When the system crosses this intersection, there is a high probability of excitation to the upper, excited state $\beta$, which leads to a very small transmission coefficient $\kappa_{\mathrm{el}}$. In this regime, the system state coincides with either of the two diabatic states throughout the entire region except at the intersection, so the ET process can be effectively described by the two diabatic FESs shown in Figure 1a. Non-adiabatic ET is particularly relevant for long-range ET where the electronic coupling constant is small. Some notable examples include e.g., ET across self-assembled monolayers to redox couples and proteins on metallic surfaces[62,63] or ET across thin insulating oxide films on metallic electrodes[64].

- Adiabatic ET at sufficiently strong electronic coupling. By "sufficiently strong", we mean that the energy splitting at the transition region is large enough such that the system has rare probability of transitioning to the excited state $\beta$ and the system evolves on the adiabatic ground state $\alpha$, where the electron adiabatically follows the nuclear configuration on the adiabatic FES of the state $\alpha$ during the whole ET process. Note that in periodic systems, the coupling between two diabatic states can be strong even if the coupling between two single electronic states (orbitals) is very small[65,66] – this indicates that it is risky to analyze adiabaticity based on orbitals. Compared with non-adiabatic ET, adiabatic ET exhibits two characteristic features. First, it is evident that the transmission coefficient $\kappa_{\mathrm{el}}$ is close to unity, which indicates that the pre-exponential factor in Eq. 1 is primarily controlled by nuclear dynamics. Second, the activation free energy is significantly lowered by the electronic coupling between the electrode surface and redox species, which is a manifestation of the electrocatalytic effect. Some typical examples of adiabatic ET include ET from the metal to adsorbed species or ET during adsorption. For traditional outer-sphere couples with bulky ligands undergoing ET without adsorption[66,67], it is still unclear whether they are electronically adiabatic or non-adiabatic.

- ET reactions with very small activation free energies, e.g., those taking place in slowly relaxing media, can also be controlled by nuclear dynamics of the reaction environment. Some notable examples include ET reactions in ionic liquids[68] with very slow solvent relaxation extending to several seconds[69]. For instance, in ionic liquids, ET can be slow despite a small activation free energy, which points to strong dynamic effects on ET rates. Surprisingly, the ET in deep eutectic solvents[70] appears less sensitive to dynamics effects than ionic liquids and similar to traditional non-polar media such as organic electrolytes where dynamics (solvent) effects are present but not dictating[71–73].

Here we provide only a qualitative description of adiabaticity using vague terms like "extremely weak" or "sufficiently strong". If the nuclei still have enough kinetic energy as they move into the transition region, there is a high probability that the system will evolve inertially to state $\beta$, as indicated by dotted arrow in Figure 1b. Therefore, the transmission coefficient $\kappa_{\mathrm{el}}$ depends not only on the electronic coupling strength but also on the nuclear dynamics. A criterion for adiabaticity can be made by comparing the timescales of electronic and nuclear motion in the transition region: the timescale for electronic motion, $\tau_{\mathrm{e}}$, can be estimated using the uncertainty principle as $\tau_{\mathrm{e}} = \hbar/4V$ while the timescale for nuclear motion, $\tau_{\mathrm{n}} = \Delta\xi/v_{\mathrm{avg}}$, is obtained from the average velocity of nuclei crossing the transition region, $v_{\mathrm{avg}}$. The stronger the coupling strength is, the smaller $\tau_{\mathrm{e}}$ is compared to $\tau_{\mathrm{n}}$. In this scenario, the electron can adjust itself more quickly to better follow the nuclear motion in the transition region, which indicates stronger electron-nuclear coupling. Consequently, the system has a higher probability of transitioning along the adiabatic FES of state $\alpha$, resulting in a larger transmission coefficient. In both the adiabatic and non-adiabatic ET, the solvent dynamics can further control reaction prefactor or even influence the activation free energy in the case of non-ergodic ET, as discussed in Section 6. Understanding the subtle differences and transitions between these different limits calls for interpolation across the non-adiabatic, adiabatic, and dynamics-controlled limits. A more detailed discussion on this will be presented in Section 6.1.7 and preceding subsections.

To conclude this subsection, we present a general picture of ET. ET involves the transition of the system from the equilibrium state of the reactant to that of the product, which requires adjustments of both classical and quantum subsystems. What we mean by the quantum subsystem includes not only the electrons but also high-frequency nuclear modes characterized by vibrational frequencies $\omega$ satisfying $\hbar\omega \gg k_B T$. Their motions are much faster than those of low-frequency nuclear modes satisfying $\hbar\omega \ll k_B T$, which are treated as part of the classical subsystem. The classical subsystem first fluctuates to a non-equilibrium state that allows the transition of quantum subsystem to occur. These fluctuation process determine the activation free energy, which can be lowered by the coupling between the quantum and classical subsystems, a feature characterized by adiabaticity. Subsequently, the quantum subsystem undergoes a transition, which determines the pre-exponential factor, followed by the relaxation of the classical subsystem toward the equilibrium state of the products. The above discussion has excluded the quantum nature of high-frequency nuclear modes , which will be discussed in Section 4.6.

### 1.3. A brief history of ET theory

Elements of a two-level ET reaction, such as a homogeneous reaction in solution have been outlined above. Theoretical consideration of such a process, as mentioned, necessitates addressing the nuclear reorganization and the electronic interactions between the electron donor and acceptor. The importance of nuclear reorganization was recognized through the role of the solvent in isotopic exchange reactions. Libby was the first to apply the Franck-Condon principle in an attempt to explain the slower isotopic exchange reaction rates observed with smaller ions[74]. Therein, it was assumed that the electron undergoes a sudden Franck-Condon transition at the equilibrium solvent configuration of the reactants. Later, Marcus realized that, at this configuration, there is no available source of energy to overcome the considerable Frank-Condon barrier for radiationless ET in solution. The solvent must reorganize into a non-equilibrium configuration to remove the Franck-Condon barrier. Marcus developed a non-equilibrium polarization theory to describe this reorganization[2] and quantitatively explain the dependence of isotopic exchange reaction rates on ion size[1,75]. For ET involving smaller ions, the stronger electric field around them makes it harder for polar solvent molecules to reorganize, thereby slowing the reaction rate. Shortly after Marcus in 1956, Hush further developed the adiabatic ET theory at electrochemical interfaces based on a two-level consideration[3,4]. It was assumed that the charging state of the system follows the solvent configuration adiabatically, such that, at the transition state, the charge distribution is intermediate between those of the initial and final states. The theories of Marcus and Hush are rooted in transition state theory. Levich and Dogonadze, in 1959, were first to develop a fully quantum mechanical theory for non-adiabatic ET reactions in homogeneous solutions[5]. In this theory, the solvent is modeled as a phonon bath, represented by a collection of harmonic oscillators with various frequencies, while the weak electronic coupling between the reactants is treated as a perturbation, such that electron transition can be described by time-dependent perturbation theory. Such a phonon bath representation refines the description of the structured solvent by accounting for its retarded and nonlocal nature of dielectric response[12,76,77].

The above fundamental understanding remains valid for ET reactions at electrochemical interfaces. The main difference from homogeneous reactions lies in the involvement of multiple electronic states of the electrode surface, between which and the valence state of the redox species ET occurs. Therefore, the electronic structure factors of the metal surface, such as the density of states (DOS) and the Fermi-Dirac distribution, which determine the population and occupancy of electronic states, must be incorporated in the theory. For non-adiabatic ET reactions, the electron transitions between each metal electronic state and the redox species can be viewed as independent events[78], such that the overall ET rate is obtained by summing or integrating over the individual contributions from all electrode electronic states. In the early 1960s, Gerischer, independently, and the group of Dogonadze, Chizmadzhev, and Kuznetsov made significant contribution to the formalism of non-adiabatic ET rates at electrochemical interfaces[6–8]. Although the formalism was well-developed at that time, it was not until several decades later that Chidsey verified the importance of incorporating the Fermi-Dirac distribution of electrons near the Fermi level from his seminal experiments on redox-active self-assembled monolayers[62]. Weaver is remembered for his incisive analysis of existing theories for non-adiabatic ET in the context of simple inorganic outer-sphere electrode reactions[79]. His list of extant anomalies is a source of inspiration for further works in this area.

For adiabatic ET reactions at metal-solution interfaces, strong electronic interactions between the metal surface and redox species may lead to the formation of a hybridized state. In this scenario, the electronic interactions between all electrode electronic states and the redox species must be considered collectively. This can be achieved in a semi-classical manner by constructing the adiabatic FES from the diabatic FESs of all electronic states using empirical valence bond (EVB) approach[80–82]. A complete quantum mechanical theory for adiabatic ET reactions was developed by Schmickler *et al.*[9,10] This theory couples the Anderson-Newns model Hamiltonian for electronic interactions in a diabatic basis, which was used to describe hydrogen chemisorption at metal-vacuum interfaces[83–85], with the phonon representation for the solvent into a unified Hamiltonian, referred to as the Anderson-Newns-Schmickler (ANS) Hamiltonian. A very important implication of ANS theory is that the electronic interactions broaden the valence electronic level of the redox species into an energy band, which may lead to the formation of partially charged chemisorbates on the metal surface[86,87]. An exact expression for the absolute ET rate within this theory was provided using time-dependent Green's function method at the end of 20th century[11].

Building on the above foundational theories and insights into ET, theoretical extensions have been developed to encompass a broader range of ET reactions at electrochemical interfaces. One such extension addresses bond-breaking ET, a process in which ET is coupled with the dissociation of a chemical bond in the redox species. This process involves not only solvent reorganization but also the reorganization of the distance between the two fragments resulting from bond breaking. Therefore, the bond dissociation energy is expected to modulate the activation free energy. Additional theoretical consideration for bond-breaking ET reactions necessitates the treatment of the chemical bond in the redox species, which was pioneered by Saveant using the semi-empirical Morse potential[88–90], or alternatively treated using Hückel molecular orbital theory by Santos *et al.*[91,92] A quantum mechanical theory of bond-breaking ET was proposed by German and Kuznetsov[59], showing that as bond dissociation evolves from the classical to the quantum limit, the pre-exponential factor rather than the activation free energy is modulated. Another important class of reactions is proton-coupled ET, which encompasses a wide range of electrocatalytic processes, including hydrogen evolution and oxidation, carbon dioxide reduction, nitrogen reduction, and oxygen reduction. The complexity of theoretical considerations for such reactions lies in the relative timescales of the electron, solvent, and proton motions, which give rise to a broad spectrum of theoretical schemes[47,93–95].

## 2. Rate theory and electrochemical systems

Reaction rate theories allow understanding of how and why different factors and processes influence (electro)chemical rate constants. While many rate theories have been developed and applied for different purposes and conditions, the most widely used rate theory for heterogeneous (electro)chemical kinetics is the transition state theory (TST), which is discussed below in detail. Our main purposes are to 1) establish a general framework deriving and computing rate constants, 2) highlight how the *separation of timescales* is deeply ingrained in how we compute and think of rate constants and 3) show that within TST only thermodynamics, *not dynamics*, defines the rate constant.

A key point in the reactive flux method is the connection between macroscopic kinetics and microscopic dynamics. This connection rests on the famous Onsager's regression hypothesis[96]: the relaxation of the average of a macroscopic observable after a small external perturbation follows the same time law as the decay of its spontaneous equilibrium fluctuations. In the context of rate theory, this means the equivalence between the (macroscopic) relaxation towards equilibrium through an irreversible process and the initial small deviations from equilibrium at the microlevel. The connection between fluctuations and relaxation is again encoded in the fluctuation-dissipation theorems and correlation functions as shown by Kubo[97], Zwanzig[98], and others. Again, for rate constants, this means that the macroscopic rate constant is related to the microscopic concentration fluctuations, which are quantified through concentration autocorrelation functions,

$$\bar{n}_{\mathrm{R}}(t) - \langle n_{\mathrm{R}} \rangle \propto \langle \delta n_{\mathrm{R}}(0) \delta n_{\mathrm{R}}(t) \rangle, \tag{2}$$

where $\bar{n}_{\mathrm{R}}(t)$ is the ensemble average of the occupation of the reactant state, $n_{\mathrm{R}}$, at time $t$, $\langle \dots \rangle$ denotes thermal averaging, $\langle n_{\mathrm{R}} \rangle$ is the equilibrium average of $n_{\mathrm{R}}$, $\delta n_{\mathrm{R}}(t) = n_{\mathrm{R}}(t) - \langle n_{\mathrm{R}} \rangle$ is the fluctuation at time $t$, and $\langle \delta n_{\mathrm{R}}(0) \delta n_{\mathrm{R}}(t) \rangle$ is the equilibrium autocorrelation function.

Based on these considerations, the regression hypothesis connects the microscopic concentration fluctuations and the macroscopic relaxation towards equilibrium through

$$C_{n_{\mathrm{R}}}^{\mathrm{eq}}(t) = \frac{\langle \delta n_{\mathrm{R}}(0) \delta n_{\mathrm{R}}(t) \rangle}{\langle \delta n_{\mathrm{R}}(0) \delta n_{\mathrm{R}}(0) \rangle} = \frac{\bar{n}_{\mathrm{R}}(t) - \langle n_{\mathrm{R}} \rangle}{\bar{n}_{\mathrm{R}}(t=0) - \langle n_{\mathrm{R}} \rangle} = \exp[-t/\tau_{\mathrm{react}}], \tag{3}$$

where $\tau_{\mathrm{react}} = k_{\rightarrow} + k_{\leftarrow}$ is the reaction timescale that depends on the forward ($\rightarrow$) and backward ($\leftarrow$) rate constants.

The precise definition of the reactant concentration is established by considering the time-dependent probability of being on the "reactant side" or "product ($P$) side" of the configurational phase space of the dividing surface, as shown in Figure 2:

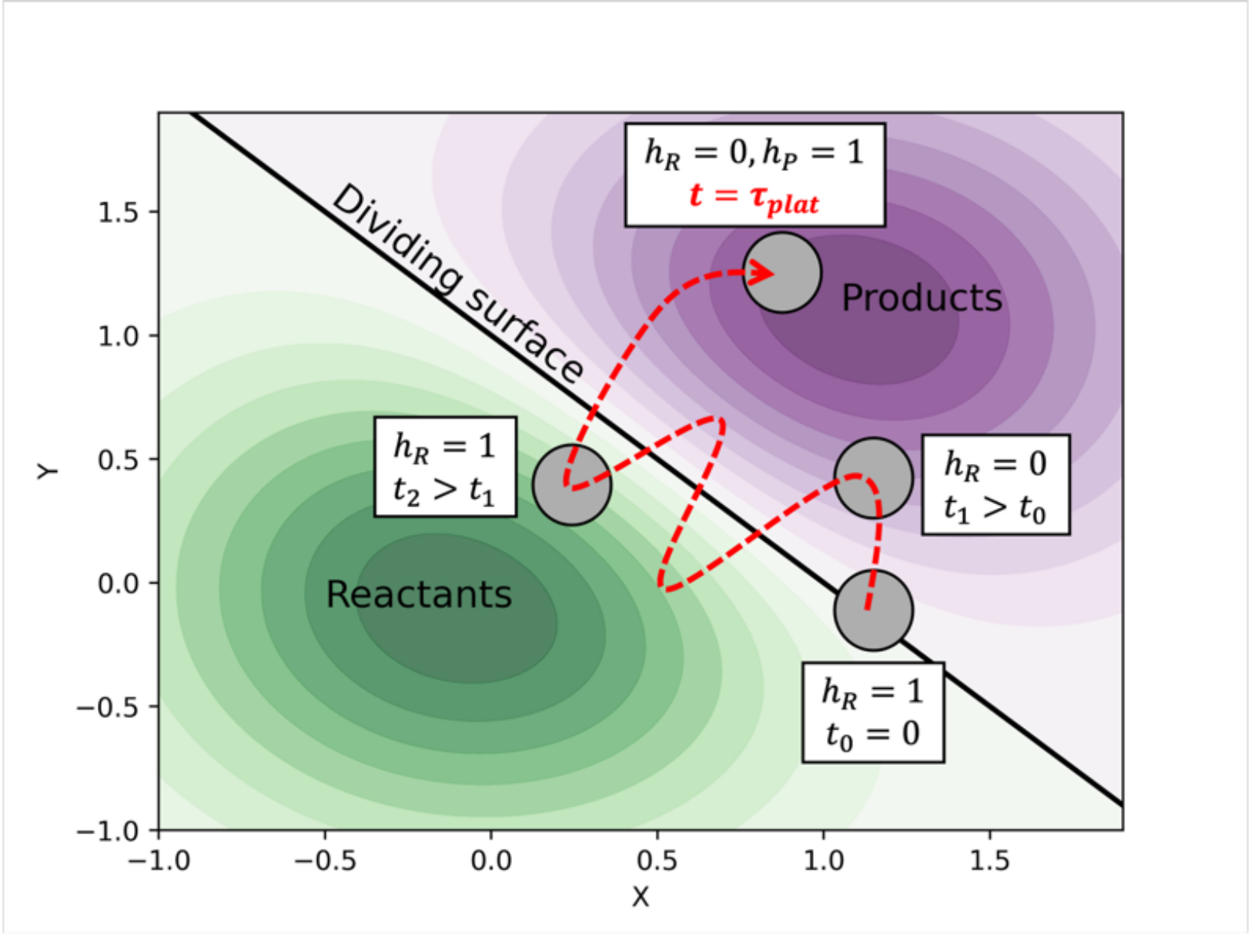

Figure 2. A schematic of a 2-dimensional XY-configurational phase space. The grey dots depict the system configuration, and the dashed red line presents the phase space trajectory. $h_{\mathrm{R}}$ and $h_{\mathrm{P}}$ denote the probability of single trajectory to be on the reactant or product side, respectively, at time when it started from the reactant side.

$$h_{\mathrm{R}}^{i}(t) = \Theta(r \leq r^{\neq}; t) = \begin{cases} 1, \boldsymbol{r} \in \boldsymbol{r}_{\mathrm{R}} \\ 0, \boldsymbol{r} \in \boldsymbol{r}_{\mathrm{P}} \end{cases}, \qquad \bar{h}_{\mathrm{R}}(t) = \sum_{i=1}^{n} h_{\mathrm{R}}^{i}(t), \tag{4}$$

where $h_{\mathrm{R}}^{i}(t)$ denotes the probability of a single trajectory to be on the reactant side at time $t$ when it started from the reactant side, $\bar{h}_{\mathrm{R}}(t)$ is the average probability that trajectory is on the reactant side at time $t$ when it started from the reactant side, $\Theta$ is a step function, $\boldsymbol{r}$ is the configurational phase space position, and $\boldsymbol{r}^{\neq}$ is the location of the dividing surface.

The rate of change of $\bar{h}_{\mathrm{P}}(t)$ is the average time-dependent probability that a trajectory initiated on the reactant sides is on the product side:

$$\frac{d\bar{h}_{\mathrm{P}}(t)}{dt} = \frac{\left\langle \delta[\boldsymbol{r}(t=0) - \boldsymbol{r}^{\neq}] \frac{d\boldsymbol{r}(t=0)}{dt} h_{\mathrm{P}}(t) \right\rangle}{Q^{\mathrm{R}}} \equiv C_{\mathrm{fs}}(t), \tag{5}$$

where $Q_{\mathrm{R}}$ is the reactant partition function and $C_{\mathrm{fs}}(t)$ is the flux-side correlation function, which measures the average probability of a trajectory starting at the dividing surface to cross to $P$ at time $t$. Time derivative of the macroscopic rate, expressed in Eqs. 2 and 3, is equal to $C_{\mathrm{fs}}(t)$

$$C_{\mathrm{fs}}(t) = \frac{d\bar{h}_{\mathrm{P}}(t)}{dt} = -\frac{d\bar{h}_{\mathrm{R}}(t)}{dt} = \frac{n_{\mathrm{R}}(t_0)}{\tau_{\mathrm{react}}} \exp\left(-\frac{t}{\tau_{\mathrm{react}}}\right). \tag{6}$$

This is, however, *not* valid for all timescales; the exponential relaxation describes long timescales and doesn't show transients at short times which are seen in $C_{\mathrm{fs}}(t)$ (see Figure 3). The macroscopic rate is valid only for timescales longer than transient relaxation of the environment where $\tau_{\mathrm{react}} \gg \tau_{\mathrm{plat}} \gg \tau_{\mathrm{env}} \rightarrow \exp\left(-\tau_{\mathrm{plat}}/\tau_{\mathrm{react}}\right) \approx 1$. Here $\tau_{\mathrm{react}} \gg \tau_{\mathrm{env}}$ means that relaxation along the reaction coordinate (see sections 2.1, 2.2, and 3.6) must be the slowest process and much slower than environment relaxation. The intermediate timescale $\tau_{\mathrm{plat}}$ corresponds to the average time it takes for the system to relax in the reactant or product region when the system starts at the dividing surface and hence allows separating the transient dynamics of the environment relaxation at short times (Figure. 3) from the reaction probability. In other words, $\tau_{\mathrm{plat}}$ presents the time that it takes for the transient dynamics to vanish, see Figure 3, and one has

$$\frac{d\bar{h}_{\mathrm{P}}\left(\tau_{\mathrm{plat}}\right)}{dt} = C_{\mathrm{fs}}\left(\tau_{\mathrm{plat}}\right) = \text{constant.} \tag{7}$$

The introduction and separation of timescales is crucial in rate theory: for the macroscopic and microscopic rate constant to be compatible, the reaction or relaxation across the dividing surface needs to take place significantly slower than the environment relaxes. This requirement will also play a substantial role in the identification of suitable reaction coordinates discussed in Section 2.1, 2.2, 2.4, and 3.6. The separation of timescales leads to Eq. 7 and after invoking the final expression connecting the microscopic fluctuations and macroscopic rate constant is given by

$$k_{\rightarrow} = \frac{d\bar{h}_{\mathrm{P}}\left(\tau_{\mathrm{plat}}\right)}{dt} = C_{\mathrm{fs}}\left(\tau_{\mathrm{plat}}\right) = \lim_{t \to \tau_{\mathrm{plat}}} \frac{\left\langle \delta[\boldsymbol{r}(t=0) - \boldsymbol{r}^{\neq}] \frac{d\boldsymbol{r}(t=0)}{dt} h_{\mathrm{P}}(t) \right\rangle}{Q_{\mathrm{R}}}. \tag{8}$$

This equation tells that the rate constant is the average flux of a trajectory starting at the dividing surface to end up in the P basin 1) after the plateau time when transients have vanished but 2) before the reaction timescale is reached (Figure 3).

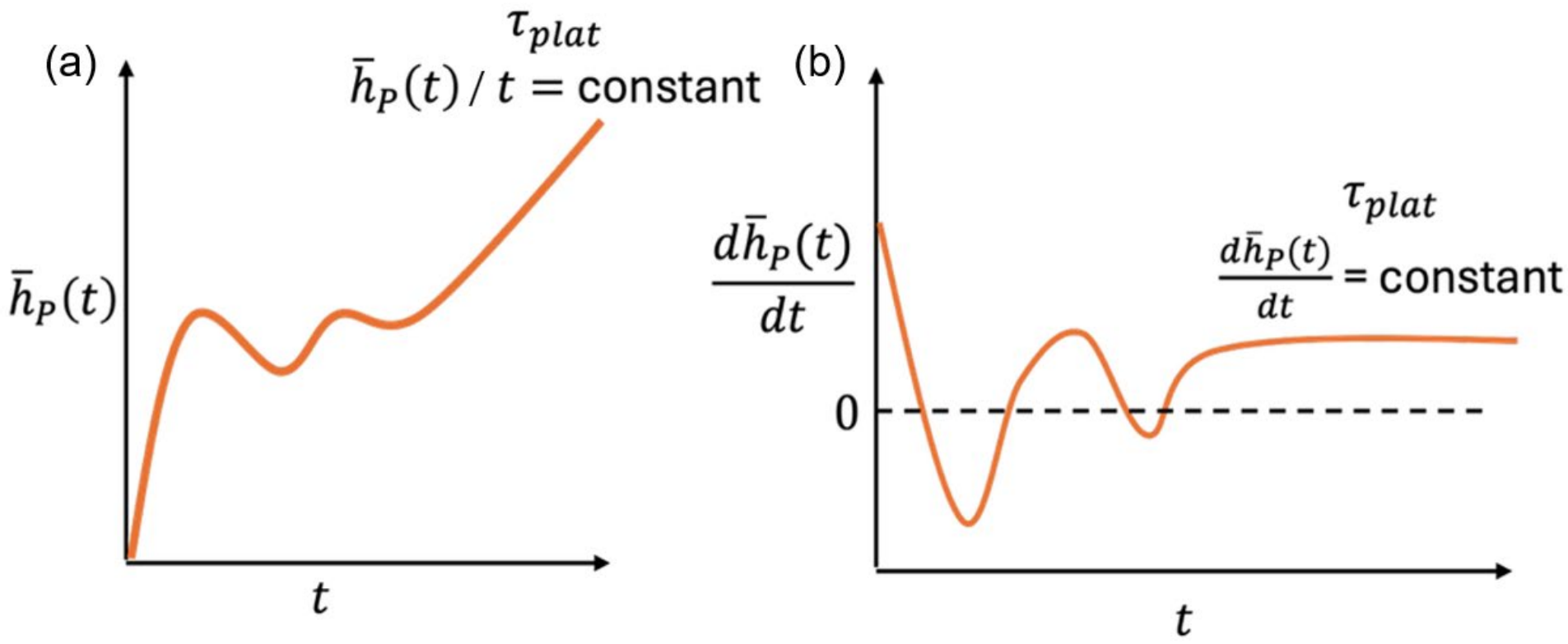

Figure 3. Schematic illustration of the dynamical evolution of (a) the averaged probability of being on the product side ($\bar{h}_{\mathrm{P}}$) and (b) its time derivative. The oscillations observed at short times in both panels correspond to the transient behaviors and environment relaxation. At longer times, the oscillations vanish, $\bar{h}_{\mathrm{P}}$ evolves linearly, and its time derivative reach a plateau. The time that it takes for the transient dynamics to vanish defines the characteristic timescale $\tau_{\mathrm{plat}}$.

The TST is obtained at the zero-time limit of $C_{\mathrm{fs}}(t)$, reflecting the initial behavior of trajectories right after crossing the dividing surface while not accounting for relaxation or recrossing dynamics,

$$k_{\rightarrow}^{\mathrm{TST}} = \lim_{t\to 0^+} \frac{\left\langle \delta[\boldsymbol{r}(t=0) - \boldsymbol{r}^{\neq}] \dfrac{d\boldsymbol{r}(t=0)}{dt} h_{\mathrm{P}}(t) \right\rangle}{Q_{\mathrm{R}}} \\ = \left\langle \frac{1}{2} v(0) \right\rangle \frac{Q^{\neq}}{Q_{\mathrm{R}}} = \left\langle \frac{1}{2} v(0) \right\rangle e^{-\beta \Delta G^{\neq}}, \tag{9}$$

where $Q^{\neq}$ is the partition function of the dividing surface, $\Delta G^{\neq}$ is the free energy difference between the reactants and the dividing surface, $\beta = 1/(k_{\mathrm{B}}T)$ , and $\left\langle \frac{1}{2} v(0) \right\rangle$ is average positive velocity (perpendicular to the dividing surface) of crossing trajectories. In an ideal dividing surface, there are no recrossing, so the TST rate (zero-time limit) equals the plateau value: $k_{\rightarrow}^{\mathrm{TST}} = k_{\rightarrow}$. In reality, trajectories can recross the dividing surface. Thus: $k_{\rightarrow}^{\mathrm{TST}} > k_{\rightarrow}$.

The general TST requires computing the full partition functions for both the initial state and the dividing surface or the corresponding free energies, which can be a formidable task. However, if simplified models for the partition functions are required, the computational cost can be significantly reduced. In ET theory, the most common approximate TSTs are the harmonic transition state theory (hTST) and the Marcus theory. In hTST, one assumes that potential energy surface (PES) at both the initial state and dividing surface are parabolic. This enables approximating the partition functions using harmonic potentials, normal mode coordinates, and the corresponding vibrational frequencies. As shown, e.g., in the supporting info of Ref.[81], the hTST rate constant is given by

$$k^{\mathrm{hTST}} = \frac{\omega_{\mathrm{n}}}{2\pi} e^{-\beta \Delta G^{\neq}_{\mathrm{harm.}}}, \tag{10}$$

where $\omega_{\mathrm{n}}$ is the effective angular frequency along the reaction coordinate in the initial state and related to the effective frequency in Eq. 1 though $\omega_{\mathrm{n}} = 2\pi\nu_{\mathrm{n}}$. $\Delta G^{\neq}_{\mathrm{harm}}$ is the free energy barrier within the harmonic approximation, which consists of the energy barrier and vibrational entropy contributions.

The simplest Marcus-like TST can be obtained from a one-dimensional, non-adiabatic hTST and further assuming that initial and final state free energy surface are given by two displaced parabolas with the same curvature. The transition state free energy is obtained from the intersection of these parabola, which leads to the Marcus equation

$$k^{\mathrm{Marcus}} = \frac{\omega_{\mathrm{n}}}{2\pi} e^{\frac{-\beta(\Delta G^0 + \lambda)^2}{4\lambda}}, \tag{11}$$

where $\Delta G^0$ and $\lambda$ are the reaction free energy and the reorganization energy, respectively.

Some notes on the above rate theory and TST are in order:

1) Only thermodynamics quantities enter TST – it does not depend on any dynamic or time-dependent quantities of the system. This means that "only" comprehensive sampling of the initial state and dividing surface is needed, and the free energy difference $\Delta G^{\neq}$ defines the rate.

2) TST assumes that all trajectories that are at the dividing surface and moving towards $P$ at time $t_0$ end up as products. This issue can be mended by computing dynamic corrections

$$\kappa_{\mathrm{dyn}} = \frac{k_{\rightarrow}}{k_{\rightarrow}^{\mathrm{TST}}}, \tag{12}$$

which measures how many trajectories starting at the dividing surface end on the product side at $\tau_{\mathrm{plat}}$. $\kappa_{\mathrm{dyn}}$ *does* depend on the system dynamics and can be computed by simulating $C_{\mathrm{fs}}(\tau_{\mathrm{plat}})$ from molecular dynamics simulations or from analytical models such Kramers-Grote-Hynes theory (see Section 6).

3) The timescale separation is pivotal to the rate theory: the reaction timescale along the reaction coordinate must be significantly longer than the environment relaxation. In practice, this sets constraints on the reaction coordinate (sections 2.1, 2.2, 2.4, and 3.6) and means that $\Delta G^{\neq}$ must be large enough to make the reaction much slower than the environment relaxation.

4) The general rate theory is not restricted to any particular reaction coordinate; the only requirement is that the timescale to cross the dividing surface must be significantly longer than the environment relaxation. This means that the chosen reaction coordinate must correspond to the slowest dynamics of the system.

5) Because dynamics across the dividing surface are much slower than other processes of the system, reactions are rare events and enhanced sampling along the reaction coordinate is required to drive the reaction away from the reactant state over the dividing surface.

6) The above treatment is based on classical statistical mechanics, and the rate equations do not include quantum or non-adiabatic effects. However, the fully quantum mechanical rate theory by Miller[99–101] follows the classical treatment very closely and the quantum and non-adiabaticity corrections to the classical (TST) rate constant can be included as a prefactor

$$\kappa_{\mathrm{el}} = \frac{k_{\rightarrow}^{\mathrm{quant}}}{k_{\rightarrow}}. \tag{13}$$

In practice, formulating and computing quantum and non-adiabaticity corrections to the classical rate can be included in several ways but this is a very difficult problem, and some possible approaches in electrochemical systems have been discussed in Ref.[102]. Non-adiabaticity corrections are discussed in Section 6.

7) The grand canonical rate theory is developed in an analogous manner and the only notable difference when moving to the open systems is the appearance of the timescale for diffusion or transfer rate of species in the system. The relevant timescales have been detailed in Ref.[81]. When the system dynamics due to electrolyte vibration, rotation, and translation are fast compared to the reaction, and the grand canonical rate theory is obtained by changing canonical partition functions and free energies with their grand canonical counterparts. If effects from dynamic are important (section 6), the dynamic corrections should be computed consistently in the chosen ensemble.

8) Marcus theory is a specific form of the classical rate theory as it fulfills all the requirements discussed above. In the absence of dynamic, quantum, and non-adiabatic corrections, Marcus theory is a form of classical TST.

### 2.1. Reaction coordinates of electrochemical systems

The discussion in the previous section shows that applying the TST or other rate theories requires the identification of a suitable reaction coordinate (RC, see sections 2.2, 2.4, and 3.6 for a detailed discussion on RCs). The choice of an RC is not unique but it needs to fulfill two key criteria[103–105]: the RC must 1) be a low-dimensional presentation or projection of the degrees of freedom (DOF) describing the advancement of a reaction and 2) the dynamics along the RC should be slower than along any other coordinates or DOF relevant to the reaction. Together, these two requirements lead to the separation of timescales for relaxation along the RC and other DOFs. For TST, this means that all other DOFs are in equilibrium with the RC and can be sampled according to Boltzmann statistics. While atomic coordinates are often used to define the RC and the DOF, it should be noted that this is not necessary; both the RC and other DOFs may also be, e.g., the solvent polarization density in continuum models, or an energetic quantity as discussed below.

When such a RC is identified, the TST rate constant may be computed by carrying out non-equilibrium sampling of phase space along the RC and equilibrium sampling along all other DOFs. The partition function along the RC is given by

$$Q(\boldsymbol{s}) = C \int \exp[-\beta H(\boldsymbol{x})]\, \delta\big(q(\boldsymbol{x}-\boldsymbol{s})\big)\, d\boldsymbol{x}, \tag{14}$$

where $\boldsymbol{s}$ is the RC, $\boldsymbol{x}$ the system DOFs, $C$ is a normalization constant, and $H(\boldsymbol{x})$ is the Hamiltonian which can be described either using an atomistic or continuum model. $q$ is a function that allows using also non-linear functions of the DOFs as the reaction coordinate. The full free energy obtained with Hamiltonian $H$, i.e., Helmholtz free energy along the RC is given by

$$F(s) = -k_{\mathrm{B}}T \ln[Q(\boldsymbol{s})]. \tag{15}$$

The probability to be at a particular $\boldsymbol{s}$ is given by

$$p(\boldsymbol{s}) = \frac{\int \exp[-\beta H(\boldsymbol{x})]\, \delta\big(q(\boldsymbol{x}-\boldsymbol{s})\big)\, d\boldsymbol{x}}{\int \exp[-\beta H(\boldsymbol{x})]\ d\boldsymbol{x}} = \big\langle \delta\big(q(\boldsymbol{x}-\boldsymbol{s})\big)\big\rangle_{H}, \tag{16}$$

where the last equation denotes the thermal averaging carried out with the Hamiltonian $H$. The probability is connected to the Gibbs free energy by

$$G(\boldsymbol{s}) = F(\boldsymbol{s}) - k_{\mathrm{B}}T\ln\big(p(\boldsymbol{s})\big)$$
$$= -k_{\mathrm{B}}T\ln\left(\int \exp\big(-\beta H(\boldsymbol{s})\big)\ d\boldsymbol{s}\right) - k_{\mathrm{B}}T\ln\big(p(\boldsymbol{s})\big). \tag{17}$$

The previous equations indicate that the partition functions, free energies, probabilities, and the TST rate constant along the RC are obtained by fixing the RC and by carrying out either equilibrium sampling of the other DOFs in explicit or implicit models, where the solvent is described either atomically or through continuum model. The explicitly atomistic model is mainly used in simulations while the implicit continuum models, such dielectric continuum or non-equilibrium polarization models, can be used in both simulations and analytical theories.

## 2.2. Explicit models and reactions coordinates

In explicit atomistic simulations, $Q(\boldsymbol{s})$ is constructed by explicitly sampling the configurational phase space. Because of the separation of timescales between the RC and the other DOFs, equilibrium sampling along the RC is not feasible; as relaxation along RC is the slowest process of the system, reactions are usually rare events, which means that reactions happen on timescale much larger than e.g. vibrations. For instance, the O-H bond vibration time sets the maximum time step that can be used in MD simulations is typically in the order of a femtosecond. When this time step is compared with the reaction timescale of ~100 ns, which corresponds to a small barrier of 0.35 eV, a single reaction would be observed once every ~$10^8$ MD steps. Even if such simulations could be performed, they would be rather uninformative. A cleverer approach is to take advantage of the times scale separation between the RC and other coordinates of the system and to use enhanced sampling methods to drive the system away from a local minimum into another one, and many efficient algorithms[106] have been developed to achieve this, and in Section 3.6 we discuss the umbrella sampling approach widely used in ET studies.

In explicit models, the RC needs to depend on the atomic coordinates, but this dependency can be either direct or indirect. The most used RCs are direct geometric quantities, such as bond lengths or angles, which work well if the identification of e.g. a reacting bond is straightforward; an example could be the adsorption of a species on a surface or bond-breaking ET. Direct RCs can also combine several geometric parameters into a single variable; an example is the use of coordination numbers to describe e.g. (de)solvation or hydrogen bonding effects on reactions. An indirect RC can, for example, be an energetic quantity or solvent polarization both of which do depend on geometry but only indirectly. Such RCs have been widely used in describing electron transfer reactions where the reaction no single bond length, angle, coordination number, or any other small number of nuclear coordinates can describe the advancement of a reaction as these reactions are driven by the overall reorganization of the reaction medium involving 3N coordinates. An example of such RC is the energy gap coordinate, discussed in Section 3.6, which describes how the reaction environment needs to reorganize itself for the electron transfer to take place iso-energetically. Technically, the energy gap coordinate is a one-dimensional projection of all system coordinates into a single energy value and while it does depend on the geometry of the system, the dependency is very indirect as many geometries may lead to the same energy gap.

## 2.3. Implicit models and reaction coordinates

In implicit models the RC does not depend on any explicit atomistic or nuclear coordinates but rather a macroscopic parameter that describes the advancement of the reaction. In practice, implicit models use free energy functionals that depend on macroscopic order parameters to describe the advancement of reaction. Some classical examples include the Landau theory of phase transitions where an order parameter describes both the advancement of the phase transition and related free energies. In ET theory, the use of implicit models is wide-spread and e.g. the Marcus theory was originally formulated using an implicit dielectric continuum model for the solvent and the non-equilibrium solvent polarization as the reaction coordinate. This is discussed in detail in Section 3.

### 2.4. Choice of reaction coordinates

While the timescale separation between the RC and the other coordinates is deeply ingrained in rate theory and free energy sampling, it needs to be re-emphasized that defining or finding RCs is not easy or even unique. However, choice of the RC is critical in practice: it not only defines the efficiency of the free energy calculations but it also dictates what we learn about the (electro)chemical reaction mechanisms[105]. In traditional theories of ET, implicit models, and the compatible EVB simulations discussed in Section 3.6, a collective nuclear reorganization coordinate or equivalently the energy gap is chosen as the RC: this choice informs how *all* nuclear coordinates together drive the ET. The participation of particular nuclear coordinates, such as bond lengths, can be analyzed a *posteriori* from simulations[107]. However, explicit atomistic models and DFT studies in particular often use only a *few* nuclear coordinates as the RC and to drive the reaction: in this case, the reaction is understood from the perspective of how the free energy depends on changes of few nuclear coordinates or bond length while the impact of the collective nuclear reorganization cannot be readily addressed. It is also possible to combine the total nuclear reorganization coordinate and some specific nuclear coordinates such as bond lengths to address e.g. bond-breaking or ion-coupled ET. Finally, the RCs may also include nuclei and treated quantum mechanically, and this is most easily achieved using path integral centroid as a proxy for position[108].

A prime example on how the choice of reaction coordinate influences our understanding of electrochemical ET is the comparison of reaction kinetics computed using Marcus-like approaches using the reorganization coordinate and explicitly potential-dependent grand canonical ensemble DFT (GCE-DFT) using specific bond lengths. In Marcus-like theories, the pivotal idea that the collective solvent/medium/nuclear reorganization or polarization drives electron transfer and is the slowest relevant degree of freedom in the system. This sets the collective nuclear reorganization as the RC while other coordinates, such the bond and angles of the solvent, are taken to be in equilibrium along this RC. When the coupling between reorganization and electron transfer is taken to be linear, i.e., when the linear response theory (or equivalently linear coupling[109–111]) between reorganization and ET is valid, the iconic Marcus rate theory is obtained, and the reaction rate has a parabolic dependency on the reaction and reorganization free energies. An analogous parabolic relation between the reaction energy and kinetics is also often obtained through typical explicit GCE-DFT simulations[112,113]: the GCE-DFT simulations have not, however, used the collective reorganization coordinate as the RC but rather a few nuclear coordinates such as bond lengths as the RC. This indicates that *while similar trends may result from the use of two different RCs, the understanding may be very different*; Marcus-like theories predict that the collective reorganization drives the reaction while GCE-DFT-MD simulations would predict changes in some specific bonds to define the kinetics. Hence, the choice of the RC dictates the insight on ET kinetics. However, GCE-DFT or explicit atomistic simulations are not restricted to using specific nuclear coordinates as the RC, and we have recently[81,114,115] shown that also (GCE-)DFT methods studies can utilize the reorganization coordinate as the RC when a diabatic DFT model is used; this is discussed in Section 3.6 and it provides a way to unify the description of implicit and explicit models of e.g. electron transfer kinetics.

## 2.5. Timescales in electrochemistry

The discussion on reaction coordinates is directly linked with the relevant timescales of the systems. In principle, theoretical and computational methods can cover all relevant timescales but in practical simulations this is not always possible.

Consider, for instance, the EDL effects on ET. First, it is well-established that resolving the EDL requires a thermodynamic treatment and that time does not explicitly appear in any thermodynamic expectation values and quantities such as free energies and structures. In practice, the simulations of (EDL) thermodynamics can be achieved through various statistical liquid state theories, including e.g. with implicit continuum models, classical DFT, and integral equation methods or with explicit atomistic models such as molecular dynamics simulations. Furthermore, modeling thermodynamics under common electrochemical reaction conditions of constant potential and electrolyte activity is achieved by using the grand canonical ensemble (GCE) which allows fixing the electron and electrolyte (electro)chemical potentials as done in experiments. Altogether, GCE theory of electrochemical thermodynamics is exact[116,117] and widely accepted[118]. Implicit EDL models naturally treat the EDL at thermodynamic equilibrium within the GCE and include all EDL timescales. Explicit EDL models on the other hand struggle to treat all timescales as sampling slow degrees of freedom requires very extensive simulations and because realizing GCE-MD simulations for nuclei is prohibitively difficult. Despite these numerical and computational challenges, the GCE treatment of electrochemical thermodynamics is well-established and the timescale issue does not, in principle, influence thermodynamic quantities.

The same is not true for reaction kinetics where the treatment of timescales is pivotal and where opposing views have been expressed on the validity of GCE in simulating reaction kinetics[119,120]. As one of the present authors has emphasized recently[120], resolving the utility of GCE in addressing electrochemical ET requires careful consideration of the system timescales as this dictates in which ensemble the kinetics simulations should be performed. While this issue has been addressed in recent works, we consider that it is not completely resolved yet and the discussion on this crucial topic is continued herein.

The disagreements[119] stem largely from the varying time- and length scales in electrochemistry: the GCE dictates that the electrode potential and the electrolyte activity within the EDL must remain in equilibrium with the electron reservoir (potentiostat) and the bulk electrolyte, respectively, which in turn means that potentiostat and electrolyte relaxation timescale must be smaller than that of the studied reaction. In other words, simulating reaction rates within GCE is valid only when the reaction rate is slower than the potentiostat or electrolyte equilibration. Schmickler and Santos[119] correctly point out that local EDL relaxation time after ET is ~1 ns[121], which is much faster than the global potentiostat response time of ~100 ns. For these reasons, Schmickler and Santos argue that because simulation timescales that can be achieved are less than 1 ns, atomistic simulations of reaction kinetics should not be carried out within GCE. These EDL and potentiostat relaxation times can be compared to reaction timescales and converted to the corresponding reaction barriers by making use of the correspondence between rate constants and relaxation times: $\tau_{\mathrm{r}} = 1/k$. By using a TST rate constant $k = \frac{k_{\mathrm{B}}T}{h} \exp\left(-\frac{\Delta G^{\neq}}{k_{\mathrm{B}}T}\right)$, at room temperature, a relaxation time of 1 ns corresponds barrier of 0.25eV, 10 ns to 0.28 eV, and 100 ns to 0.35 eV, all which can be considered rather small barriers and fast kinetics at room temperature. Importantly, if the reaction barrier is larger than 0.35 eV the electrode potential remains in both local and global equilibrium and therefore constant during a reaction and the GCE is expected to applicable for describing the ET kinetics. For small enough barriers or slow system dynamics, the simulation of ET rates needs to account for non-ergodicity—this is discussed in Section 6.2.

In this context, it must be emphasized that both the timescale and rate constant must be considered as ensemble averages of reactive events from an initial state to a final state, not the rate of a single reactive event of e.g. electron tunnelling at the transition state. In fact, the electron tunnelling at the transition state can be considered instantaneous and the relaxation rate to the final (equilibrium) state is indeed controlled by (solvent) dynamics characterized by the reorganization time $\tau_s$; this relaxation following ET was studied by Amatore *et al*[121], who showed that the EDL relaxes within $\sim$1 ns after ET. It should, however, be noted that this relaxation timescale is not indicative of the *reaction* timescale computed from the rate constant; *the reaction rate constant from the initial to the final state must also account for the timescale of reaching the transition state where the instantaneous tunnelling and following relaxation can take place*. If the timescale is must larger than the EDL relaxation timescale, i.e., $\tau_r \gg \tau_s$, the rate constant depends mostly on the reaction barrier and can be described through TST while the relaxation dynamics control the prefactor, as discussed in Sections 2 and 6. For reactions with a barrier above $\sim$0.3 eV, the EDL relaxation and potentiostat can be considered at thermal equilibrium on the reaction timescale and that GCE can be used for describing the reaction kinetics.

Further insight on whether the electrode potential remains constant during an electrochemical reaction can be obtained by analyzing the reaction turn-over-frequency (TOF), which measures how many reactions take place at a given active site in a second. To our knowledge, the highest TOF reported for an electrocatalysis is ~$10^4$ / (site*s) as observed for acidic hydrogen evolution reaction (HER) on Pt nanoparticles at high overpotentials[122]. While HER consists of two PCET steps, it facilitates a qualitative discussion on how often a very fast reaction takes place on a single active site. As a typical surface model used in density functional theoretical (DFT) calculations consists of ~10 HER active sites, the TOF corresponds to ~$10^5$ reactions / (cell*s) in DFT cell. Therefore, a single HER reaction in a DFT cell takes place once every ~1 $\mu s$ on average. This indicates that even macroscopically very fast and frequent electrocatalytic reactions 1) are rare events at the microscopic scale relevant to atomistic simulations and 2) take ~100 times longer to occur in a DFT cell -sized system than it takes for potentiostat to relax to a constant potential. These points indicate that macroscopically fast reactions are slower than the potential equilibration time and that GCE provides a valid description of their electrochemical kinetics.

It should nevertheless be noted at the small time and length scales used in DFT simulations, notable fluctuations around the average electrode potential take place during MD simulations. The potential fluctuations during constant potential MD are analogous to the temperature or pressure fluctuations in constant temperature or pressure MD. In all these cases the fluctuations are due to the natural response or equilibration time of the system[123] and can be correctly treated by a careful choice of the potentio-, thermo-, or barostat algorithm, which does not alter the natural dynamics of the system. In constant potential MD the potential fluctuations are enforced through the fluctuation-dissipation theorem[124], which links the fluctuations with the *microscopic* potentiostat timescale and capacitance of the system, and which can be implemented in computational potentiostats. It is important to note that the fluctuations impact only the instantaneous values and their variance; the fluctuations do not impact the system dynamics or thermodynamics, or invalidate the applicability of GCE, when potentiostat accounts for them correctly.

Finally, it should also be pointed that if the barrier is low enough and the reaction time is comparable to ($\tau_{\mathrm{react}} \sim \tau_{\mathrm{env}}$) or smaller than the environment relaxation time, the assumptions of transition state theory are no longer valid and TST should not be used for computing the reaction rate constant. This in turn means that one must explicitly simulate and sample the system dynamics, not only the free energy, to compute the reaction rate as discussed in Section 6.

## 3. Marcus theory

In this section, we focus on the formulation of Marcus theory for ET between an oxidized state (diabatic state $k$) and a reduced state (diabatic state $a$), as defined in Section 1.2.2. To this end, we first consider the thermal fluctuations of the nuclear configuration, in which the inner-sphere nuclear motions are treated as a set of harmonic vibrational modes, whereas the outer-sphere solvent is described as a dielectric continuum. The solvent configurational fluctuations are implicitly reflected in fluctuations of the slow or inertial polarization, and the resulting free-energy fluctuations are treated using non-equilibrium polarization theory in Section 3.2. Two multi-dimensional diabatic FESs for the oxidized and reduced states can then be obtained. With the constraint of energy conservation, the intersection of the two diabatic FESs with minimized activation energy in the non-adiabatic ET regime can be identified, and an effective nuclear coordinate can be introduced. Finally, we discuss how the Marcus theory can be parametrized using explicit MD simulations.

### 3.1. Thermal fluctuations of the nuclear configuration

As discussed in Section 1.2, the FESs are fluctuated by both the nuclear coordinates or DOFs of the inner-sphere redox species and those of the outer-sphere solvent. The motions along these nuclear coordinates can be combined into a set of nuclear modes. For the inner sphere, these modes, e.g., the low-frequency metal-ligand stretching modes, are discrete in frequency and are specific to a given molecular structure, whereas for the outer-sphere solvent they form a continuum of frequencies. Accordingly, the inner sphere modes are treated explicitly, while the outer-sphere modes are treated using continuum theory in the following. It should be noted that solvent molecules in aqua-complexes, e.g., $[\mathrm{Fe(H_2O)_6}]^{3+}$, are considered part of the inner sphere, as they are tightly coordinated to the metal center.

For the inner-sphere modes, we consider the simplest case, in which the nuclear modes remain unchanged during ET and only small displacements about the equilibrium normal coordinates occur. Under this assumption, the inner-sphere free energies in the oxidized and reduced states, $G_{\mathrm{ox}}^{\mathrm{in}}$ and $G_{\mathrm{red}}^{\mathrm{in}}$, can be represented as sums of displaced harmonic potentials about their respective equilibrium normal coordinates, i.e.,

$$G_{\mathrm{ox}}^{\mathrm{in}}(\{Q_j\}) = G_{\mathrm{ox}}^{\mathrm{in,eq}} + \frac{1}{2}\sum_j \mu_j \omega_{j,\mathrm{ox}}^2 \left(Q_j - Q_{j,\mathrm{ox}}\right)^2, \tag{18}$$

$$G_{\mathrm{red}}^{\mathrm{in}}(\{Q_j\}) = G_{\mathrm{red}}^{\mathrm{in,eq}} + \frac{1}{2}\sum_j \mu_j \omega_{j,\mathrm{red}}^2 \left(Q_j - Q_{j,\mathrm{red}}\right)^2, \tag{19}$$

where $\mu_j$ and $Q_j$ are the effective mass and normal coordinate of the $j_{\mathrm{th}}$ nuclear mode, respectively; $\omega_{j,\mathrm{ox}}$ and $\omega_{j,\mathrm{red}}$ are the corresponding vibrational frequencies in the oxidized and reduced states; and $Q_{j,\mathrm{ox}}$ and $Q_{j,\mathrm{red}}$ denote the equilibrium normal coordinates. The terms, $G_{\mathrm{ox}}^{\mathrm{in,eq}}$ and $G_{\mathrm{red}}^{\mathrm{in,eq}}$, denote the inner-sphere free energies at the equilibrium nuclear configurations of the oxidized and reduced states, respectively. As mentioned, these nuclear modes refer specifically to low-frequency modes that can be separated from the electronic motion. The above considerations are provided for illustrative purposes in the formulation of the theory. In more general cases, the nuclear modes may also change; for example, a Jahn-Teller induced symmetry change occurs when $[\mathrm{Cr(H_2O)_6}]^{3+}$ is converted to $[\mathrm{Cr(H_2O)_6}]^{2+}$. In some cases, anharmonic potentials can be expected to be important, e.g., for the nuclear mode associated with bond dissociation in the bond-breaking ET. This will be discussed in Section 5.3.

For the outer sphere, in response to the electric field generated by the redox species, the solvent molecules form a net dipole distribution around the redox species, effectively screening the electrostatic field and thereby reducing the electrostatic energy of the system, a phenomenon known as polarization. The polarization can be initially classified into two categories based on the response microscopic mechanisms associated with the electronic motion and nuclear motion of solvent molecules, denoted as $\boldsymbol{P}^{\mathrm{e}}$ and $\boldsymbol{P}^{\mathrm{n}}$, respectively. The former, arising from the displacement of electron cloud relative to the nuclei of solvent molecules, occurs on a timescale of $10^{-16}\sim10^{-15}$ s, so that it responds almost instantaneously to charge redistribution in ET. The latter involves the response of solvent molecules through reorientation of the entire molecule (orientational polarization) and distortion of its internal structure (intramolecular polarization) in an external field, occurring on a time scale of $10^{-14}\sim10^{-10}$ s, and typically lags far behind the nearly instant charge redistribution associated with ET. Hence, we can refer to $\boldsymbol{P}^{\mathrm{e}}$ as the fast polarization, while $\boldsymbol{P}^{\mathrm{n}}$ can be termed slow or inertial polarization. The slow polarization is linked to the solvent (nuclear) configuration, defined by a set of solvent nuclear coordinates that fluctuates due to the thermal motions of the solvent molecules. Fluctuations in the slow polarization modify the electrostatic interactions with the redox species and therefore lead to fluctuations in the solvation free energy. It is also important to note that the timescale difference between large timescale separation between $\boldsymbol{P}^{\mathrm{e}}$ and $\boldsymbol{P}^{\mathrm{n}}$ makes $\boldsymbol{P}^{\mathrm{n}}$ a good outer-sphere or solvent reaction coordinate for ET kinetics according to the discussion in Section 2.1.

### 3.2. Non-equilibrium polarization theory

In this subsection, we will derive the fluctuating solvation free energy of the redox species using non-equilibrium polarization theory developed by Marcus[2]. As discussed, the solvation free energy is expected to fluctuate as the solvent nuclear coordinates deviates from its equilibrium nuclear coordinates due to thermal motion. We consider a dielectric medium with a charge distribution, $\varrho_{\mathrm{ox}}$, in the oxidized state. In principle, the charge distribution in the oxidized state comprises both the charge on the oxidized species and that on the electrode surface (see Ref.[125]). However, considering that ET is a local phenomenon near the metal surface, its effect on the charging state of the electrode surface before and after an electron transfer is negligible and there is only significant charge redistribution occurs only on the redox species. Another way to view this is that ET is a rare event and the electrode surface charge will on average remain unchanged. Therefore, when considering ET between the electrode surface and redox species, we only consider $\varrho_{\mathrm{ox}}$ as the charge distribution in the oxidized state, while the electrode surface charge is held fixed and enters implicitly through its influence on the solvent's dielectric properties. However, the image charge induced by the redox species on the electrode surface should be included in the charge density distributions of both the oxidized and reduced states, as it adjusts instantaneously to changes in the charge distribution of the redox species during the ET process. The image charge can modulate the electronic energy levels of the redox species as well as influence the solvent reorganization process through electrostatic interactions. In computational approaches that explicitly treat metal electrons, either through Kohn-Sham DFT[41] or orbital-free DFT[37,40], the electrostatic interactions between the redox species and the metal electron are accounted for directly, and no additional treatment of image-charge effects is required. The electric displacement $\boldsymbol{D}_{\mathrm{ox}}$ corresponding to $\varrho_{\mathrm{ox}}$ is uniquely determined by the fundamental electrostatic relation,

$$\nabla\cdot\boldsymbol{D}_{\mathrm{ox}}=\varrho_{\mathrm{ox}}. \tag{20}$$

At the equilibrium nuclear coordinates, the electrostatic free energy of the system achieves its minimum with the corresponding fast polarization $\boldsymbol{P}_{\mathrm{ox}}^{\mathrm{e}}$ and slow polarization $\boldsymbol{P}_{\mathrm{ox}}^{\mathrm{n}}$, represented by point A in Figure 4. As the solvent deviates from its equilibrium configuration, the fast polarization remains unchanged at $\boldsymbol{P}_{\mathrm{ox}}^{\mathrm{e}}$, while the slow polarization $\boldsymbol{P}^{\mathrm{n}}$ varies, pushing the system to a non-equilibrium state with higher electrostatic free energy. Point C in Figure 4 represents such a non-equilibrium state. Thus, the following electrostatic relation holds for all states in Figure 4,

$$\boldsymbol{D}_{\mathrm{ox}} = \epsilon_0 \boldsymbol{\mathcal{E}} + \boldsymbol{P}_{\mathrm{ox}}^{\mathrm{e}} + \boldsymbol{P}^{\mathrm{n}}, \tag{21}$$

where $\epsilon_0$ is the vacuum permittivity, $\boldsymbol{\mathcal{E}}$ the electric field. In equilibrium state A, we have $\boldsymbol{\mathcal{E}} = \boldsymbol{\mathcal{E}}_{\mathrm{ox}}$ and $\boldsymbol{P}^{\mathrm{n}} = \boldsymbol{P}_{\mathrm{ox}}^{\mathrm{n}}$. If the solvent fluctuations are small, we can assume that the linear response of polarization to the electric field holds: this linear-response assumption is a key step for all Marcus-type theories in both macroscopic, implicit[109] or microscopic, explicit[110] models. Under this assumption, the solvent can be equivalently modeled as a set of harmonic oscillators, based on the fact that its dynamical behavior is identical to that of a linear system[126,127]. For the fast and slow polarization responses, they are respectively given by

$$\boldsymbol{P}_{\mathrm{ox}}^{\mathrm{e}} = \chi^{\mathrm{e}} \varepsilon_0 \boldsymbol{\mathcal{E}}_{\mathrm{ox}}, \tag{22}$$

$$\boldsymbol{P}_{\mathrm{ox}}^{\mathrm{n}} = \chi^{\mathrm{n}} \varepsilon_0 \boldsymbol{\mathcal{E}}_{\mathrm{ox}}, \tag{23}$$

where $\chi^{\mathrm{e}}$ and $\chi^{\mathrm{n}}$ are the electrical polarizability of fast and slow polarization modes, respectively. Based on Eqs. 21-23, we can obtain the following relations for the equilibrium state,

$$\boldsymbol{D}_{\mathrm{ox}} = \varepsilon_{\mathrm{s}} \boldsymbol{\mathcal{E}}_{\mathrm{ox}}, \tag{24}$$

$$\boldsymbol{P}_{\mathrm{ox}}^{\mathrm{e}} = \left(\frac{\varepsilon_\infty - \varepsilon_0}{\varepsilon_{\mathrm{s}}}\right) \boldsymbol{D}_{\mathrm{ox}}, \tag{25}$$

$$\boldsymbol{P}_{\mathrm{ox}}^{\mathrm{n}} = \left(\frac{\varepsilon_{\mathrm{s}} - \varepsilon_\infty}{\varepsilon_{\mathrm{s}}}\right) \boldsymbol{D}_{\mathrm{ox}}, \tag{26}$$

with the optical dielectric permittivity $\varepsilon_\infty$,

$$\varepsilon_\infty = (1 + \chi^{\mathrm{e}}) \varepsilon_0, \tag{27}$$

and the static dielectric permittivity $\varepsilon_{\mathrm{s}}$,

$$\varepsilon_{\mathrm{s}} = (1 + \chi^{\mathrm{e}} + \chi^{\mathrm{n}}) \varepsilon_0. \tag{28}$$

The corresponding relations in Eqs. 24-26 hold not only for equilibrium state A, but for any non-equilibrium states with the electric field $\boldsymbol{\mathcal{E}}$, electric displacement $\boldsymbol{D}$ and polarization responses $\boldsymbol{P}^{\mathrm{e}}$ and $\boldsymbol{P}^{\mathrm{n}}$.

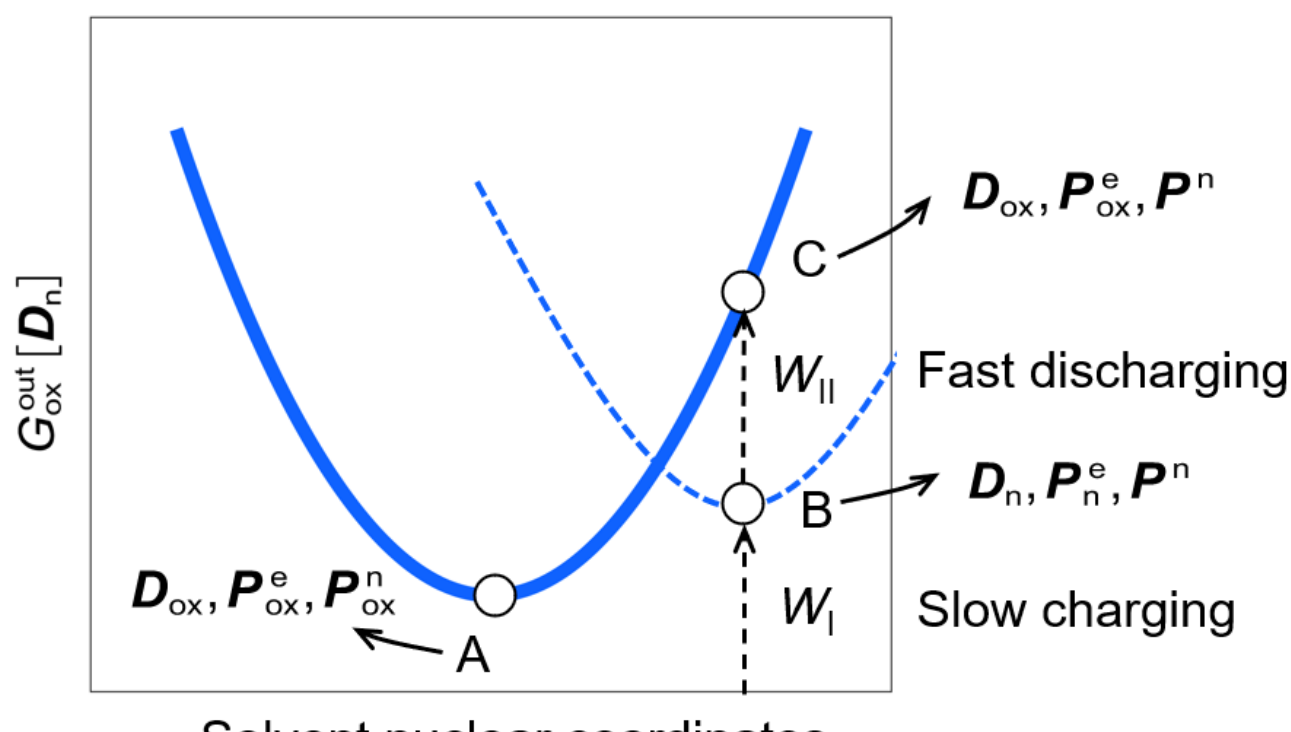

Figure 4. Thermal fluctuations of free energy for the dielectric medium in the charging state $\boldsymbol{D}_{\mathrm{ox}}$ or $\rho_{\mathrm{ox}}$. In equilibrium state A, corresponding to the system at equilibrium solvent nuclear coordinates, the fast polarization $\boldsymbol{P}_{\mathrm{ox}}^{\mathrm{e}}$ and slow polarization $\boldsymbol{P}_{\mathrm{ox}}^{\mathrm{n}}$ are in equilibrium with $\boldsymbol{D}_{\mathrm{ox}}$.

Due to thermal motions of solvent molecules, the system may fluctuate into a non-equilibrium state C with the non-equilibrium solvent configuration characterized by the slow polarization $\boldsymbol{P}^{\mathrm{n}}$. To determine the electrostatic free energy in the non-equilibrium state, the system can be constructed by two consecutive charging-discharging steps: first, a slow charging to state B where the electric displacement $\boldsymbol{D}_{\mathrm{n}}$ is in equilibrium with $\boldsymbol{P}^{\mathrm{n}}$, then followed by a fast discharging to state C. State B corresponds to the minimum of the electrostatic free energy (dashed curve) in the fictious charging state $\boldsymbol{D}_{\mathrm{n}}$. The electrostatic free energy in state C is then the sum of the reversible work $W_{\mathrm{I}}$ and $W_{\mathrm{II}}$ from the two steps. For any state with specific solvent configuration, there exists a fictitious charging state $\boldsymbol{D}_{\mathrm{n}}$ in equilibrium with it. The electrostatic free energy, $G_{\mathrm{ox}}^{\mathrm{out}}$, can be expressed as a functional of $\boldsymbol{D}_{\mathrm{n}}$.

To obtain the electrostatic free energy in the non-equilibrium state C, we need to evaluate the reversible work required to charge the dielectric medium to that state. For this, we can conceptualize the process as two consecutive steps: first, a slow charging process to state B where the electric displacement $\boldsymbol{D}_{\mathrm{n}}$ is in equilibrium with $\boldsymbol{P}^{\mathrm{n}}$, then followed by a fast discharging to state C, as shown in Figure 4. State B corresponds to the minimum of the electrostatic free energy (dashed curve in Figure 4) in the fictious charging state $\boldsymbol{D}_{\mathrm{n}}$. The electrostatic free energy in state C is then the sum of the reversible work from these two steps.

The reversible work required to charge an electrostatic system from state $i$ to state $j$ can be given by

$$W = \int \left( \int_{\boldsymbol{D}_i}^{\boldsymbol{D}_j} \boldsymbol{\mathcal{E}} \cdot \delta \boldsymbol{D} \right) dV, \tag{29}$$

where $\boldsymbol{D}_i$ and $\boldsymbol{D}_j$ are the electric displacements in states $i$ and $j$, respectively, and $V$ is the spatial volume. This expression holds for a quasistatic charging process in which the electric field varies linearly with the electric displacement. A detailed derivation is provided in Appendix 10.1. The first charging step is sufficiently slow to allow both the fast and slow polarization modes to remain in quasi-equilibrium with the electric displacement field, which ensures that the relation between the electric displacement and electric field remains in the same form as described in Eq. 24. The reversible work required in the first step is then obtained as,

$$W_{\mathrm{I}} = \int \left( \int_0^{\boldsymbol{D}_{\mathrm{n}}} \boldsymbol{\mathcal{E}} \cdot \delta \boldsymbol{D} \right) dV = \int \left( \int_0^{\boldsymbol{D}_{\mathrm{n}}} \frac{\boldsymbol{D}}{\varepsilon_{\mathrm{s}}} \cdot \delta \boldsymbol{D} \right) dV = \int \left( \frac{\boldsymbol{D}_{\mathrm{n}}^2}{2\varepsilon_{\mathrm{s}}} \right) dV. \tag{30}$$

In the second step, the slow polarization cannot react in time and thus remain fixed at $\boldsymbol{P}^{\mathrm{n}}$, while only the fast polarization responds linearly to the electric field, i.e.,

$$\boldsymbol{D} = \varepsilon_\infty \boldsymbol{\mathcal{E}} + \boldsymbol{P}^{\mathrm{n}}. \tag{31}$$

Then the reversible work required for this process can be evaluated as follows,

$$\begin{aligned} W_{\mathrm{II}} &= \int \left( \int_{\boldsymbol{D}_{\mathrm{n}}}^{\boldsymbol{D}_{\mathrm{ox}}} \boldsymbol{\mathcal{E}} \cdot \delta \boldsymbol{D} \right) dV \\ &= \int \left[ \int_{\boldsymbol{D}_{\mathrm{n}}}^{\boldsymbol{D}_{\mathrm{ox}}} \left( \frac{\boldsymbol{D} - \boldsymbol{P}^{\mathrm{n}}}{\varepsilon_\infty} \right) \cdot \delta \boldsymbol{D} \right] dV \\ &= \int \frac{1}{2\varepsilon_\infty} (\boldsymbol{D}_{\mathrm{ox}}^2 - \boldsymbol{D}_{\mathrm{n}}^2) dV - \int \frac{1}{\varepsilon_\infty} \boldsymbol{P}^{\mathrm{n}} \cdot (\boldsymbol{D}_{\mathrm{ox}} - \boldsymbol{D}_{\mathrm{n}}) dV \\ &= \int \frac{1}{2\varepsilon_\infty} (\boldsymbol{D}_{\mathrm{ox}}^2 - \boldsymbol{D}_{\mathrm{n}}^2) dV - \int c \boldsymbol{D}_{\mathrm{ox}} \cdot \boldsymbol{D}_{\mathrm{n}} dV + \int c \boldsymbol{D}_{\mathrm{n}}^2 dV, \end{aligned} \tag{32}$$

with $c = 1/\varepsilon_\infty - 1/\varepsilon_s$. The electrostatic free energy in the non-equilibrium state C is then the sum of $W_\mathrm{I}$ in Eq. 30 and $W_\mathrm{II}$ in Eq. 32,

$$G_\mathrm{ox}^\mathrm{out}[\boldsymbol{D}_\mathrm{n}] = W_\mathrm{I} + W_\mathrm{II} = \int \frac{1}{2\varepsilon_\mathrm{s}} \boldsymbol{D}_\mathrm{ox}^2 dV + \int \frac{c}{2} (\boldsymbol{D}_\mathrm{n} - \boldsymbol{D}_\mathrm{ox})^2 dV, \tag{33}$$

which is a free energy functional of $\boldsymbol{D}_\mathrm{n}$. Similarly, the electrostatic free energy in the reduced state can be obtained as a functional of $\boldsymbol{D}_\mathrm{n}$:

$$G_\mathrm{red}^\mathrm{out}[\boldsymbol{D}_\mathrm{n}] = \int \frac{1}{2\varepsilon_\mathrm{s}} \boldsymbol{D}_\mathrm{red}^2 dV + \int \frac{c}{2} (\boldsymbol{D}_\mathrm{n} - \boldsymbol{D}_\mathrm{red})^2 dV. \tag{34}$$

For each distinct solvent configuration, there is a corresponding fictitious equilibrium distribution of the electric displacement that aligns with the solvent nuclear polarization of that configuration. We observe that the first terms in $G_\mathrm{ox}^\mathrm{out}$ and $G_\mathrm{red}^\mathrm{out}$ are the electrostatic free energies in the equilibrium solvent configurations of the oxidized and reduced species, respectively. They differ from the Born solvation free energy by the electrostatic energy of the same charge distribution in the vacuum[128], i.e.,

$$G_\mathrm{Born} = \int \frac{1}{2} \left( \frac{1}{\varepsilon_\mathrm{s}} - \frac{1}{\varepsilon_0} \right) \boldsymbol{D}_i^2 dV, \qquad i = \mathrm{ox}, \mathrm{red}. \tag{35}$$

Hereafter, the term solvation energy specifically refers to the free energy of electrostatic interactions between the redox species and the surrounding solvent molecules, rather than the difference relative to the electrostatic energy in vacuum.

The second terms in $G_\mathrm{ox}^\mathrm{out}$ and $G_\mathrm{red}^\mathrm{out}$ account for non-equilibrium thermal fluctuations of the electrostatic free energy arising from deviations of the solvent configuration from their equilibrium configurations in the oxidized and reduced states, respectively. The fluctuation in electrostatic energy exhibits the characteristics of a harmonic oscillator, with $c$ acting as the local force constant. This harmonic behavior arises from the linear response of the solvent polarization to the electric field, analogous to the relationship between force and displacement in classical mechanics and widely used in statistical thermodynamics. The first terms in Eqs. 33 and 34 describe the thermodynamic aspects of solvation, while the second terms capture the non-equilibrium aspects.

### 3.3. Diabatic free energy surfaces

As mentioned, the electronic states on the metal surface form a continuous energy spectrum, with a certain probability for electrons transition between each of these states and the electronic state of the redox species. Let the metal electronic states be represented by a set of unperturbed, one-electron eigenstates of the metal-vacuum interface $|\psi_k\rangle$, with the corresponding energy $\epsilon_k$. The electronic state of the redox species involved in the electron transfer, i.e., valence electronic state is described by an unperturbed state in vacuum $|\psi_a\rangle$ with energy $\epsilon_a$. This state corresponds to either the lowest unoccupied electronic orbital of the oxidized species or the highest occupied electronic orbital of the reduced species. The energies of these two orbitals often differ slightly due to orbital relaxation and electronic correlation effects[129]. As this energy difference contributes only a constant shift to the reaction free energy and does not affect the rate constant expressions derived below, it is neglected here for simplicity. In other words, we assume the Koopmans's theorem holds. At the metal-solution interface, $\epsilon_k$ and $\epsilon_a$ consist of a chemical component, denoted as $\epsilon_k^0$ and $\epsilon_a^0$, along with an electrostatic component related to the electrostatic potential of the metallic electrons and the valence electron of the redox species in the presence of the interfacial electric field, i.e.,

$$\epsilon_k = \epsilon_k^0 - e_0\phi_\mathrm{M}, \tag{36}$$

$$\epsilon_a = \epsilon_a^0 - e_0\phi_a, \tag{37}$$

where $\phi_\mathrm{M}$ is the inner potential of the metallic phase, $\phi_a$ the electrostatic potential at the site of the redox species during the reaction, i.e., the electrostatic potential at the reaction plane.

We now consider an electron transferring between the metal state $k$ and the state $a$ of the redox species, i.e.,

$$\mathrm{ox} + \mathrm{e}^-(\epsilon_k) \rightleftharpoons \mathrm{red}. \tag{38}$$

If there no overlap or electronic interactions between these two states exists, the electron will remain either in state $k$ or state $a$, and the ET is forbidden. Since electrons move much faster than nuclei, electronic motion can be separated from the nuclear held motion while the nuclei move within the effective potential energy surface generated by the electrons and nuclei together—this is the Born-Oppenheimer approximation. Under this approximation, the free energies of the oxidized and reduced states are given by the sum of the gas-phase electronic energy of the electrons involved and the corresponding nuclear free energies contributed by both the inner and outer spheres. These free energies, corresponding to the cases where the electron occupies the one-electron states k and a, respectively, can be written as follows:

$$\begin{aligned} G_k\left[\{Q_j\}, \boldsymbol{D}_\mathrm{n}\right] = \epsilon_\mathrm{ox} + \epsilon_k + G_\mathrm{ox}^\mathrm{eq} + \frac{1}{2}\sum_j \mu_j \omega_{j,\mathrm{ox}}^2 \left(Q_j - Q_{j,\mathrm{ox}}\right)^2 \\ + \int \frac{c}{2}(\boldsymbol{D}_\mathrm{n} - \boldsymbol{D}_\mathrm{ox})^2 dV, \end{aligned} \tag{39}$$

$$G_a\left[\{Q_j\}, \boldsymbol{D}_\mathrm{n}\right] = \epsilon_\mathrm{red} + G_\mathrm{red}^\mathrm{eq} + \frac{1}{2}\sum_j \mu_j \omega_{j,\mathrm{red}}^2 \left(Q_j - Q_{j,\mathrm{red}}\right)^2 + \int \frac{c}{2}(\boldsymbol{D}_\mathrm{n} - \boldsymbol{D}_\mathrm{red})^2 dV, \tag{40}$$

with the nuclear free energies of the oxidized and reduced states at their respective equilibrium nuclear configurations,

$$G_\mathrm{ox}^\mathrm{eq} = G_\mathrm{ox}^\mathrm{in,eq} + \int \frac{1}{2\varepsilon_\mathrm{s}} \boldsymbol{D}_\mathrm{ox}^2 dV, \tag{41}$$

$$G_\mathrm{red}^\mathrm{eq} = G_\mathrm{red}^\mathrm{in,eq} + \int \frac{1}{2\varepsilon_\mathrm{s}} \boldsymbol{D}_\mathrm{red}^2 dV, \tag{42}$$

where $\epsilon_\mathrm{ox}$ and $\epsilon_\mathrm{red}$ are the total electronic energies of the oxidized and reduced species, respectively. With the assumption of Koopmans theorem, we have $\epsilon_\mathrm{red} = \epsilon_\mathrm{ox} + \epsilon_a$. To reduce the complexity, we assume the same average vibrational frequency $\omega_{j,\mathrm{av}}$ for the inner-sphere nuclear mode $j$ of oxidized and reduced species. As suggested by Marcus[130], the average frequency is

$$\omega_{j,\mathrm{av}} = \frac{2\omega_{j,\mathrm{ox}}\omega_{j,\mathrm{red}}}{\omega_{j,\mathrm{ox}} + \omega_{j,\mathrm{red}}}. \tag{43}$$

As shown in Eqs. 39 and 40, the free energies $G_k$ and $G_a$ are the functions of nuclear coordinates, as reflected in their dependence on a set of configurational variables consisting of normal coordinates of the inner-sphere nuclear modes $\{Q_j\}$ and the slow solvent polarization $\boldsymbol{D}_\mathrm{n}$. The variations of $G_k$ and $G_a$ with respect to the configurational variables constitute the multi-dimensional FESs for the oxidized and reduced states, respectively. On these two surfaces, the electronic states of the oxidized or reduced species remain unchanged as the nuclei move along the FESs. As mentioned, we refer to these FESs as the diabatic FESs. The terms "diabatic" and "adiabatic" in electron transfer originate from the adiabatic approximation in quantum mechanics, which differs from their use in thermodynamics, where it indicates no heat exchange between the thermodynamic system and its environment[131].

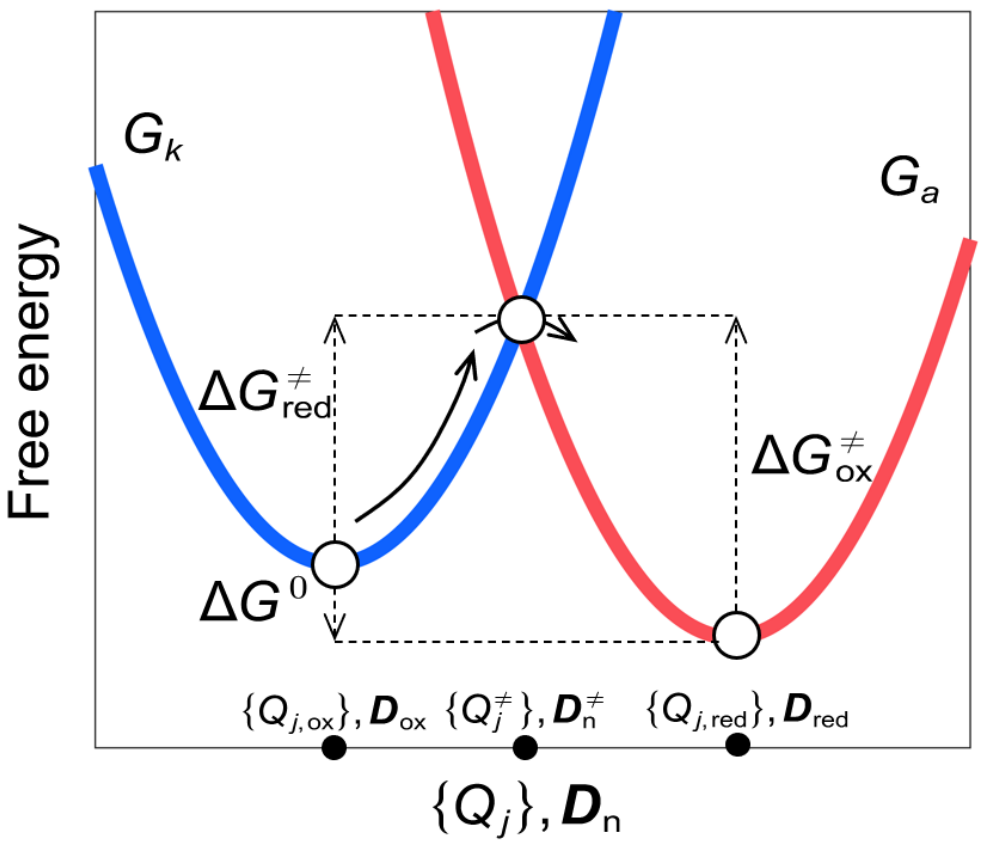

Figure 5. One-dimensional schematic representation of multi-dimensional diabatic free energy surfaces (FESs), $G_k$ and $G_a$, for the ET systems with the electron residing in states $k$ (blue line) and $a$ (red line), respectively. The nuclear configuration is described by a set of configurational variables consisting of the normal coordinates $\{Q_j\}$, and the slow polarization $\boldsymbol{D}_{\mathbf{n}}$. Their values at the minima of the FESs and at their intersections are indicated by the dots on the horizontal axis. For the non-adiabatic ET, the activation states can be found at their intersections, and the reaction proceeds via: (1) nuclear reorganization; (2) electron transition at the intersections; (3) nuclear relaxation. $\Delta G^0$, $\Delta G^{\neq}_{\mathrm{red}}$, and $\Delta G^{\neq}_{\mathrm{ox}}$ denote the reaction free energy, the activation free energy of the reduction reaction, and that of the reverse (oxidation) reaction, respectively.

For electron transfer to occur, electronic interactions, i.e., coupling between electronic states $k$ and $a$, is required. When the coupling is very weak, non-adiabatic ET is operational, and the diabatic FESs are only minimally perturbed. For an almost instantaneous electron transition, the nuclei are frozen and their kinetic energy remains unchanged, in accordance with the Franck–Condon principle, in which the electron transition occurs at a fixed nuclear configuration. Given that the entropy change associated with the electron transition at a fixed nuclear configuration—arising primarily from variations in electronic occupations—is negligible[1]; the fundamental law of energy conservation requires that the free energies of the two diabatic states be equal for the electron transition to occur with appreciable probability. The transition state or activated state can then be found at the intersections of the diabatic FESs, as shown in Figure 5. After the electron transition at the transition state, the nuclear configuration will naturally relax to the equilibrium configuration of the reduced state.

### 3.4. Activation free energy

Assuming the configurational variables in the activated state are $\{Q_j^{\neq}\}$ and $\boldsymbol{D}_{\mathrm{n}}^{\neq}$, the activation free energy $\Delta G^{\neq}_{\mathrm{red}}$ of reduction reaction in Eq. 38 is the difference between the free energies of the oxidized state with the configurational variables in the activated state and that in the equilibrium state:

$$\Delta G^{\neq}_{\text{red}} = G_k[r^{\neq}, \boldsymbol{D}^{\neq}_{\text{n}}] - G_k[r_{\text{ox}}, \boldsymbol{D}_{\text{ox}}] = \frac{1}{2}\sum_j \mu_j \omega^2_{j,\text{av}}\left(Q^{\neq}_j - Q_{j,\text{ox}}\right)^2 + \int \frac{c}{2}(\boldsymbol{D}^{\neq}_{\text{n}} - \boldsymbol{D}_{\text{ox}})^2 dV, \tag{44}$$

where $\{Q^{\neq}_j\}$ and $\boldsymbol{D}^{\neq}_{\text{n}}$ are subject to energy conservation, namely,

$$G_k\left[\{Q^{\neq}_j\}, \boldsymbol{D}^{\neq}_{\text{n}}\right] = G_a\left[\{Q^{\neq}_j\}, \boldsymbol{D}^{\neq}_{\text{n}}\right]. \tag{45}$$

By substituting Eqs. 39 and 40 into the above equation, we have,

$$\frac{1}{2}\sum_j \mu_j \omega^2_{j,\text{av}}\left(2Q^{\neq}_j - Q_{j,\text{ox}} - Q_{j,\text{red}}\right)\left(Q_{j,\text{red}} - Q_{j,\text{ox}}\right) + \int \frac{c}{2}(2\boldsymbol{D}^{\neq}_{\text{n}} - \boldsymbol{D}_{\text{ox}} - \boldsymbol{D}_{\text{red}})(\boldsymbol{D}_{\text{red}} - \boldsymbol{D}_{\text{ox}})dV - \Delta G^0 = 0, \tag{46}$$

with the reaction free energy $\Delta G^0$ for a specific metal level $k$,

$$\Delta G^0(\epsilon_k) \equiv G^0_{\text{red}} - G^0_{\text{ox}} - \epsilon_k = \epsilon^0_a + \Delta G_{\text{n}} - e_0\phi_a - \epsilon_k, \tag{47}$$

and the free energies of the reduced and oxidized species at their respective equilibrium nuclear configurations, including both electronic and nuclear contributions,

$$G^0_{\text{red}} = \epsilon_{\text{red}} + G^{\text{eq}}_{\text{red}}, \tag{48}$$

$$G^0_{\text{ox}} = \epsilon_{\text{ox}} + G^{\text{eq}}_{\text{ox}}. \tag{49}$$

The term $\Delta G_{\text{n}} = G^{\text{eq}}_{\text{red}} - G^{\text{eq}}_{\text{ox}}$ represents the difference in the equilibrium nuclear free energies of the reduced and oxidized states. As shown in Eqs. 41 and 42, this difference arises from structural distortion of the redox species as well as from the corresponding change in solvation free energy. The energy conservation constraint (Eq. 45) implies that electron transfer can occur only at the intersections of the two diabatic FESs and in Figure 5 we plot the diabatic FESs in a one-dimensional case as an illustration of this. In this scenario, there is only one single intersection point that corresponds to a unique set of $\{Q^{\neq}_j\}$ and $\boldsymbol{D}^{\neq}_{\text{n}}$. However, it should be noted that the diabatic FESs are inherently multi-dimensional, resulting in infinitely many intersection points, each of which corresponds to a distinct set of $\{Q^{\neq}_j\}$ and $\boldsymbol{D}^{\neq}_{\text{n}}$ that satisfies the energy conservation constraint. Given that the reaction rate decreases with increasing activation energy, the objective is to identify the set of $\{Q^{\neq}_j\}$ and $\boldsymbol{D}^{\neq}_{\text{n}}$ that minimizes the activation free energy (Eq. 44) while satisfying the energy conservation constraint. This minimization corresponds to finding the most favorable reaction pathway, i.e., the minimum free energy path. To this end, we construct the following Lagrange functional,

$$\mathcal{L}\left[\{Q^{\neq}_j\}, \boldsymbol{D}^{\neq}_{\text{n}}\right] = \frac{1}{2}\sum_j \mu_j \omega^2_{j,\text{av}}\left(Q^{\neq}_j - Q_{j,\text{ox}}\right)^2 + \int \frac{c}{2}(\boldsymbol{D}^{\neq}_{\text{n}} - \boldsymbol{D}_{\text{ox}})^2 dV - \xi\left(\frac{1}{2}\sum_j \mu_j \omega^2_{j,\text{av}}\left(2Q^{\neq}_j - Q_{j,\text{ox}} - Q_{j,\text{red}}\right)\left(Q_{j,\text{red}} - Q_{j,\text{ox}}\right) + \int \frac{c}{2}(2\boldsymbol{D}^{\neq}_{\text{n}} - \boldsymbol{D}_{\text{ox}} - \boldsymbol{D}_{\text{red}})(\boldsymbol{D}_{\text{red}} - \boldsymbol{D}_{\text{ox}})dV - \Delta G^0\right), \tag{50}$$

where $\xi$ is the Lagrange multiplier enforcing the constraint of energy conservation (Eq. 46). The minimum of this Lagrangian is located at a point where its differential with respect to $\{Q_j^{\neq}\}$ and its variation with respect to $\boldsymbol{D}_\mathrm{n}^{\neq}$ is both zero. This leads to

$$\frac{\partial \mathcal{L}}{\partial Q_j^{\neq}} = \mu_j \omega_{j,\mathrm{av}}^2 \left(Q_j^{\neq} - Q_{j,\mathrm{ox}}\right) - \xi \mu_j \omega_{j,\mathrm{av}}^2 \left(Q_{j,\mathrm{red}} - Q_{j,\mathrm{ox}}\right) = 0, \tag{51}$$

$$\frac{\delta \mathcal{L}}{\delta \boldsymbol{D}_\mathrm{n}^{\neq}} = c(\boldsymbol{D}_\mathrm{n}^{\neq} - \boldsymbol{D}_\mathrm{ox}) - \xi c(\boldsymbol{D}_\mathrm{red} - \boldsymbol{D}_\mathrm{ox}) = 0, \tag{52}$$

By combining Eqs. 46, 51, and 52, we can solve for $\xi$, $\{Q_j^{\neq}\}$ and $\boldsymbol{D}_\mathrm{n}^{\neq}$ at the intersections and find the minimized activation free energy,

$$\xi = \frac{1}{2}\left(\frac{\Delta G^0(\epsilon_k)}{\lambda} + 1\right), \tag{53}$$

$$Q_j^{\neq} = Q_{j,\mathrm{ox}} + \xi\left(Q_{j,\mathrm{red}} - Q_{j,\mathrm{ox}}\right), \tag{54}$$

$$\boldsymbol{D}_\mathrm{n}^{\neq} = \boldsymbol{D}_\mathrm{ox} + \xi(\boldsymbol{D}_\mathrm{red} - \boldsymbol{D}_\mathrm{ox}), \tag{55}$$

with,

$$\lambda = \lambda_\mathrm{in} + \lambda_\mathrm{out}, \tag{56}$$

$$\lambda_\mathrm{in} = \frac{1}{2}\sum_j \mu_j \omega_{j,\mathrm{av}}^2 \left(Q_{j,\mathrm{red}} - Q_{j,\mathrm{ox}}\right)^2, \tag{57}$$

$$\lambda_\mathrm{out} = \frac{1}{2}\int \left(\frac{1}{\varepsilon_\infty} - \frac{1}{\varepsilon_\mathrm{S}}\right)(\boldsymbol{D}_\mathrm{red} - \boldsymbol{D}_\mathrm{ox})^2 dV, \tag{58}$$

where $\lambda$ is the nuclear reorganization energy, consisting of contributions from both the inner sphere $\lambda_\mathrm{in}$ and outer sphere $\lambda_\mathrm{out}$ components. By substituting Eqs. 53-55 into Eq. 44, we obtain the minimized activation free energy for ET at the state $k$ of the metal surface,

$$\Delta G_\mathrm{red}^{\neq}(\epsilon_k) = \lambda \xi^2 = \frac{\left(\lambda + \Delta G^0(\epsilon_k)\right)^2}{4\lambda}. \tag{59}$$

The reaction free energy given in Eq. 47 can be connected with the electrode potential by introducing the Fermi level $\epsilon_\mathrm{F}$,

$$\Delta G^0(\epsilon_k) = \epsilon_a^0 + \Delta G_\mathrm{n} - e_0\phi_a - \epsilon_k + (\epsilon_\mathrm{F} - \epsilon_\mathrm{F}). \tag{60}$$

At the Fermi level, $\epsilon_\mathrm{F} = \tilde{\mu}_\mathrm{e}$, where $\tilde{\mu}_\mathrm{e}$ is the electrochemical potential of metal electrons. $\tilde{\mu}_\mathrm{e}$ is In turn related to the chemical potential of metal electrons $\mu_\mathrm{e}$, and the inner potential of the metal, $\phi_\mathrm{M}$, by,

$$\tilde{\mu}_\mathrm{e} = \mu_\mathrm{e} - e_0\phi_\mathrm{M}. \tag{61}$$

The reaction free energy in Eq. 60 is locally defined at the energy level $\epsilon_k$ of the metal, as it depends on the local environment that influences the local electrostatic potential $\phi_a$ and the local equilibrium solvation difference, the latter contributing to the nuclear free energy difference $\Delta G_\mathrm{n}$. Here, we first consider the reaction free energy defined using bulk solution properties, specifically, the inner potential of the solution, $\phi_\mathrm{S}$, and the nuclear free energy difference $\Delta G_\mathrm{n}^\mathrm{bulk}$. The difference between this bulk-defined reaction free energy and the locally defined one is incorporated into the work terms and is discussed in detail in Section 7. Using bulk solution properties, the reaction free energy in Eq. 60 can be rearranged into,

$$\Delta G^0(\epsilon_k) = \epsilon_a^0 + \Delta G_{\mathrm{n}}^{\mathrm{bulk}} + e_0\varphi_{\mathrm{abs}} + \epsilon_{\mathrm{F}} - \epsilon_k, \quad (62)$$

where $\varphi_{\mathrm{abs}} = \Delta\phi_{\mathrm{S}}^{\mathrm{M}} - \mu_{\mathrm{e}}/e_0$ is the absolute electrode potential[132], with $\Delta\phi_{\mathrm{S}}^{\mathrm{M}} = \phi_{\mathrm{M}} - \phi_{\mathrm{S}}$ as the difference between the inner potentials of metal and solution. A standard equilibrium value of the absolute electrode potential, $\varphi_{\mathrm{abs}}^0$, can be defined as $-\left(\epsilon_a^0 + \Delta G_{\mathrm{sol}}^{\mathrm{bulk}}\right)/e_0$, at which the reaction energy at the Fermi level is zero. The term in the parentheses refers to the free energy difference between the reduced and oxidized states at their respective equilibrium nuclear configuration in the solution bulk, comprising the differences in both electronic energy in the vacuum (the electrostatic potential energy of the valence electron in the interfacial electric field has been eliminated from $\epsilon_a^0$) and equilibrium nuclear free energy. It is evident that the absolute value of standard equilibrium electrode potential of the redox couple is independent of the metal properties as it depends only on the nature of the redox species and the solvent. With this definition, Eq. 62 can be recast into,

$$\Delta G^0(\epsilon_k) = e_0\left(\varphi_{\mathrm{abs}} - \varphi_{\mathrm{abs}}^0\right) + \epsilon_{\mathrm{F}} - \epsilon_k = e_0\eta + \epsilon_{\mathrm{F}} - \epsilon_k, \quad (63)$$

with the overpotential $\eta$ defined as,

$$\eta = \varphi_{\mathrm{abs}} - \varphi_{\mathrm{abs}}^0 = \varphi - \varphi^0, \quad (64)$$

where $\varphi$ and $\varphi^0$ are the electrode potentials relative to a chosen reference electrode[132]. The overpotential defined here is referenced to the standard equilibrium electrode potential rather than to the equilibrium electrode potential that depends on the redox-species concentrations through the Nernst equation. By substituting Eq. 63 into Eq. 59, we obtain the activation free energy for ET at the metal state $k$ as,

$$\Delta G_{\mathrm{red}}^{\neq}(\epsilon_k) = \frac{(\lambda + e_0\eta + \epsilon_{\mathrm{F}} - \epsilon_k)^2}{4\lambda}. \quad (65)$$

For the inverse reaction of Eq. 38, i.e., the oxidation reaction, as shown in Figure 5, the corresponding activation free energy is,

$$\Delta G_{\mathrm{ox}}^{\neq}(\epsilon_k) = \Delta G_{\mathrm{red}}^{\neq}(\epsilon_k) - \Delta G^0(\epsilon_k) = \frac{(\lambda - e_0\eta + \epsilon_k - \epsilon_{\mathrm{F}})^2}{4\lambda}. \quad (66)$$

The solvent effects on the activation free energy are two-fold. First, the equilibrium thermodynamics dictate the extent of equilibrium solvation of the redox species, thereby influencing the reaction free energy, or driving force. Specifically, the difference between the equilibrium solvation free energies of the oxidized and reduced species controls the standard equilibrium potential of the reaction. Second the nuclear non-equilibrium fluctuations determine the energetic penalty for reorganizing the solvent into the transitions state configuration that permits electron transfer; a smaller reorganization energy indicates a more favorable process and a lower free energy barrier.

Figure 6 shows the activation free energies at different electronic state energies for the reduction (solid lines) and oxidation reactions (dashed lines) at overpotentials of -0.2 V, 0 V, and 0.2 V, with line colors transitioning from light to dark. At the standard equilibrium potential, where the overpotential referenced to this value is zero, the activation free energies for both the reduction and oxidations are equal, each being one-fourth of the reorganization energy $\lambda$. As the overpotential increases, the activation free energy for the reduction rises at each electronic level, while decreasing for oxidation. At the absolute zero temperature, the metal electronic states are filled up to the Fermi level, and therefore reduction reactions are prohibited above the Fermi level because there are no occupied states to donor electrons, while oxidation reactions are prohibited below the Fermi level due to the absence of unoccupied states to accept electrons. At finite temperatures, electrons occupy the electrode states following the Fermi-Dirac distribution such that the electronic states with energies above the Fermi level can become partly occupied and states below the Fermi level may be partially unoccupied. While the occupation of the electrode's electronic states change due to thermal effects, in practice only the states near the Fermi level make significant contributions to the reaction rate as only these states have a significant contribution on the reaction barrier as discussed in Section 4.4. Therefore, accounting for thermal effects on the electrode's energy level occupations does not significantly alter the fact that reduction is unfavorable above the Fermi level, while oxidation is unfavorable below it. Hence, we can attribute the ET rate primarily to contributions from the Fermi level, with the corresponding activation energies at this level being,

$$\Delta G^{\neq}_{\mathrm{red}}(\epsilon_{\mathrm{F}}) = \frac{(\lambda + e_0\eta)^2}{4\lambda}, \qquad \Delta G^{\neq}_{\mathrm{ox}}(\epsilon_{\mathrm{F}}) = \frac{(\lambda - e_0\eta)^2}{4\lambda}. \tag{67}$$

When the overpotential is small, the second-order terms of $\eta$ can be neglected, resulting in activation free energies that resemble the form found in the Butler-Volmer equation,

$$\Delta G^{\neq}_{\mathrm{red}}(\epsilon_{\mathrm{F}}) = \frac{\lambda}{4} + \frac{1}{2}e_0\eta, \qquad \Delta G^{\neq}_{\mathrm{ox}}(\epsilon_{\mathrm{F}}) = \frac{\lambda}{4} - \frac{1}{2}e_0\eta \;\; (\text{small } \eta), \tag{68}$$

which give the transfer coefficients (or symmetry factors) of 0.5 for both reduction and oxidation, where the transfer coefficient is defined as the derivative of the activation free energy with respect to the driving force $\Delta G^0$, corresponding to $e_0\eta$ for reduction and $-e_0\eta$ for oxidation[36]. The Marcus formulas in Eq. 67 provide a driving force-dependent transfer coefficient, $0.5(1 + \Delta G^0/\lambda)$, according to its definition. As the driving force varies from $-\lambda$ to $\lambda$, the transfer coefficient ranges between zero and unity, with these two limits corresponding to the activationless and barrierless reaction regimes, respectively.

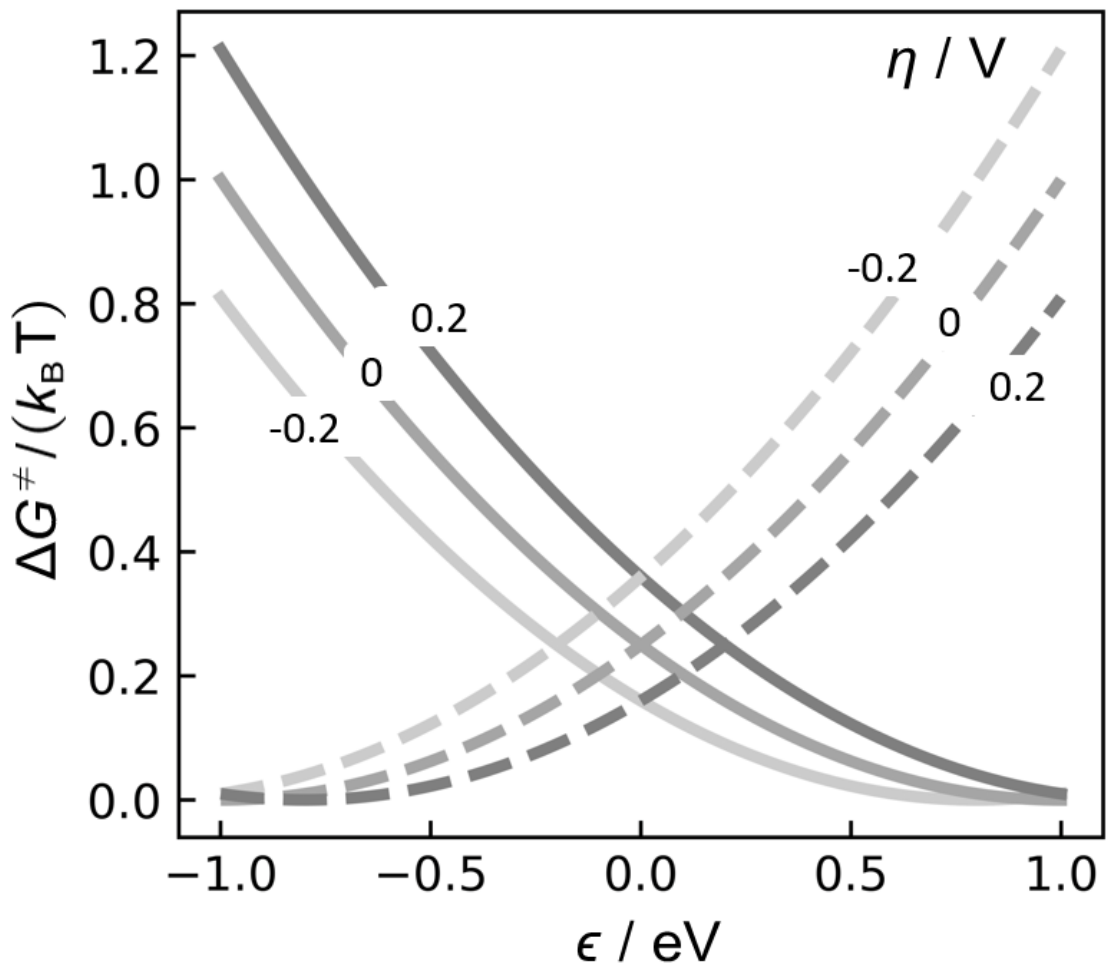

Figure 6. The activation free energies at different electronic state energies for the reduction (solid lines) and oxidation reactions (dashed lines) at overpotentials of -0.2 V, 0 V, and 0.2 V, with line colors transitioning from light to dark. The reorganization energy is set to 1 eV.

The most essential concept introduced by Marcus theory is the nuclear reorganization energy. It represent the free energy required to reorganize the nuclear modes from the equilibrium configuration of the reactant (or product) to that of the product (or reactant) while the electronic state remains that of the reactant (or product), as defined in Eqs. 56-58. The inner-sphere reorganization energy can be estimated using Eq. 57, with the relevant parameters, such as vibrational frequencies and changes in interatomic distances, obtained from X-ray crystal diffraction, extended X-ray absorption fine structure (EXAFS) measurements, and structure-optimized electronic-structure calculations. The outer-sphere, or solvent, reorganization energy can be estimated using Eq. 58 within a continuum model. Beyond the simple Born-sphere solvation model, the calculation can be refined by invoking the nonlocal electrostatic description of the solvent developed by Kornyshev *et al*[133,134]. Further discussions on the solvent reorganization energy in Eq. 58 will be given in Section 7.2. Alternatively, atomistic simulations can be performed to obtain the reorganization energy and to separate it into inner- and outer-sphere contributions, as will be discussed in Section 3.6. As shown in Eqs. 56-58, when the parameters entering the reorganization energies are assumed to remain constant, the reorganization energies for the reduction and oxidation reactions are identical, since they depend only on the squared displacements of configurational variables and are therefore invariant with respect to the ET direction. In this case of symmetric reorganization, the diabatic FESs for the reduced and oxidized states can be represented by parabolas of identical curvature, as shown in Figure 1a. Asymmetric reorganization is expected when these assumptions are no longer valid. For the inner-sphere reorganization, instead of assuming the same average frequency as in Eq. 57, such asymmetry may arise from shifts in the vibrational frequencies between the reduced and oxidized states. The inner-sphere asymmetry has been estimated for reactions of large transition-metal complexes[135], e.g., chromium(III) ethylenediaminetetraacetate (EDTA), in which the inner-sphere reorganization is prominent; such asymmetry plays a critical role in the theoretical analysis of the activation energy[136]. The outer-sphere asymmetry arises from different coupling strengths of the reactant and product with the solvent polarization, as a consequence of the distinct intrinsic properties of the reduced and oxidized species, e.g., the polarizability[137,138].

### 3.5. Effective nuclear coordinate

Here, we give some further insight on the meaning of the Lagrange multiplier $\xi$ in Eq. 53. As discussed, each non-equilibrium nuclear configuration in the oxidized state corresponds to a non-equilibrium state, with a fluctuating free energy representing the differences between the free energy of this non-equilibrium state and the equilibrium state. As shown in Eq. 59, all fluctuating free energies associated with different nuclear configurations of the oxidized state can be identified as activation free energies at specific values of $\xi$, ranging from negative infinity to positive infinity. In other words, as $\xi$ varies in this rage, the corresponding activation free energies effectively account for all fluctuating free energies of the oxidized state. Eq. 59 thereby allows us to describe the fluctuating free energy of the oxidized state using a single, dimensionless parameter $\xi$, which effectively represents the nuclear coordinate. With this, the diabatic FES of the oxidized state can be described as,

$$G_k(\xi) = \epsilon_{\text{ox}} + \epsilon_k + G_{\text{ox}}^{\text{eq}} + \lambda\xi^2, \tag{69}$$

where $\xi = 0$ represents the equilibrium nuclear configuration of the oxidized state. As shown in Eqs. 53 and 63, $\xi$ varies linearly as function of the energy of the metal state $k$. The statement around Eq. 69 indicates that each non-equilibrium states of the oxidized species can serve as an activated state for electron transfer at a specific metal electronic state $k$ with its energy varying from negative infinity to positive infinity. Under real conditions, the metal states are confined to a specific energy range, while the density of states (DOS) vanishes outside this range, making no contribution to the reaction rate. This effect is captured in the pre-exponential factor of the rate expression, which will be discussed in Section 4. Similarly, we can find that the diabatic FES for the reduced state can be also effectively described as a function of $\xi$,

$$G_a(\xi) = \epsilon_{\text{red}} + G_{\text{red}}^{\text{eq}} + \lambda(\xi - 1)^2, \tag{70}$$

where $\xi = 1$ represents the equilibrium nuclear configuration of the reduced state.

### 3.6. MD simulations of diabatic FESs

Molecular dynamics simulations can be used to explain and understand the microscopic atomistic details of the nuclear reorganization and non-equilibrium nuclear fluctuations Marcus theory. Achieving this requires three key steps: 1) the definition and construction of diabatic states, 2) finding a reaction coordinate to project the 3N-dimensional explicit solvent coordinates onto the 1D nuclear reorganization coordinate, and 3) sampling diabatic states along this reaction coordinate. Below, each step is discussed separately. Note, that we do not separate inner- and outer-sphere reorganization as both are treated as nuclear rearrangements. Hence, the reorganization contributions obtained from the methods described in this section refer to the nuclear reorganization and total reorganization energy.

#### 3.6.1. Defining diabatic states

A diabatic state can be generally and qualitatively defined as an electronic state that does not change its physical character along the reaction coordinate[139]. Mathematically, this means that the diabatic states $\{\Psi_i\}$ satisfy

$$\left\langle \Psi_i(\boldsymbol{r},\boldsymbol{R}) \middle| \frac{\partial}{\partial R} \Psi_j(\boldsymbol{r},\boldsymbol{R}) \right\rangle = \int \Psi_i^*(\boldsymbol{r},\boldsymbol{R}) \frac{\partial}{\partial \boldsymbol{R}} \Psi_j(\boldsymbol{r},\boldsymbol{R}) \, d\boldsymbol{r} = 0, \tag{71}$$

at all electronic ($\boldsymbol{r}$) and nuclear ($\boldsymbol{R}$) coordinates. This condition means two different diabatic states remain orthogonal (zero overlap) to each other even as the nuclear position of the other is change. Effectively, this means that the diabatic states do not change when nuclei move but in practice it cannot be met for systems with more than a few nuclei[140].

Therefore, in electron transfer studies the practical approach to generate diabatic states is to build states that approximate Eq. 71 as well as possible, i.e., minimize the coupling between two diabatic states. Various schemes to achieve this have been developed for different purposes[139,141] but they share a few key similarities: the diabatic states are generally smooth functions of the nuclear and reaction coordinates, which makes them ideal for investigating chemical dynamics and kinetics, and the coupling between the states is minimized.

In the context of Marcus theory, the diabatic states are chosen to present the reactant and product states both of which have a well-defined and localized charge at all solvent geometries. In practice, this is most often achieved through the use of empirical valence bond (EVB) approaches pioneered by Warshel *et al.*[142–144]. EVB describes a reacting system as a superposition of two resonance structures, corresponding to the reaction and product states, and the energy of the system is described as a combination of the reactant and product Hamiltonians and their coupling terms. Notably, the resonance structures maintain their character in all nuclear arrangements and differ only in their charge localization. To achieve a connection with Marcus theory that describes the ET between a donor and acceptor states, $\mathrm{D} + \mathrm{A}^+ \rightarrow \mathrm{D}^+ + \mathrm{A}$, and the resonance structures presenting these two states are used to define the EVB states of reactant $|\Psi_1(\boldsymbol{r},\boldsymbol{R})\rangle$ and product $|\Psi_2(\boldsymbol{r},\boldsymbol{R})\rangle$, which in the Born-Oppenheimer approximation are given as

$$|\Psi_1(\boldsymbol{r},\boldsymbol{R})\rangle = |\psi_{\mathrm{DA}^+}(\boldsymbol{r};\boldsymbol{R})\rangle |\chi_{\mathrm{DA}^+}(\boldsymbol{R})\rangle, \tag{72}$$

$$|\Psi_2(\boldsymbol{r},\boldsymbol{R})\rangle = |\psi_{\mathrm{D}^+\mathrm{A}}(\boldsymbol{r};\boldsymbol{R})\rangle |\chi_{\mathrm{D}^+\mathrm{A}}(\boldsymbol{R})\rangle, \tag{73}$$

where $\chi_{\mathrm{DA^+}}(\boldsymbol{R})$ and $\chi_{\mathrm{D^+A}}(\boldsymbol{R})$ denotes the nuclear wave function corresponding to the reactant and product states. In ET theory, $\chi_{\mathrm{DA^+}}(\boldsymbol{R})$ and $\chi_{\mathrm{D^+A}}(\boldsymbol{R})$ are often used to characterize the nuclear structures in reactant and product states. The total adiabatic wave function is written as a linear combination of these two resonance structures within a minimal configuration interaction model:

$$\left|\Psi_i^{\mathrm{adia}}(\boldsymbol{r},\boldsymbol{R})\right\rangle = c_1^i|\Psi_1(\boldsymbol{r},\boldsymbol{R})\rangle \pm c_2^i|\Psi_2(\boldsymbol{r},\boldsymbol{R})\rangle. \tag{74}$$

$\left|\Psi_i^{\mathrm{adia}}(\boldsymbol{r},\boldsymbol{R})\right\rangle$ is the adiabatic state obtained as the solution of a $2\times 2$ non-orthogonal Schrödinger equation in the basis of the reactant and product EVB states through the corresponding secular equation

$$\begin{bmatrix} H_{11} & H_{12} \\ H_{21} & H_{22} \end{bmatrix}\begin{bmatrix} c_1 \\ c_2 \end{bmatrix} = E\begin{bmatrix} 1 & S_{12} \\ S_{21} & 1 \end{bmatrix}\begin{bmatrix} c_1 \\ c_2 \end{bmatrix} \rightarrow \begin{vmatrix} H_{11}-E & H_{12}-ES_{12} \\ H_{21}-ES_{21} & H_{22}-E \end{vmatrix} = 0, \tag{75}$$

where $H_{ij} = \langle\Psi_i(\boldsymbol{r},\boldsymbol{R})|\widehat{H}\left|\Psi_j(\boldsymbol{r},\boldsymbol{R})\right\rangle$ are the Hamiltonian matrix elements for the Hamiltonian $\widehat{H}$ and $S_{ij} = \left\langle\Psi_i(\boldsymbol{r},\boldsymbol{R})\middle|\Psi_j(\boldsymbol{r},\boldsymbol{R})\right\rangle$ are the overlap matrix elements which are non-diagonal because $|\Psi_i(\boldsymbol{r},\boldsymbol{R})\rangle$ are not eigenfunctions of the $\widehat{H}$ but two different diabatic Hamiltonians as discussed below. The diagonal elements of the Hamiltonian matrix correspond to the diabatic energies, and the off-diagonal elements are the coupling matrix elements. The diagonalization of the previous equations gives two adiabatic states: the ground and the first excited state. If the coupling element, $H_{12}$, is small, the resulting ground and excited states strongly resemble the diabatic FESs in Figure 1a. If $H_{12}$ is large, the adiabatic FES in Figure 1b is obtained – this is discussed in detail in Section 5.

The diagonal elements of Eq. 75, the diabatic energies, can be obtained either using a classical force field, QM/MM (quantum mechanics/molecular mechanics), or diabatic DFT methods. In all these cases, one follows the Born-Oppenheimer approximation and then considers the electronic and nuclear parts separately. Specifically, owing to the much faster motion of electrons compared to nuclei, the electronic states and corresponding energies can be obtained by solving the electronic Schrödinger equation at fixed nuclear coordinates $\boldsymbol{R}$, i.e.,

$$\langle\psi_{\mathrm{DA^+}}(\boldsymbol{r};\boldsymbol{R})|H_{\mathrm{el}}|\psi_{\mathrm{DA^+}}(\boldsymbol{r};\boldsymbol{R})\rangle = E_1^{\mathrm{el}}(\boldsymbol{R}), \tag{76}$$

$$\langle\psi_{\mathrm{D^+A}}(\boldsymbol{r};\boldsymbol{R})|H_{\mathrm{el}}|\psi_{\mathrm{D^+A}}(\boldsymbol{r};\boldsymbol{R})\rangle = E_2^{\mathrm{el}}(\boldsymbol{R}), \tag{77}$$

where $H_{\mathrm{el}}$ denotes the electronic Hamiltonian, which also incorporates the Coulomb interactions between nuclei in the computational simulations. The electronic energy as a function of nuclear coordinates $\boldsymbol{R}$, together with the interaction energy between nuclei, defines the potential energy surface for the nuclei, whose statistical average contributes to FES. The nuclear dynamics in the PESs is then obtained from

$$\left[T_{\mathrm{n}} + E_1^{\mathrm{el}}\right]|\chi_{\mathrm{DA^+}}(\boldsymbol{R})\rangle = E_1^{\mathrm{T}}\chi_{\mathrm{DA^+}}(\boldsymbol{R}), \tag{78}$$

$$\left[T_{\mathrm{n}} + E_2^{\mathrm{el}}\right]|\chi_{\mathrm{D^+A}}(\boldsymbol{R})\rangle = E_2^{\mathrm{T}}\chi_{\mathrm{DA^+}}(\boldsymbol{R}), \tag{79}$$

where $T_{\mathrm{n}}$ is the nuclear kinetic energy, and $E_i^{\mathrm{T}}$ is the total energy of the system in state $|\Psi_i\rangle$. It should be noted that the PES or FES does not include contributions from nuclear kinetic energy. In classical force field and QM/MM EVB methods, the energies are computed separately for both acceptor and donor states as the interactions between redox species and the solvent as well as the solvent-solvent interactions depend on the redox species' charge [142–144]. The reacting part of the system is formally described using the electronic gas-phase Hamiltonian, which also defines the charge state of the redox species ($\mathrm{D}+\mathrm{A}^+$ or $\mathrm{D}^+ + \mathrm{A}$). In QM/MM methods the electronic gas-phase Hamiltonian in the presence of the external potential created by the ligands and solvent molecules is explicitly evaluated while in classical force field models this interaction is treated with an effective force field[145–147]. In both cases the nuclear-nuclear interactions are described through a classical force field and typically force field "calibration"[146,148,149] with experiments or quantum chemical calculations is needed in the EVB simulations.

The construction of diabatic states through DFT requires modifications to the DFT Hamiltonian to achieve the charge localization at all nuclear positions. The most common method to achieve is constrained DFT (cDFT)[150–154] which has been implemented in several electronic structure codes[155–161] that can model 2D periodic systems, such as electrochemical interfaces, and explicitly account for the electrode potential through GCE-cDFT[81]. In cDFT, the diabatic states are generated by predefining the charge that a certain group of atoms should have. This is achieved by defining an extended energy functional

$$E[n(\boldsymbol{r}), V_{\mathrm{c}}]_{\mathrm{cDFT}} = E[n(\boldsymbol{r})]_{\mathrm{DFT}} + V_{\mathrm{c}}\left[\int w_{\mathrm{c}}(\boldsymbol{r})n(\boldsymbol{r})\,d\boldsymbol{r} - N_{\mathrm{c}}\right], \tag{80}$$

where $n(\boldsymbol{r})$ is the electron density, $w_{\mathrm{c}}$ is the weight function which defines how the charge is to be partitioned, i.e., the regions where charge is to be localized, $N_{\mathrm{c}}$ is the desired number of excess electrons within the constrained region, $V_{\mathrm{c}}$ is the Lagrange multiplier enforcing the charge localization, and $E[n(\boldsymbol{r})]_{\mathrm{DFT}}$ is the standard DFT energy functional. The introduction of the constraining terms adds a new effective local potential to the DFT equations $V_{\mathrm{c}}w_{\mathrm{c}}(\boldsymbol{r})$ and the Lagrange multiplier, i.e., the strength of the local potential, is solved self-consistently so that the convergence criterion for the charge localization is satisfied.

### 3.6.2. Microscopic description of the nuclear reorganization coordinate

The polarization coordinate used in the continuum models used in the preceding section cannot be directly adopted for explicit MD simulations because the polarization corresponds to the collective movement of tens if not hundreds of solvent molecules which would result in a very complex and inefficient sampling of a very high-dimensional FES. Furthermore, it is difficult to identify suitable direct geometric reaction coordinates, such as bond lengths and angles, which can describe ET and achieve the needed timescale separation between the reaction coordinate and other degrees of freedom. Both of these issues can be circumvented by reversing the role of nuclear reorganization and the energy gap between the reactant and product diabatic states; in the continuum theory, the solvent reorganization leads to the disappearance of the energy gap between the diabatic states, as shown in Figure 1a, and thereby electron transfer while in explicit MD simulations we are forcing electron transfer by closing the energy gap and this then leads to solvent reorganization and charge transfer. This change from "the nuclear reorganization closing the energy gap" to "energy gap closing leading to nuclear reorganization" is valid because in computing reversible work one is free to choose the most convenient reversible path[162].

The energy gap $\Delta E(\boldsymbol{R})$ can be explicitly specified as the energy difference between the reaction and product diabatic states at a given nuclear arrangement

$$\Delta E(\boldsymbol{R}) = H_{11}(\boldsymbol{R}) - H_{22}(\boldsymbol{R}) = E_1(\boldsymbol{R}) - E_2(\boldsymbol{R}). \tag{81}$$

Because $\Delta E(\boldsymbol{R})$ depends on all nuclear coordinates $(\boldsymbol{R})$ of the system, it can be seen as projection of all nuclear coordinates onto a single reaction coordinate. The energy gap has a long history in spectroscopy, the description relaxation processes, and it was first used in the MD simulations of condensed phase reactions by Warshel[142–144] more than four decades ago; now, the energy gap coordinate can be considered as a universal reaction coordinate as it achieves efficient sampling in condensed phases and allows a more localized description of the transition state than many geometrical reaction coordinate[163].

### 3.6.3. Constructing diabatic FESs

Using the energy gap as the reaction coordinate means that the partition functions, free energies, and/or probabilities as function of $\Delta E(\boldsymbol{R})$ need to be constructed—this is achieved using Eqs. 14-17 in Section 2.1. However, the direct computation of these quantities through normal MD is not feasible, as discussed in Section 2.2, enhanced sampling methods[106] need to be used. The most common way to sample the energy gap coordinate is to use a mapping Hamiltonian, which linearly interpolates between the reactant and product Hamiltonians

$$H_{\alpha_i}(\boldsymbol{R}) = \alpha_i H_{11}(\boldsymbol{R}) + (1-\alpha_i) H_{22}(\boldsymbol{R}), \alpha_i \in [0,1], \tag{82}$$

where $\alpha_i$ is the $i^{\text{th}}$ point along the discretized path connecting the reactant and product diabatic states. With this Hamiltonian one can sample the energy gap coordinate and compute the free energy changes through free energy perturbation (FEP) theory[164] or umbrella sampling[165]. Changes in the diabatic free energy ($\delta G_i(\Delta E)$) at $i^{\text{th}}$ sampling point for the reactant (R, with Hamiltonian $H_{11}$) and product (P, with Hamiltonian $H_{22}$) state is given by[144]

$$\delta G_i^{\text{R}}(\Delta E) = -k_{\text{B}} T \ln \left[ \left\langle \frac{\delta\left[\Delta E(\boldsymbol{R}) - \left(H_{11}(\boldsymbol{R}) - H_{22}(\boldsymbol{R})\right)\right]}{\exp\left[-\beta\left(H_{11}(\boldsymbol{R}) - H_{\alpha_i}(\boldsymbol{R})\right)\right]} \right\rangle_{\alpha_i} \right], \tag{83}$$

$$\delta G_i^{\text{P}}(\Delta E) = -k_{\text{B}} T \ln \left[ \left\langle \frac{\delta\left[\Delta E(\boldsymbol{R}) - \left(H_{11}(\boldsymbol{R}) - H_{22}(\boldsymbol{R})\right)\right]}{\exp\left[-\beta\left(H_{22}(\boldsymbol{R}) - H_{\alpha_i}(\boldsymbol{R})\right)\right]} \right\rangle_{\alpha_i} \right], \tag{84}$$

$$\Delta G_i = \sum_{j=0}^{i-1} \delta G_j(\Delta E), \tag{85}$$

where $\langle \ldots \rangle_{\alpha_i}$ indicates that the sampling is carried out $i^{\text{th}}$ sampling point using the corresponding mapping Hamiltonian. The free energies from the previous equations are completely general and allow constructing the reactant and product diabatic FESs from MD simulations. Note that the obtained diabatic FES do not need to be parabolic as required or predicted by the Marcus theory, and this general approach can also capture non-linear coupling or response between ET and nuclear reorganization.

### 3.6.4. Connecting with Marcus theory

The connection between the energy gap coordinate and the diabatic FES from Eqs. 83-85 can be established in several ways. An elegant way to achieve this is re-writing the diabatic free energies as the corresponding probabilities, which then leads to the energy gap probability distribution

$$p_{\text{R}}(\Delta E) = \left\langle \delta\left(\Delta E - \Delta E(\boldsymbol{R})\right) \right\rangle_{\text{R}}, \tag{86}$$

for the reactant state and similarly for the product. The subscript R tells that the sampling is done using the reactant Hamiltonian $H_{11}$. A key assumption in Marcus theory is that the energy gap has a Gaussian distribution; this assumption is equal to the linear response theory[111] which is equivalent to the linear coupling model used in *e.g.* Marcus theory and Anderson-Newns-Schmickler model Hamiltonian in Section 5.1.

$$p_{\mathrm{R}}(\Delta E)=\frac{1}{\sqrt{2\pi\langle(\Delta E-\langle\Delta E_{\mathrm{R}}\rangle)^{2}\rangle_{\mathrm{R}}}}\exp\left[-\frac{(\Delta E-\langle\Delta E_{\mathrm{R}}\rangle)^{2}}{2\langle(\Delta E-\langle\Delta E_{\mathrm{R}}\rangle)^{2}\rangle_{\mathrm{R}}}\right], \tag{87}$$

where $\langle\Delta E_{\mathrm{R}}\rangle$ is the average value of the energy gap in the reactant state and $\sigma_{\mathrm{R}}^{2}=\langle(\Delta E-\langle\Delta E_{\mathrm{R}}\rangle)^{2}\rangle_{\mathrm{R}}$ is the energy gap variance in the reactant state. The corresponding free energy is obtained using Eq. 17

$$G_{\mathrm{R}}(\Delta E)=G_{\mathrm{R}}+\frac{k_{\mathrm{B}}T(\Delta E-\langle\Delta E_{\mathrm{R}}\rangle)^{2}}{2\sigma_{\mathrm{R}}^{2}}+\frac{k_{\mathrm{B}}T}{2}\ln(2\pi\sigma_{\mathrm{R}}^{2}). \tag{88}$$

And similarly for the product state

$$G_{\mathrm{P}}(\Delta E)=G_{\mathrm{P}}+\frac{k_{\mathrm{B}}T(\Delta E-\langle\Delta E_{\mathrm{P}}\rangle)^{2}}{2\sigma_{\mathrm{P}}^{2}}+\frac{k_{\mathrm{B}}T}{2}\ln(2\pi\sigma_{\mathrm{P}}^{2}). \tag{89}$$

Both the reactant and product diabatic FES are quadratic functions of $\Delta E$ and variances define the curvature of the parabola. To arrive at the Marcus equation, one uses the exact linear free energy relation[166]

$$G_{\mathrm{P}}(\Delta E)-G_{\mathrm{R}}(\Delta E)=\Delta E, \tag{90}$$

which restricts the gap fluctuations to be equal[111,167]: $\sigma_{\mathrm{R}}^{2}=\sigma_{\mathrm{P}}^{2}$. From this the following relations can be derived[111]

$$\Delta G^{0}=\frac{1}{2}(\langle\Delta E_{\mathrm{R}}\rangle+\langle\Delta E_{\mathrm{P}}\rangle), \tag{91}$$

$$\lambda=\frac{1}{2}(\langle\Delta E_{\mathrm{R}}\rangle-\langle\Delta E_{\mathrm{P}}\rangle)=\frac{\beta\sigma_{\mathrm{R}}^{2}}{2}. \tag{92}$$

These allow writing the diabatic FESs as[168]

$$G_{\mathrm{R}}(\Delta E)=\frac{\left(\Delta E-(\Delta G^{0}+\lambda)\right)^{2}}{4\lambda}, \tag{93}$$

$$G_{\mathrm{P}}(\Delta E)=\frac{\left(\Delta E-(\Delta G^{0}-\lambda)\right)^{2}}{4\lambda}+\Delta G_{0}. \tag{94}$$

At the transition state, the energy gap $\Delta E=0$, i.e., $G_{\mathrm{R}}(\Delta E)=G_{\mathrm{P}}(\Delta E)$. The Marcus barrier in Eq. 11 is thus obtained from Eqs. 91-92.

### 3.6.5. Computational aspects

The relations in Eqs. 91-92 enable computing the Marcus barriers through MD simulations once the diabatic states have been defined and one only needs to carry out sampling in the reactant and product states. It is also important to note that we have not specified the ensemble in which the sampling is carried out—the same formalism is equally suitable for both canonical, constant charge as well as grand canonical, constant potential calculations. This means that the microscopic version of the Marcus theory and the sampling of Eqs. 83-87 can be readily applied to simulate electrochemical ET kinetics either under constant potential or charge conditions. A theoretical and practical computational difference is that in constant potential calculations the diabatic Hamiltonian (Eq. 82) corresponds to a grand canonical EVB Hamiltonian[81] while in the constant charge calculations it corresponds to a canonical EVB Hamiltonian such that the former leads to grand free energies and reorganization energies in Eqs. 91-92 while the latter canonical free energies in Eqs. 91-92.

It should also be noted that Eqs. 83-94 contain only time-independent thermodynamic expectation values, which means that constant charge calculations can be converted to constant potential results and *vice versa* through a Legendre transform. This is due to ensemble equivalency[169,170] which holds for most electrochemical systems (thin slit systems and thin semiconductor electrodes are known exceptions) and which indicates that the thermodynamic expectation values are independent of the ensemble. In the current context, this indicates that sampling in the grand canonical ensemble can be substituted by sampling in the canonical ensemble and then weighting the computed quantities using the grand canonical version of the Boltzmann weight. For instance, the grand canonical expectation values are obtained from canonical values through $\langle O(\tilde{\mu}_\mathrm{e})\rangle = \sum_{N_\mathrm{e}}\langle O(N_\mathrm{e})\rangle \exp(-\beta N_\mathrm{e}\tilde{\mu}_\mathrm{e})$ for a general observable $O$ either as a function of the number of electrons ($N_\mathrm{e}$) or as function of the electrochemical potential of electrons ($\tilde{\mu}_\mathrm{e}$), i.e., the electrode potential. To reverse this, the canonical expectation values are obtained from the grand canonical sampling by choosing phase space points that have the desired charge. Note also that when explicitly time-dependent dynamic effects are included, the canonical and grand canonical description of ET kinetics are no longer equal as discussed in Section 6.

Eqs. 83-94 serves as the foundational equations for simulating ET kinetics. The atomistic simulations using these equations were realized and accomplished with the classical EVB simulations in the 1980s first for molecular charge transfer reactions in polar solvents[142,143] followed by biomolecular systems[171]. At this time, the interactions and diabatic states were treated using classical force fields and classical MD simulations. These early studies on molecular systems were quickly followed by the first simulations of electrochemical ET[172] still using classical MD force fields and simulations in early 1990s. The approaches based on classical MD simulations of electrochemical ET are still in use. They have been extended from simple outer-sphere reactions to inner-sphere reactions[173], modified with the inclusion of the electrode potential[174,175] for constant potential simulations, and further developed with improved force fields and efficient EVB simulations[146]. While the classical MD simulations achieve very efficient simulation of the solvent environment and allow addressing the long time and length scales needed for comprehensive phase space sampling, the force field parameters still require careful "calibration" or fitting against experiments to achieve quantitative accuracy[146].

Starting from the early 2000s, the application of classical MD simulations of ET has been accompanied by various realizations of EVB-like, diabatic DFT simulations[168], such as embedding methods, constrained DFT, and fragment methods. Again, the earliest simulations were done for molecules in polar solvents[176] followed by biomolecular simulations within DFT/MM-MD approaches[168], and most recently by cDFT-MD simulations of electrochemical ET[114,115]. While DFT-based methods improve the accuracy of energy and force evaluation with respect to classical force fields and remove the calibration with respect to experiments, the achievable sampling efficiency or time and length scales are compromised. Recent advances in DFT simulations of electrochemical ET kinetics include e.g. constant potential GCE-cDFT simulations[81] and the use of tight-binding DFT[177] or QM/MM approaches[178–180] to extending the sampling time and length scales.

Avoiding the compromise between accuracy and sampling efficiency can potentially be achieved by the development and application of machine learning (ML) techniques that have recently started to influence the simulation of ET kinetics. Currently, the ML simulations within the Marcus theory and EVB methods are limited to the molecules in polar solvents[181]. However, by projecting on the previous advances in the simulation of ET with classical and DFT methods as well as the fast developments of ML methods for the simulation of solid-liquid interfaces[182] and constant potential conditions[183–185], it is likely that the ML-based simulations of ET in biochemical and electrochemical systems will be realized rather sooner than later. Nevertheless, simulating the electrified solid-liquid interfaces within the GCE conditions of constant electrode potential are still outstanding challenges for ML methods.

The main pitfall in EVB simulations is the definition of the EVB states themselves as they are not unique but depend on the user's choice, understanding, and intuition of the studied system. Another potential issue in the simulations is using a fine enough discretization ($\alpha$) in the mapping Hamiltonian Eq. 82 for the Hamiltonian such that the FES or umbrella sampling calculations converges (typically ~10 discretization). Regarding the sampling timescales to converge the mapping Hamiltonian for a single $\alpha$ value there are no universal measures to estimate the needed sampling time. Some rough estimates can be obtained from e.g., the solvent rotation correlation times which for water are at least 5-10 ps. More generally, the statistical convergence of the EVB simulations and sampling can be checked by studying e.g., the energy gap correlation and relaxation times so that the gap sampling is robust, confirming that the energy gap distribution and e.g., radial distribution functions are stationary and do not vary with time, and that the free energy is converged. Such tests for EVB-MD seem to be rare as most EVB-MD simulations use classical force fields where the sampling time is not as pressing as in DFT-MD. In Ref.[186] it is concluded that for PCET the free energy converges on times typical for DFT-MD, i.e., within ~10-100 ps. Finally, direct computation of the electronic coupling constant has not been widely tested for periodic systems or in the grand canonical ensemble. While k-point sampling is possible, it is not widely supported or tested[155]. However, the effective coupling constant can be estimated indirectly from the energy gap between the diabatic and adiabatic ground state free energy surfaces through Eq. 172 at the transition state[81]: $H_{12} = E^{\text{diabatic}} - E^{\text{ground}}$. The needed energies are obtained from the EVB-FES simulations for both canonical and grand simulations[81].

In addition to developments in the computational methods for parametrizing Marcus theory for various systems, it should be remembered that the relations underlying the simple evaluation of Marcus parameters (Eqs. 91-92) are valid only within the linear response theory; if non-linear solvent effects are present, enhanced sampling along the energy gap, Eqs. 83-85 need to be employed. Such enhanced sampling simulations would be highly beneficial to study the validity of the linear response theory and to possibly develop non-linear Marcus theories. Finally, it should also be noted that the explicit simulation approaches only yield the total reorganization energy. If the separation to inner- and outer-sphere contributions is desired, the approach in Ref.[187] can be employed, see more details in Appendix 10.4. Furthermore, computational methods to address e.g. non-adiabatic effects and solvent dynamics, discussed in Section 6, should also be advanced.

## 4. Non-adiabatic rate theory

Section 3 provides the central concepts and approaches for understanding the non-adiabatic activation free energy in ET reactions. However, computing the absolute rate of ET reactions through Eq. 1 needs also the prefactors. In non-adiabatic ET reactions the prefactor $\nu_{\mathrm{n}}\kappa_{\mathrm{el}}$ is needed to account for probability of the electron transfer in the transition region, where the transferring electron suddenly jumps or tunnels between the diabatic FESs of the oxidized and reduced states at their intersection, as required by Frank-Condon principle and energy conservation. In this section, we will present a quantum mechanical description of such electron transitions at the intersection point, as this is essential in formulating an absolute rate theory of ET.

### 4.1. The master equation

In the non-adiabatic limit, no hybridized states or covalent bonds are formed, allowing us to treat electron transitions between each metal state and the redox species as independent events[78]. In this case the coupling between the diabatic states is weak and this assumption, the quantum transition of interest at the intersection occurs directly between the diabatic states $k$ and $a$ without the involvement of other states. For each diabatic state, the potential energy surfaces can be obtained within the Born-Oppenheimer approximation, as shown in Eqs. 76 and 77. In particular, these modes include the inner-sphere nuclear vibrational modes and the constrained orientational motions of solvent molecules, commonly referred to as librations[188]. Translational modes of the solvent are neglected, as they are much slower than vibrational modes, and typically frozen during the whole ET process. The oxidized and reduced states each consist of a set of microscopic vibronic states, labeled by the quantum numbers $km$ and $an$, respectively. The corresponding total wavefunctions are denoted as $|\Psi_{km}^0\rangle$ and $|\Psi_{an}^0\rangle$, with the energy eigenvalues given by

$$E_{km} = E_k^{\min} + \epsilon_m, \qquad E_{an} = E_a^{\min} + \epsilon_n, \tag{95}$$

where $E_k^{\min}$ and $E_a^{\min}$ represent the minima of the PES for diabatic states $k$ and $a$, i.e., electronic energies at the equilibrium nuclear configurations. It is important to note that $E_{km}$ and $E_{an}$ are not the PES energies as they also include the kinetic energy of the solvent nuclei and therefore represent the total energy of the system in each diabatic state. Under the Born-Oppenheimer approximation the vibronic wavefunctions can be written as the product of electronic wavefunctions and nuclear wavefunctions:

$$|\Psi_{km}^0\rangle = |\psi_k(\boldsymbol{r};\boldsymbol{R})\rangle|\chi_{km}(\boldsymbol{R})\rangle, \qquad |\Psi_{an}^0\rangle = |\psi_a(\boldsymbol{r};\boldsymbol{R})\rangle|\chi_{an}(\boldsymbol{R})\rangle, \tag{96}$$

where $|\chi_{km}\rangle$ and $|\chi_{an}\rangle$ represent the nuclear wavefunctions in vibrational states $m$ and $n$, respectively.

When the electronic states $k$ and $a$ are coupled, quantum transitions between the vibronic states $|\Psi_{km}^0\rangle$ and $|\Psi_{an}^0\rangle$ may occur. On the timescale of ET, the total system, consisting of the electronic subsystem and all nuclear modes of the redox species and solvent, the latter serving as a heat bath, can be regarded to remain at constant energy and is therefore microcanonical. The corresponding master equation for probability of being in the vibronic state $km$ is given by[98]

$$\frac{dP_{km}(t)}{dt} = \sum_{an} W_{an,km}P_{an}(t) - \sum_{km} W_{km,an}P_{km}(t), \tag{97}$$

where $P_{km}$ and $P_{an}$ are the time-dependent probabilities of the system being in states $|\Psi_{km}^0\rangle$ and $|\Psi_{an}^0\rangle$, respectively. $W_{an,km}$ is the rate of transition from $|\Psi_{an}^0\rangle$ to $|\Psi_{km}^0\rangle$, and $W_{km,an}$ is the rate for the reverse transition. All nuclear modes are considered as a heat bath for the electronic subsystem, which remains in thermal equilibrium regardless of the electronic subsystem. Then $P_{km}(t)$ and $P_{an}(t)$ can be factored into a non-equilibrium probability $P_k(t)$ and $P_a(t)$ for the electronic subsystem and a thermal equilibrium probability $\rho_m$ and $\rho_n$ for the heat bath, i.e.[98],

$$P_{km}(t) \cong P_k(t)\rho_m, \qquad P_{an}(t) \cong P_a(t)\rho_n. \tag{98}$$

Substituting this into the above master equation, and summing over $m$, we obtain

$$\frac{dP_k(t)}{dt} = \sum_m \frac{dP_{km}(t)}{dt} = \sum_a P_a(t) \sum_{mn} W_{an,km}\rho_n - \sum_k P_k(t) \sum_{mn} W_{km,an}\rho_m$$
$$= \sum_a W_{ak} P_a(t) - \sum_k W_{ka} P_k(t), \tag{99}$$

with

$$W_{ak} = \sum_{mn} W_{an,km}\rho_n \,, W_{ka} = \sum_{mn} W_{km,an}\rho_m, \tag{100}$$

where $W_{ka}$ denotes the transition rate of an electron from electronic state $k$ to $a$ , and $W_{ak}$ represents the rate of the reverse transition. Eq. 99 resembles the form of a chemical kinetics equation, with the concentrations of reactants and products replaced by the probabilities of electronic states. In the case of first-order ET, e.g., Eq. 38, $W_{ka}$ and $W_{ak}$ correspond to the oxidation and reduction rate constants[12], respectively, when considering ET between electronic state $k$ on the metal surface and valence state $a$ of a redox species. If multiple electronic states on the metal surface contribute to the ET, the overall rate constant is obtained by summing the transition rates over all pairs of state $k$ and $a$, since transitions between each pair of electronic states are independent under the weak-coupling condition.

In the following subsections we show how the quantities entering Eq. 100 are obtained. In Section 4.2, the transition rates $W_{an,km}$ and $W_{km,an}$ are derived using time-dependent perturbation theory. In Section 4.3, the equilibrium thermal populations $\rho_m$ and $\rho_n$ are described using Boltzmann statistics, where all nuclear modes are treated as a bath of harmonic oscillators with an effective frequency. In Section 4.4, the transition rate between the electronic states, $W_{ak}$ and $W_{ka}$ is formulated. Section 4.5 details the rate constants of ET between the metal surface and redox species. The non-adiabatic ET rate constant is derived in the high-temperature limit, in which all nuclear modes behave classically. This approximation, is valid for most cases of non-adiabatic ET, in which the electronic transition is much slower than the nuclear (vibrational) dynamics. In Appendix 10.3 we complete the classical picture and present an alternative derivation of the non-adiabatic ET rate constant based on Franck-Condon factors, which can incorporate also non-classical high-frequency vibrations and which gives a microscopic expression for the reorganization energy.

### 4.2. The time-dependent perturbation theory

We assume that the ET can be described using one-electron metal states $k$ and that correlations between them can be neglected. This assumption enables treating the electron transition between each metal state $k$ and state $a$ as independent events which is generally true at the weak coupling limit, i.e., when the coupling $V_k$ is small. The time evolution of the system state $|\Psi\rangle$ with the total perturbation Hamiltonian $H = H^0 + V_k$ can be described by the time-dependent Schrödinger equation,

$$i\hbar\big|\dot{\Psi}(t)\big\rangle = H|\Psi(t)\rangle, \tag{101}$$

where the dot over the physical quantity denotes its differential or derivative with respect to time $t$. In the absence of coupling between electronic states, the system Hamiltonian is unperturbed $H = H^0$ and the system starts in a pure or superposition state of the diabatic eigenstates of $H^0$:

$$|\Psi(0)\rangle = c_{km}(0)\big|\Psi_{km}^0\big\rangle + c_{an}(0)|\Psi_{an}^0\rangle, \tag{102}$$

where $c_{km}(0)$ and $c_{an}(0)$ are the expansion coefficients at $t = 0$, and the squares of their moduli represent the probabilities of finding the system in the diabatic states $|\Psi_{km}^0\rangle$ and $|\Psi_{an}^0\rangle$, respectively. With this initial condition, $|\Psi(t)\rangle$ can be generally solved as,

$$|\Psi(t)\rangle = c_{km}(0)|\Psi_{km}^0\rangle \exp\left(-\frac{iE_{km}t}{\hbar}\right) + c_{an}(0)|\Psi_{an}^0\rangle \exp\left(-\frac{iE_{an}t}{\hbar}\right), \quad (103)$$

where $\hbar$ is the reduced Planck constant. Without coupling, the probability of finding the system in the state $|\Psi_{km}^0\rangle$ or $|\Psi_{an}^0\rangle$ at time $t$ is then given by,

$$P_i(t) = \langle\Psi_i^0|\Psi(t)\rangle^2 = |c_i(0)|^2 \; (i = km, an), \quad (104)$$

which remains constant at their initial values. For instance, if the system initially is in the diabatic state $k$, it will not be found in the diabatic state $a$ at any subsequent time. This implies that no transitions between the diabatic states $k$ and $a$ can take place which is expected as $H^0$ excludes the electronic coupling between the metal state $k$ and the valence state $a$, thereby preventing the electron exchange between these two electronic states.

The electron transition in the system may be described by including a very small time-independent electronic coupling potential $V_k$ between the states $k$ and $a$. The small $V_k$ can be treated as a perturbation acting on the system starting after $t = 0$, and the system Hamiltonian $H = H^0 + V_k$. With inclusion of perturbation the expansion coefficients in Eq. 103 can no longer remain at their initial values and are expected to be time-dependent, i.e.,

$$|\Psi(t)\rangle = c_{km}(t)|\Psi_{km}^0\rangle \exp\left(-\frac{iE_{km}t}{\hbar}\right) + c_{an}(t)|\Psi_{an}^0\rangle \exp\left(-\frac{iE_{an}t}{\hbar}\right). \quad (105)$$

Substituting the above expression into Eq. 101, we have,

$$\begin{aligned} i\hbar\left(\dot{c}_{km}(t)|\Psi_{km}^0\rangle e^{-\frac{iE_{km}t}{\hbar}} + \dot{c}_{an}(t)|\Psi_{an}^0\rangle e^{-\frac{iE_{an}t}{\hbar}}\right) \\ = c_{km}(t)V_k|\Psi_{km}^0\rangle e^{-\frac{iE_{km}t}{\hbar}} + c_{an}(t)V_k|\Psi_{an}^0\rangle e^{-\frac{iE_{an}t}{\hbar}}. \end{aligned} \quad (106)$$

Multiplying from the left with the complex conjugate of $|\Psi_{an}^0\rangle$, i.e., $\langle\Psi_{an}^0|$, on both sides of the above equation, and integrating over the electronic coordinates, gives

$$i\hbar\dot{c}_{an}(t)e^{-\frac{iE_{an}t}{\hbar}} = c_{km}(t)\langle\Psi_{an}^0|V_k|\Psi_{km}^0\rangle\, e^{-\frac{iE_{km}t}{\hbar}} + c_{an}(t)\langle\Psi_{an}^0|V_k|\Psi_{an}^0\rangle e^{-\frac{iE_{an}t}{\hbar}}. \quad (107)$$

Since the perturbation $V_k$ is very weak, the variations of $c_{km}(t)$ and $c_{an}(t)$ are minimal. Thus, as a first-order approximation, we can replace them with their initial values in the above equation. Further assuming that the system is initiated from the vibronic state $|\Psi_{km}^0\rangle$, we have,

$$i\hbar\dot{c}_{an}(t) = V_{an,km}e^{i\omega_{an,km}t}, \quad (108)$$

with the matrix element and frequency

$$V_{an,km} = \langle\Psi_{an}^0|V_k|\Psi_{km}^0\rangle = \langle\psi_a|V_k|\psi_k\rangle\langle\chi_{an}|\chi_{km}\rangle = H_{ak}S_{an,km}, \quad (109)$$

$$\omega_{an,km} = (E_{an} - E_{km})/\hbar, \quad (110)$$

where $S_{an,km}^2 = |\langle\chi_{an}|\chi_{km}\rangle|^2$ is the Franck-Condon factor, $H_{ak} = \langle\psi_a|V_k|\psi_k\rangle$ is the electronic matrix element. In writing Eq. 109, we have invoked the Condon approximation, by which the coupling matrix element between any two vibronic states was written as a product of the electronic

matrix element and a nuclear overlap function. Integrating the above differential equation Eq. 108 with the initial condition $c_{an}(0) = 0$, we obtain,

$$c_{an}(t) = \frac{1}{i\hbar}\int_0^t V_{an,km} e^{i\omega_{an,km}t} dt. \tag{111}$$

The probability of the system transitioning from the state $|\Psi_{km}^0\rangle$ to $|\Psi_{an}^0\rangle$ is then the probability of finding the system in the initial diabatic state $|\Psi_{an}^0\rangle$, i.e.,

$$P_{km,an}(t) = |c_{an}(t)|^2 = \frac{1}{\hbar^2}\left|\int_0^t V_{ak} e^{i\omega_{an,km}t} dt\right|^2. \tag{112}$$

If the perturbation $V_k$ is time-independent at $t > 0$, the above integral can be calculated as,

$$\begin{aligned} P_{km,an}(t) &= \frac{|V_{an,km}|^2}{\hbar^2}\left|\frac{e^{i\omega_{an,km}t}-1}{i\omega_{ak}}\right|^2 = \frac{|V_{an,km}|^2}{\hbar^2}\frac{2-2\cos(\omega_{an,km}t)}{\omega_{an,km}^2} \\ &= \frac{|V_{an,km}|^2 t}{\hbar^2}\frac{\sin^2\left(\frac{1}{2}\omega_{an,km}t\right)}{\left(\frac{1}{2}\omega_{an,km}\right)^2 t}. \end{aligned} \tag{113}$$

As $t \to \infty$, we have $\lim_{t\to\infty}\frac{\sin^2(\omega t)}{\omega^2 t} = \pi\delta(\omega)$, where $\delta(\omega)$ is the Dirac delta function and the preceding equation becomes

$$P_{km,an}(t) = \frac{|V_{an,km}|^2 t}{\hbar^2}\pi\delta\left(\frac{1}{2}\omega_{an,km}\right) = \frac{2\pi}{\hbar}|V_{an,km}|^2 t\cdot\delta(E_{km}-E_{an}). \tag{114}$$

Here, the transition probability is proportional to time. The transition rate from $km$ to $an$ is obtained as the transition probability per unit time:

$$W_{km,an} = \frac{dP_{km,an}(t)}{dt} = \frac{2\pi}{\hbar}|V_{an,km}|^2\delta(E_{km}-E_{an}). \tag{115}$$

This equation is the Fermi golden rule between two vibronic states, $an$ and $km$.

### 4.3. Thermal statistics of a harmonic oscillator

In this section, we derive the thermal distribution of the nuclear modes over its vibrational states in the oxidized ($\rho_m$) and reduced states ($\rho_n$), respectively, as appearing in Eqs. 98 and 100. We consider one-dimension case, where the FESs of the oxidized states are given by Eqs. 69 and 70. All nuclear modes are modeled using an effective vibrational mode with the same frequency $\omega_{\text{eff}} = \sqrt{2\lambda/\mu}$ for both oxidized and reduced states, where $\mu$ denotes the reduced mass of the motion along the effective nuclear coordinate $\xi$. The corresponding effective vibrational energies are then given by

$$\epsilon_m = \left(m+\frac{1}{2}\right)\hbar\omega_{\text{eff}}, \qquad \epsilon_n = \left(n+\frac{1}{2}\right)\hbar\omega_{\text{eff}}. \tag{116}$$

Assuming that all nuclear modes serve as a heat bath that remains in thermal equilibrium, the Boltzmann statistics hold in the FESs and the probability of finding the oxidized state with the vibrational state $m$ is given by

$$\rho_m = \frac{e^{-\beta\epsilon_m}}{Q_\mathrm{c}} = \frac{e^{-\beta\epsilon_m}}{\sum_{m=0}^{\infty} e^{-\beta\epsilon_m}} = \frac{e^{-\beta\epsilon_m}}{\left[2\sinh\left(\frac{1}{2}\beta\hbar\omega_\mathrm{eff}\right)\right]^{-1}}, \tag{117}$$

where $Q_\mathrm{c}$ is the canonical partition function approximated as an effective bath here.

At the high-temperature limit, the nuclear motions behave classically. The classical nuclei approximation is valid when the relevant nuclear motions driving the ET reaction include e.g. the intramolecular vibrations of the redox species and solvent molecules, and orientational librations of solvent molecules. For water, its orientational librations fall within a broad microwave band, ranging from $10^{10}$ Hz to $10^{12}$ Hz, with the corresponding energy spacing of these modes ranging from $0.000041$ eV to $0.0041$ eV. Since this energy spacing is much smaller than the thermal energy at the room temperature ($k_\mathrm{B}T = 0.025$ eV at $T = 298.15$ K), these modes can be treated as *classical*. On the other hand, the vibrational frequencies of the water are found within the narrow, high-frequency infrared band, centered around $10^{14}$ Hz[12]. Such fast vibrations can be assumed to respond instantaneously, i.e., adiabatically, to the change in the electronic state of the system. In this high-temperature limit, where the classical solvent treatment is valid, the discreteness of the vibrational states vanishes which means that the influence of solvent's vibrational properties on ET can be incorporated in $\varepsilon_\infty$, with a suggested value of $4.2\,\varepsilon_0$ for water[12].

At the high-temperature limit the state of harmonic oscillator presenting the effective classical nuclei environment is described by the corresponding vibrational momenta and configurations or coordinates. The calculation of thermal averaged quantities is instead performed in the phase space. Since the classical Hamiltonians of the vibrational modes in the oxidized and reduced states appear within the $\delta$-function in the transition rate $W_{km,an}$, the kinetic energy terms cancel out. As a result, $W_{km,an}$ becomes independent of the momenta, and its thermal average is obtained from over the configurational distribution function, which is determined by the potential energy of harmonic oscillators. Given the one-dimensional harmonic potential in Eq. 69 for the solvent vibrations in the oxidized state, the corresponding configurational distribution function is expressed as

$$\rho_\mathrm{ox}(\xi) = \frac{e^{-\beta\lambda\xi^2}}{Q_\mathrm{c}}, \tag{118}$$

where $Q_\mathrm{c}$ is the canonical configuration partition function and is obtained by requiring that $\rho_\mathrm{ox}$ is normalized:

$$Q_\mathrm{c} = \int_{-\infty}^{+\infty} e^{-\beta\lambda\xi^2} d\xi = \sqrt{\frac{\pi}{\beta\lambda}}. \tag{119}$$

Similarly, the configurational distribution function for the one-dimensional harmonic potential in Eq. 70 can be obtained as,

$$\rho_\mathrm{red}(\xi) = \frac{e^{-\beta\lambda(\xi-1)^2}}{Q_\mathrm{c}}. \tag{120}$$

It can be readily demonstrated that this yields the same coordinate partition function, owing to the similar quadratic dependence of the fluctuating energy on $\xi$ in both the oxidized and reduced states.

### 4.4. Electronic transition rate

The treatment in Section 3.3. achieves a classical, effective harmonic treatment of the nuclear modes. Because the harmonic frequencies of the nuclear modes are assumed to be equal in both the oxidized and reduced states (Eq. 116), the effective nuclear wave functions are also the same. As a result, the nuclear overlap term in the Golden rule, i.e., Eq. 115 equals unity and the difference in the harmonic energies in the $\delta$-function can be replaced with the harmonic potentials due to the cancellation of the kinetic energy contributions. The thermal averaged transition rate $W_{ka}$ in Eq. 100 then takes the form

$$W_{ka} = \frac{2\pi}{\hbar} |H_{ak}|^2 \int \rho_{\mathrm{ox}}(\xi) \delta\left(E_k^{\min} + \lambda\xi^2 - E_a^{\min} - \lambda(\xi - 1)^2\right) d\xi. \tag{121}$$

The argument in the $\delta$-function coincides with the difference between the FESs shown in Eqs. 69 and 70, which vanishes at the crossing the two FESs, i.e., at the nuclear coordinate shown in Eq. 53. Then the above integral is calculated using the sifting property of the $\delta$-function, leading to the Levich-Dogonadze formula[5] for the two-level ET rate

$$W_{ka} = \frac{2\pi}{\hbar} \sqrt{\frac{\beta}{4\pi\lambda}} |H_{ak}|^2 e^{-\frac{\beta\left(\lambda + \Delta G^0(\epsilon_k)\right)^2}{4\lambda}}. \tag{122}$$

By substituting Eq. 63 into the above equation, we have

$$W_{ka} = \frac{2\pi}{\hbar} \sqrt{\frac{\beta}{4\pi\lambda}} |H_{ak}|^2 e^{-\frac{\beta(\lambda + e_0\eta + \epsilon_{\mathrm{F}} - \epsilon_k)^2}{4\lambda}}. \tag{123}$$

The electrons further follow the Fermi-Dirac distribution that describes the probability $f(\epsilon_k)$ that a state with energy $\epsilon_k$ is occupied by an electron,

$$f(\epsilon_k) = \frac{1}{e^{\beta(\epsilon_k - \epsilon_{\mathrm{F}})} + 1}. \tag{124}$$

Considering this thermal distribution on the occupancy probability of the state $k$, the electron transfer probability at this state $k$ should be corrected by multiplying $W_{ka}$ by the Fermi-Dirac distribution function, namely,

$$W_{ka}^{\mathrm{T}}(\epsilon_k) = f(\epsilon_k) W_{ka} = \frac{2\pi}{\hbar} \sqrt{\frac{\beta}{4\pi\lambda}} |H_{ak}|^2 f(\epsilon_k) e^{-\frac{\beta(\lambda + e_0\eta + \epsilon_{\mathrm{F}} - \epsilon_k)^2}{4\lambda}}. \tag{125}$$

In the oxidation reaction, electrons are transferred from the reduced species to the metal surface. It is evident that Eqs. 115 and 121 remain valid in this case, as we only need to exchange the labels of $a$ and $k$ in these equations and replace $\rho_{\mathrm{ox}}$ in Eq. 121 with $\rho_{\mathrm{red}}$, which does not alter the form of final expressions. Nevertheless, two distinct points should be noticed in the oxidation case as compared to the reduction reaction. First, we need to consider the distribution probability of the reduced states over the nuclear coordinate rather than the oxidized species, as the reduced species serves as the reactant in the oxidation process. Second, we must account for the unoccupancy probability of state $k$, since empty electrode states are required to accept electrons during oxidation. The temperature-dependent electron transfer probability $W_{ak}^{\mathrm{T}}(\epsilon_k)$ from the reduced species to a metal state $k$ is then expressed as,

$$W_{ak}^{\mathrm{T}}(\epsilon_k) = \frac{2\pi}{\hbar}\sqrt{\frac{\beta}{4\pi\lambda}}|H_{ka}|^2\big(1 - f(\epsilon_k)\big)e^{-\frac{\beta(\lambda - e_0\eta + \epsilon_k - \epsilon_{\mathrm{F}})^2}{4\lambda}}. \tag{126}$$

The above equation for oxidation could also be directly obtained from the corresponding reduction rate expression by invoking detailed balance between the oxidation and reduction rate constants and the equilibrium constant for the reaction in Eq. 38.

### 4.5. Rate constant

As we treat all ET events as independent processes, the total probability of electrons transferring from the metal surface to an oxidized species per unit time, namely the reduction rate constant $k_{\mathrm{red}}$, is the sum of $W_{ka}^{\mathrm{T}}(\epsilon_k)$ over all metal electronic states,

$$k_{\mathrm{red}} = \sum_k W_{ka}^{\mathrm{T}}(\epsilon_k) = \frac{2\pi}{\hbar}\sqrt{\frac{\beta}{4\pi\lambda}}\sum_k |H_{ak}|^2 f(\epsilon_k)e^{-\frac{\beta(\lambda + e_0\eta + \epsilon_{\mathrm{F}} - \epsilon_k)^2}{4\lambda}}. \tag{127}$$

By applying the shifting property of the Dirac delta function, we have,

$$\begin{aligned} k_{\mathrm{red}} &= \int \sum_k W_{ak}^{\mathrm{T}}(\epsilon)\delta(\epsilon - \epsilon_k)\, d\epsilon \\ &= \frac{2\pi}{\hbar}\sqrt{\frac{\beta}{4\pi\lambda}}\int f(\epsilon)e^{-\frac{\beta(\lambda + e_0\eta + \epsilon_{\mathrm{F}} - \epsilon)^2}{4\lambda}}\sum_k |H_{ak}|^2\delta(\epsilon - \epsilon_k)\, d\epsilon. \end{aligned} \tag{128}$$

The integral is evaluated from negative infinity to the positive infinity, going through each electronic states at the metal surface. As alluded in Section 3.4, appreciable contributions to the electron transfer rate are mostly confined to the energy region around the Fermi level, spanning over a few $k_{\mathrm{B}}T$ around the Fermi level. Therefore, for practical purposes, the integration can be limited to this specific energy region near the Fermi level, hereafter referred to as the active energy region of the metal surface. If the properties of the metal electrons are nearly identical within this region, it is reasonable to assume that the coupling strength remains constant across the considered metal states: $|H_{ak}|^2 = |H_{ka}|^2 \approx |V|^2$. This leads to the Gerischer formula[6,36,61],

$$\begin{aligned} k_{\mathrm{red}} &= \frac{2\pi|V|^2}{\hbar}\sqrt{\frac{\beta}{4\pi\lambda}}\int f(\epsilon)e^{-\frac{\beta(\lambda + e_0\eta + \epsilon_{\mathrm{F}} - \epsilon)^2}{4\lambda}}\sum_k \delta(\epsilon - \epsilon_k)\, d\epsilon \\ &= \frac{2\pi|V|^2}{\hbar}\sqrt{\frac{\beta}{4\pi\lambda}}\int f(\epsilon)\rho(\epsilon)e^{-\frac{\beta(\lambda + e_0\eta + \epsilon_{\mathrm{F}} - \epsilon)^2}{4\lambda}} d\epsilon, \end{aligned} \tag{129}$$

with the DOS of electrons at the metal surface,

$$\rho(\epsilon) = \sum_k \delta(\epsilon - \epsilon_k). \tag{130}$$

The DOS integration over a specific energy range yields the total number of electronic states within that interval. Based on Eq. 129, we can also define the coupling strength between a specific energy level $\epsilon$ of the metal surface and the redox species,

$$\Delta(\epsilon) = \pi \sum_k |H_{ak}|^2 \delta(\epsilon - \epsilon_k). \tag{131}$$

We should be careful to distinguish between $H_{ak}$ and $\Delta(\epsilon)$: $H_{ak}$ represents the coupling strength between a specific metal state and the redox species, while $\Delta(\epsilon)$ refers to the coupling strength at a specific energy level, which may contain degenerate metal states. In other words, $\Delta(\epsilon)$ represents the coupling strength weighted by the DOS of the metal electrons. If we further assume $\Delta(\epsilon)$ that is independent of the energy level in the active energy region, namely $\Delta(\epsilon) \approx \Delta$, we reach the wide-band approximation, which assumes that both the DOS of the metal electrons and its coupling with the redox species remain relatively constant over the active energy region[189]. With this, then Eq. 129 can be simplified into,

$$k_{\mathrm{red}} = \frac{\Delta}{\hbar}\sqrt{\frac{\beta}{\pi\lambda}} \int f(\epsilon) e^{-\frac{\beta(\lambda + e_0\eta + \epsilon_{\mathrm{F}} - \epsilon)^2}{4\lambda}} d\epsilon. \tag{132}$$

This equation is commonly referred to as the Marcus-Hush-Chidsey formula[190–192]. It also shows the Fermi-Dirac distribution can be regarded as a modification to the Marcus energy barrier due to the thermal effects of metal electrons. The modified barrier at the energy level $\epsilon$ is,

$$\Delta G^{\neq}_{\mathrm{red,T}}(\epsilon) = \frac{(\lambda + e_0\eta + \epsilon_{\mathrm{F}} - \epsilon)^2}{4\lambda} - \frac{\ln f(\epsilon)}{\beta}. \tag{133}$$

Figure 7 shows the dependence of reduction energy barrier modified by the thermal effects of metal electrons on the energy level at various overpotentials. Compared with the results in Figure 6, the thermal distribution of metal electrons causes the energy barrier above the Fermi level to increase significantly as the energy deviates further from the Fermi level, resulting in a pronounced minimum free energy barrier occurring near the Fermi level.

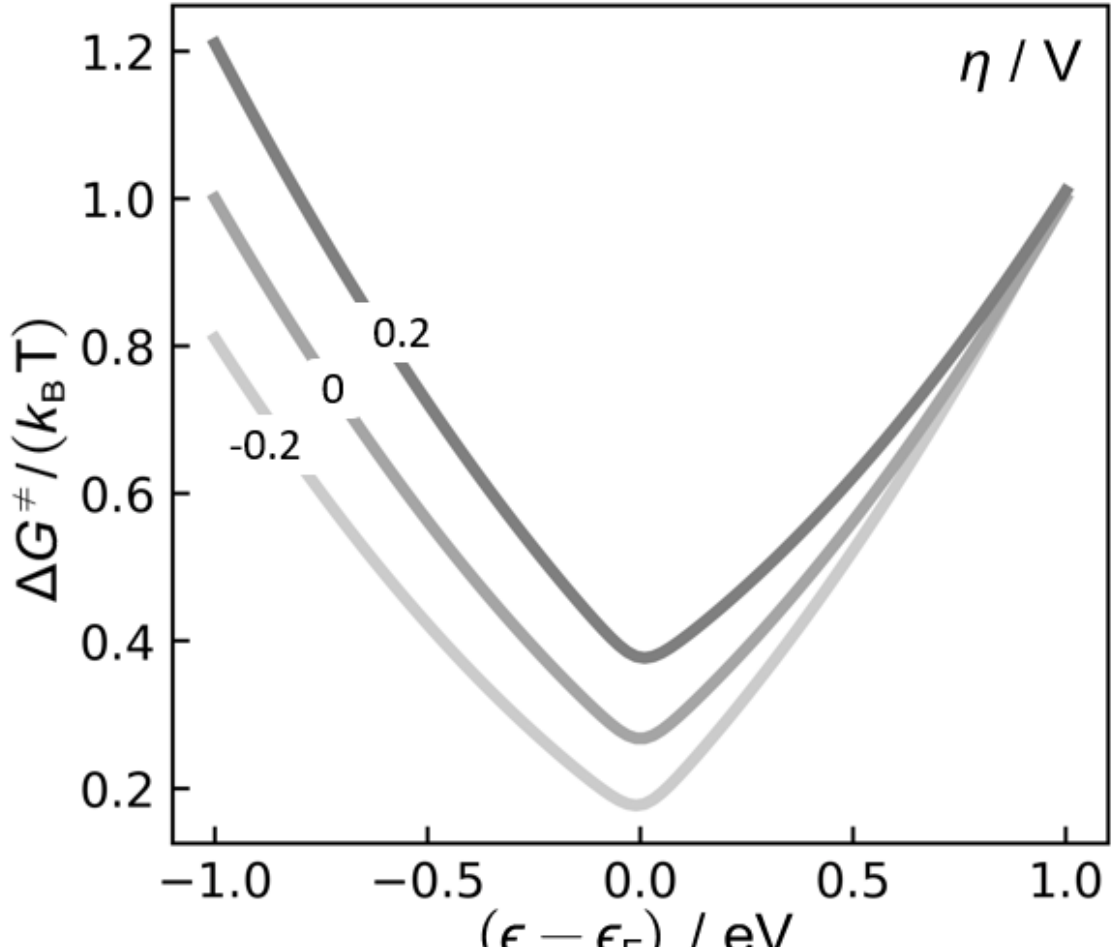

Figure 7. The dependence of the free energy barrier modified by the thermal effects of metal electrons on the energy level for the reduction reaction at overpotentials of -0.2 V, 0 V, and 0.2 V, with line colors transitioning from light to dark. The reorganization energy is set to 1 eV.

Given that the free energy barrier for electron transfer reaches a clear minimum around the Fermi level, we can focus exclusively on electron transfer at this energy level. By approximating the Fermi-Dirac distribution function as a Dirac delta function at the Fermi level, we have,

$$k_{\text{red}} = \frac{\Delta}{\hbar}\sqrt{\frac{\beta}{\pi\lambda}}e^{-\frac{\beta(\lambda+e_0\eta)^2}{4\lambda}}. \tag{134}$$

A more accurate approximation of the Marcus-Hush-Chidsey formula, obtained through asymptotic matching over the entire realistic parameter range, is provided in Ref.[192]. As for the oxidation rate constant $k_{\text{ox}}$, we sum the probability of electrons transferring from a reduced species to the metal surface per unit time over all metal states,

$$k_{\text{ox}} = \sum_k W_{ak}^{\text{T}}(\epsilon_k) = \frac{2\pi}{\hbar}\sqrt{\frac{\beta}{4\pi\lambda}}\sum_k |H_{ak}|^2\big(1 - f(\epsilon_k)\big)e^{-\frac{\beta(\lambda-e_0\eta+\epsilon_k-\epsilon_{\text{F}})^2}{4\lambda}}. \tag{135}$$

By applying the sifting property of the Dirac delta function, we have

$$\begin{aligned} k_{\text{ox}} &= \int \sum_k W_{ak}^{\text{T}}(\epsilon)\delta(\epsilon - \epsilon_k)\, d\epsilon \\ &= \frac{2\pi}{\hbar}\sqrt{\frac{\beta}{4\pi\lambda}}\int \big(1 - f(\epsilon)\big)e^{-\frac{\beta(\lambda-e_0\eta+\epsilon_k-\epsilon_{\text{F}})^2}{4\lambda}}\sum_k |H_{ak}|^2\delta(\epsilon - \epsilon_k)\, d\epsilon. \end{aligned} \tag{136}$$

Similarly, we can readily derive corresponding approximate forms of Eqs. 130, 133 and 135 but these are not presented here. Finally, we connect the rate constants $k_{\text{ox}}$ and $k_{\text{red}}$ with the corresponding reaction rates, which are

$$v_{\text{red}} = k_{\text{red}}N_A c_{\text{ox}},\ \ v_{\text{ox}} = k_{\text{ox}}N_A c_{\text{red}}, \tag{137}$$

where $c_{\text{ox}}$ and $c_{\text{red}}$ are the local molar concentrations of the oxidized and reduced species at the reaction site of the redox species, to be determined in Section 7. $N_A$ is the Avogadro constant.

$$k_{\text{red}} = \frac{2\pi}{\hbar}\sqrt{\frac{\beta}{4\pi\lambda}}\int f(\epsilon)e^{-\frac{\beta(\lambda+e_0\eta+\epsilon_{\text{F}}-\epsilon)^2}{4\lambda}}\sum_k |H_{ak}|^2\delta(\epsilon - \epsilon_k)\, d\epsilon. \tag{138}$$

It should be noted that Eq. 128 remains applicable for ET reactions at semiconductor electrodes, which are usually considered to be non-adiabatic[45]. However, in semiconductors the Fermi level lies within the band gap between the conduction and valence bands. Consequently, the energy integration in Eq. 128, performed around the Fermi level for metal electrodes, must instead be separated into contributions from the conduction band above the Fermi level and the valence band below it. The presence of the band gap at semiconductor electrodes significantly reduces the reaction currents compared with those at metal electrodes. Another feature of ET reactions at semiconductor electrodes is that, due to the very low charge-carrier concentration, most of the interfacial potential drop occurs on the electrode side[44]. This is opposite to the situation at metal electrodes, where the potential drop mainly occurs on the electrolyte side. The electrostatic potential variation on the electrode side leads to band bending near the electrode surface. Therefore, instead of correcting for the potential drop on the solution side as in ET reactions at metal electrodes, the rate constant at semiconductor electrodes should include a correction for the electrostatic potential difference between the electrode bulk and the surface. The modified Marcus theory for semiconductor electrode-liquid interface has been developed in Ref.[193].

As mentioned in Section 1.2.6, another type of non-adiabatic ET reaction is long-range electrochemical ET, in which electrons tunnel between the electrode surface and redox species over relatively long distances. For long-range electrochemical ET, the distance between the redox species and electrode surface becomes an important parameter affecting the ET rate, as the coupling matrix element, $|H_{ak}|^2$, is sensitive to this distance and determines the tunnelling probability. Specifically, a longer distance, corresponding to a smaller overlap between the wavefunctions of the electrode electrons and that of the valence electron of the redox species, leads to weaker coupling strength. Therefore, a particular feature of long-range electrochemical ET is that its rate potentially depends on the electrode surface charge, which influence the lability of the surface electronic profile Specifically, as the electrode surface becomes more negatively charged, electron spillover extends further into the solution, increasing the overlap between the electronic wavefunctions of the electrode and the redox species and thus facilitating tunnelling. In this sense, it is beneficial to approximate the coupling matrix element by the overlap of the electron densities of the redox species and the electrode, rather than by the overlap of their wavefunctions, in order to explore the lability effects on long-range chemical ET reactions. By doing so, Kornyshev, Kuznetsov and Ulstrup recast Eq. 128 and obtained an approximate expression[194], in which the rate constant is proportional to the overlap integral between the electronic densities of the electrode and the redox species. A jellium model for the metal electrons and an exponential form for the valence electron density of the redox species were then used to calculate this overlap integral[194]. They found that the current-potential relationship can be significantly affected by the lability of the surface electronic profile, e.g., the asymmetry in anodic and cathodic current-potential relationships, with stronger effects observed when the electron localization on the redox species is greater and the metal electron density is smaller. This calculation can be further improved using the recently developed density potential functional theory by Huang[37,40], in which the responses of the metal electrons and electrolyte particles to the interfacial potential are incorporated self-consistently through orbital-free DFT and statistical field theory. It is also attractive for developing ET theories applicable to more general ET reactions, formulated in terms of electronic density rather than wavefunctions, thereby enabling its integration with density potential functional theory to describe reaction reactivity at electrochemical interfaces.

### 4.6. High-frequency nuclear modes

Up to now, we have considered only low-frequency nuclear modes, which are treated purely classically. The nuclear coordinates associated with these modes fluctuate toward a prerequisite configuration at which the quantum subsystem, e.g., the electrons considered in the preceding sections, can undergo transitions via tunnelling. However, the inclusion of high-frequency nuclear modes is expected to be of great importance for ET processes at low temperatures and for reactions involving the transfer of light particles, e.g., protons. The high-frequency nuclear modes may include local modes within the redox species, i.e., intramolecular vibrations, and the environmental modes associated with the high-frequency tails of the solvent dielectric dispersion spectrum, typically the infrared band, as discussed in Section 4.3. With these modes included, quantum transitions in the transition region, as shown in Figure 1, involves both electron tunnelling and nuclear tunnelling corresponding to the reorganization of high-frequency nuclear modes. In other words, at the prerequisite configuration of classical nuclei, quantum transitions occur between vibronic states of the reactant and those of the product. Thus, ET between a specific metal electronic state $k$ and the valence state $a$ of the redox species can be viewed as transitions between manifolds of high-frequency vibrational states of the reactant and that of the product. Assuming that these vibrational states for the reactant and product are described by nuclear wavefunctions, $\chi_{kv}$ and $\chi_{aw}$, respectively, with vibrational energies, $\epsilon_v$ and $\epsilon_w$, the Levich-Dogonadze formula in Eq. 122 is modified as[12,46,195]

$$W_{ka} = \frac{2\pi}{\hbar}\sqrt{\frac{\beta}{4\pi\lambda}}|H_{ak}|^2\frac{1}{Q_\mathrm{q}}\sum_{vw}|\langle\chi_{aw}|\chi_{kv}\rangle|^2 e^{-\frac{\epsilon_v}{k_\mathrm{B}T}} e^{-\frac{\beta\left(\lambda+\Delta G_{vw}^0(\epsilon_k,\epsilon_v,\epsilon_w)\right)^2}{4\lambda}}, \tag{139}$$

with the canonical partition function of the high-frequency vibrational states of the reactant,

$$Q_\mathrm{q} = \sum_v e^{-\frac{\epsilon_v}{k_\mathrm{B}T}}, \tag{140}$$

and the reaction free energy between the vibronic states of the reactant and product,

$$\Delta G_{vw}^0(\epsilon_k,\epsilon_v,\epsilon_w) = \Delta G^0(\epsilon_k) + \epsilon_w - \epsilon_v = e_0\eta + \epsilon_\mathrm{F} - \epsilon_k + \epsilon_w - \epsilon_v. \tag{141}$$

Equation 138 represents a statical average of quantum transition rates over a manifold of reactant vibrational states. The coupling strength between the vibrionic states of the reactant and product is determined by the electronic coupling constant, $|H_{ak}|^2$, and the overlap factor of the nuclear wavefunctions, $S_{aw,kv} = |\langle\chi_{aw}|\chi_{kv}\rangle|^2$. The effect of intramolecular quantum modes on the relationship between the rate constant and the reaction free energy is discussed in detail in Ref.[195].

## 5. Adiabatic ET rate theory

In the preceding two Sections 3 and 4 we have focused on the non-adiabatic ET. The electronic coupling between the diabatic states has been assumed small, which allows treating the transition state as the crossing point between the diabatic FES as well as treating electron transition probabilities as independent events between individual electronic states of the metal surface and the valence electronic state of the redox species. In adiabatic ET, where the coupling between the metal states and the valence state of the redox species is strong, such treatments are no longer valid as all electronic states and transitions between them are coupled and need to be considered collectively. The strong electronic coupling is often observed when the distance between the electrode surface and the reactants is short or when a covalent bond is formed. Typical examples include inner-sphere electrostatic pure ET or decoupled proton-electron transfer reactions taking place on the electrode surface[196] and adsorption coupled with charge transfer[114]. Also several outer-sphere ET reactions can be adiabatic on different metallic electrodes[197]. From an experimental point of view, understanding whether the ET reaction is non-adiabatic or adiabatic is pivotal when selecting the correct model for analyzing the results and thereby drawing the conclusions[198]. In this section, we discuss the case of adiabatic ET and focus on how stronger coupling necessitates several modifications to the theory and simulation of ET as we shall show below by using model Hamiltonian and EVB approaches.

### 5.1. Model Hamiltonian

Instead of using the exact Hamiltonian and carrying out full quantum mechanical calculations, the model Hamiltonian approach adopts a simplified or effective Hamiltonian treatment to capture the essential physics of the system; this is particularly advantageous for achieving conceptual clarity and saving computational cost in the study of (strongly) interacting many-body systems. A well-known example is the Anderson impurity model, which describes the interaction between the localized electronic state of an impurity atom and the valence electrons in the metallic host[83]. This model was later extended by Newns to study the chemisorption of a hydrogen atom on the metal surface in vacuum[85]. The Anderson-Newns model describe the combined systems of the metal surface and adsorbate in the basis of the one-electron metal and adsorbate states, which are obtained when the metal and adsorbate are infinitely far apart. These one-electron states are a set of unperturbed and orthonormal electronic state with a continuous energy spectrum $\epsilon_k$ for the metal electronic states and a localized energy level $\epsilon_a$ for the valence state of adsorbate. As the adsorbate approaches the metal surface, they interact and their electronic interaction is characterized by coupling matrix elements $H_{ka}$ between the metal states $k$ and the valence state $a$. The magnitude of $H_{ka}$ depends exponentially on the distance between the metal surface and adsorbate, and is typically larger at shorter distances[199].

In the basis of the diabatic states, the electronic Hamiltonian, $H_{\text{el}}$, in the Anderson-Newns model can be expressed in the particle number representation, i.e., the second quantization form, as,

$$H_{\text{el}} = \epsilon_a n_a + \sum_k \epsilon_k n_k + \sum_k \left(H_{ka} c_k^\dagger c_a + H_{ak} c_a^\dagger c_k\right), \tag{142}$$

where $c_i^\dagger$, $c_i$ and $n_i = c_i^\dagger c_i$ are the creation, annihilation, and number particle operators for the one-electron state $i$ $(i = k, a)$, respectively. Here, the interaction between the two spin states in the valence orbital is not considered, which is reasonable in cases where no valence electrons are initially present, for example, in the adsorption of a hydrogen ion and when the electron correlation effects are expect small. However, if necessary, the spin interaction can be taken into account by incorporating an additional term into $\epsilon_a$ within the Hartree-Fork approximation[85].

When considering the electron transfer between the metal surface and the redox species embedded in the solvent, an additional Hamiltonian, $H_{\text{n}}$, needs to be introduced to describe the interactions between the electrons and nuclear modes. As introduced in Section 3, these nuclear modes include discrete, localized modes within the redox species and a continuum of solvent modes. In this context, Schmickler combined the Anderson-Newns model for the electronic subsystem with a representation of the nuclear modes as a phonon bath or a set of harmonic oscillators at the metal-solution interface. A key assumption in Schmickler's treatment is that the coupling between the electronic subsystem and nuclear modes is assumed to vary linearly with the occupation number in state $a$[9–11,200,201]. This assumption implies that the nuclear free energy of the redox species depends *linearly* on the occupation number $n_a$ and presents another version of the linear response theory. In this case, $H_{\text{n}}$ can be expressed as a switching function that interpolates between the nuclear free energies associated with the oxidized and reduced states, weighted by $n_a$:

$$H_{\text{n}} = (1 - n_a)\left(G_{\text{ox}}^{\text{eq}} + \lambda\xi^2\right) + n_a\left(G_{\text{red}}^{\text{eq}} + \lambda(\xi - 1)^2\right). \tag{143}$$

In the oxidized and reduced states, where the expectation value of $n_a$ is zero and unity, the expectation value of $H_{\text{n}}$ yields the nuclear free energies given in Eqs. 69 and 70.

The total Hamiltonian $H$ of the system is given by the sum of the electronic part, $H_{\text{el}}$, and nuclear part, $H_{\text{n}}$, namely,

$$H = H_{\text{el}} + H_{\text{n}} = H_{\text{el}}' + G_{\text{ox}}^{\text{eq}} + \lambda\xi^2, \tag{144}$$

with the modified electronic Hamiltonian $H_{\text{el}}'$,

$$H_{\text{el}}' = \epsilon_a' n_a + \sum_k \epsilon_k n_k + \sum_k \left(H_{ka} c_k^\dagger c_a + H_{ak} c_a^\dagger c_k\right), \tag{145}$$

and the modified electronic energy,

$$\epsilon_a' = \epsilon_a + \Delta G_{\text{n}} + \lambda(1 - 2\xi), \tag{146}$$

where $\Delta G_{\text{n}}$ is the difference in the equilibrium nuclear free energies of the reduced and oxidized states, as defined after Eq. 49. Eq. 144 describes the modification of interaction potential from nuclear modes on the electronic Hamiltonian at a specific nuclear coordinate. Eq. 144 decouples the electronic motion from the nuclear motion, which is the essence of the Born-Oppenheimer approximation. The decoupling enables us to consider the time evolution of the electronic state at a specific nuclear coordinate. After the electronic couplings between the metal surface and redox species are switched on, the electronic state would relax to its equilibrium state at a given nuclear coordinate, with the corresponding electronic energy reaching its equilibrium value. The metal surface serves as an electron reservoir, with its electronic structure and occupation remaining almost completely undisturbed by the interaction with the redox species, so the electronic state of the system can be specifically referred to as the occupation number in state $a$.

### 5.2. Time evolution of the electronic state

At a specific nuclear coordinate $\xi$, the time evolution of the electronic state of the system can be described by the equations of motion (EOMs) of $H'_{\mathrm{el}}$ in the Heisenberg picture, as follows[202],

$$i\hbar\dot{c}_a(t) = \epsilon'_a c_a(t) + \sum_k H^*_{ka} c_k(t), \tag{147}$$

$$i\hbar\dot{c}_k(t) = \epsilon_k c_k(t) + H_{ka} c_a(t), \tag{148}$$

with the initial values of $c_a(t)$ and $c_k(t)$ at $t = 0$,

$$c_a(0) = c_a, c_k(0) = c_k, \tag{149}$$

where $c_a(t)$ and $c_k(t)$ are the time-dependent Heisenberg operators, corresponding to the Schrödinger operators $c_a$ and $c_k$, respectively. We distinguish the Heisenberg operators by explicitly indicating their time-dependence following the operators. The detailed derivation of the EOMs is provided in the Appendix 10.2. Eq. 147 is a first-order differential equation, which can be solved as,

$$c_k(t) = -\frac{i}{\hbar}\int_0^t H_{ka} c_a(\tau) e^{\frac{i\epsilon_k(\tau-t)}{\hbar}} d\tau + c_k e^{-\frac{i\epsilon_k t}{\hbar}}. \tag{150}$$

Inserting the above result into Eq. 146 gives

$$i\hbar\dot{c}_a(t) = \epsilon'_a c_a(t) - \frac{i}{\hbar}\int_0^t c_a(\tau)\sum_k |H_{ka}|^2 e^{\frac{i\epsilon_k(\tau-t)}{\hbar}} d\tau + \sum_k H^*_{ka} c_k e^{-\frac{i\epsilon_k t}{\hbar}}. \tag{151}$$

By using Eq. 131 we obtain,

$$\begin{aligned}\sum_k |H_{ka}|^2 e^{\frac{i\epsilon_k(\tau-t)}{\hbar}} &= \int_{-\infty}^{+\infty} \sum_k |H_{ka}|^2 \delta(\epsilon-\epsilon_k)\, e^{\frac{i\epsilon(\tau-t)}{\hbar}} d\epsilon \\ &= \frac{1}{\pi}\int_{-\infty}^{+\infty} \Delta(\epsilon) e^{\frac{i\epsilon(\tau-t)}{\hbar}} d\epsilon.\end{aligned} \tag{152}$$

Again, we apply the sifting property of the Dirac delta function in the first equality. As previously discussed, under the wide-band approximation, where $\Delta(\epsilon) = \Delta$, we have,

$$\sum_k |H_{ka}|^2 e^{\frac{i\epsilon_k(\tau-t)}{\hbar}} = \frac{\Delta}{\pi}\int_{-\infty}^{+\infty} e^{\frac{i\epsilon(\tau-t)}{\hbar}} d\epsilon = 2\hbar\Delta\delta(\tau-t), \tag{153}$$

where we observe that the integral in the above equation is exactly the Fourier transform of the Dirac delta function, $2\pi\hbar\delta(\tau - t)$. Inserting the above identity into Eq. 150, we obtain,

$$i\hbar\dot{c}_a(t) = (\epsilon_a' - i\Delta)c_a(t) + \sum_k H_{ka}^* c_k e^{-\frac{i\epsilon_k t}{\hbar}}, \tag{154}$$

which can be explicitly solved as,

$$c_a(t) = \alpha(t)c_a + \sum_k H_{ka}^* \beta_k(t) c_k, \tag{155}$$

with,

$$\alpha(t) = e^{-\frac{i(\epsilon_a' - i\Delta)t}{\hbar}}, \tag{156}$$

$$\beta_k(t) = -\frac{i}{\hbar} e^{-\frac{i(\epsilon_a' - i\Delta)t}{\hbar}} \int_0^t e^{\frac{i(\epsilon_a' - \epsilon_k - i\Delta)\tau}{\hbar}} d\tau. \tag{157}$$

The particle number operator in the Heisenberg picture $n_a(t)$ is then,

$$\begin{aligned} n_a(t) &= c_a^\dagger(t) c_a(t) \\ &= |\alpha(t)|^2 n_a + \sum_k \sum_{k'} H_{ka} H_{k'a}^* \beta_k^*(t) \beta_{k'}(t) c_k^\dagger c_{k'} \\ &+ \sum_k H_{ka} \beta_k^*(t) \alpha(t) c_k^\dagger c_a + \sum_k H_{ka}^* \beta_k(t) \alpha^*(t) c_a^\dagger c_k. \end{aligned} \tag{158}$$

Given the system state $|\Psi_0\rangle$ at the initial time, where the metal states $k$ are fully occupied and state $a$ may be either occupied or empty, the time-dependent expectation value of $n_a(t)$ can be evaluated as follows,

$$\begin{aligned} \langle n_a(t)\rangle &= \langle \Psi_0 | n_a(t) | \Psi_0 \rangle \\ &= \langle n_a(0)\rangle |\alpha(t)|^2 + \sum_k |H_{ka}|^2 |\beta_k(t)|^2 \\ &= \langle n_a(0)\rangle e^{-\frac{2\Delta t}{\hbar}} + \frac{1}{\hbar^2} e^{-\frac{2\Delta t}{\hbar}} \sum_k |H_{ka}|^2 \left| \int_0^t e^{\frac{i(\epsilon_a' - \epsilon_k - i\Delta)\tau}{\hbar}} d\tau \right|^2 \\ &= \langle n_a(0)\rangle e^{-\frac{2\Delta t}{\hbar}} + \sum_k \langle n_{ka}\rangle, \end{aligned} \tag{159}$$

with,

$$\langle n_{ka}\rangle = \frac{|H_{ka}|^2}{(\epsilon_a' - \epsilon_k)^2 + \Delta^2} \left( e^{-\frac{2\Delta t}{\hbar}} - 2e^{-\frac{\Delta t}{\hbar}} \cos\left( \frac{(\epsilon_a' - \epsilon_k)t}{\hbar} \right) + 1 \right). \tag{160}$$

In the second identity, the $k \neq k'$ components in the second term of $n_a(t)$, as well as the third and fourth terms, involve operators that change the particle occupation of the initial state. These operators map the system state into states that are orthogonal to the original state, and thus their contributions vanish when evaluating the expectation value of $n_a(t)$. $\langle n_{ka}\rangle$ can be taken as the probability of finding the electron occupying in the state $a$ at a later time $t$ from a state $k$ prepared at $t = 0$ [201]. Considering the Fermi-Dirac distribution of metal electrons at finite temperatures, we have,

$$
\begin{aligned}
\langle n_a(t)\rangle &= \langle n_a(0)\rangle e^{-\frac{2\Delta t}{\hbar}} + \sum_k f(\epsilon_k)\,\langle n_{ka}\rangle \\
&= \langle n_a(0)\rangle e^{-\frac{2\Delta t}{\hbar}} + \int f(\epsilon) \sum_k \langle n_{ka}(\epsilon)\rangle \delta(\epsilon-\epsilon_k)\, d\epsilon.
\end{aligned}
\tag{161}
$$

As a simplification we can set the Fermi level as the energy zero and approximate the Fermi-Dirac distribution function using the Heaviside step function at the Fermi level. Considering that there is no electron initially occupying state $a$, Eq. 160 simplifies to,

$$
\langle n_a(t)\rangle = \int_{-\infty}^{0} \frac{\Delta}{\pi} \frac{1+e^{-\frac{2\Delta t}{\hbar}} - 2e^{-\frac{\Delta t}{\hbar}} \cos\left(\frac{(\epsilon_a'-\epsilon)t}{\hbar}\right)}{(\epsilon_a'-\epsilon)^2+\Delta^2}\, d\epsilon. \tag{162}
$$

At short times, the occupation number in state $a$ exhibits oscillatory behavior. However, at long times, specifically for $t \gg \hbar/(2\Delta)$, the oscillatory term in Eq. 161 becomes negligible. In this case, the rate of the occupation number changes can be described as,

$$
\frac{d\langle n_a(t)\rangle}{dt} = \frac{2\Delta}{\hbar}\big(\langle n_a(t\to\infty)\rangle - n_a(t)\big), \tag{163}
$$

with the relaxation time of the electronic state being $\tau_{\mathrm{e}} = \hbar/(2\Delta)$.

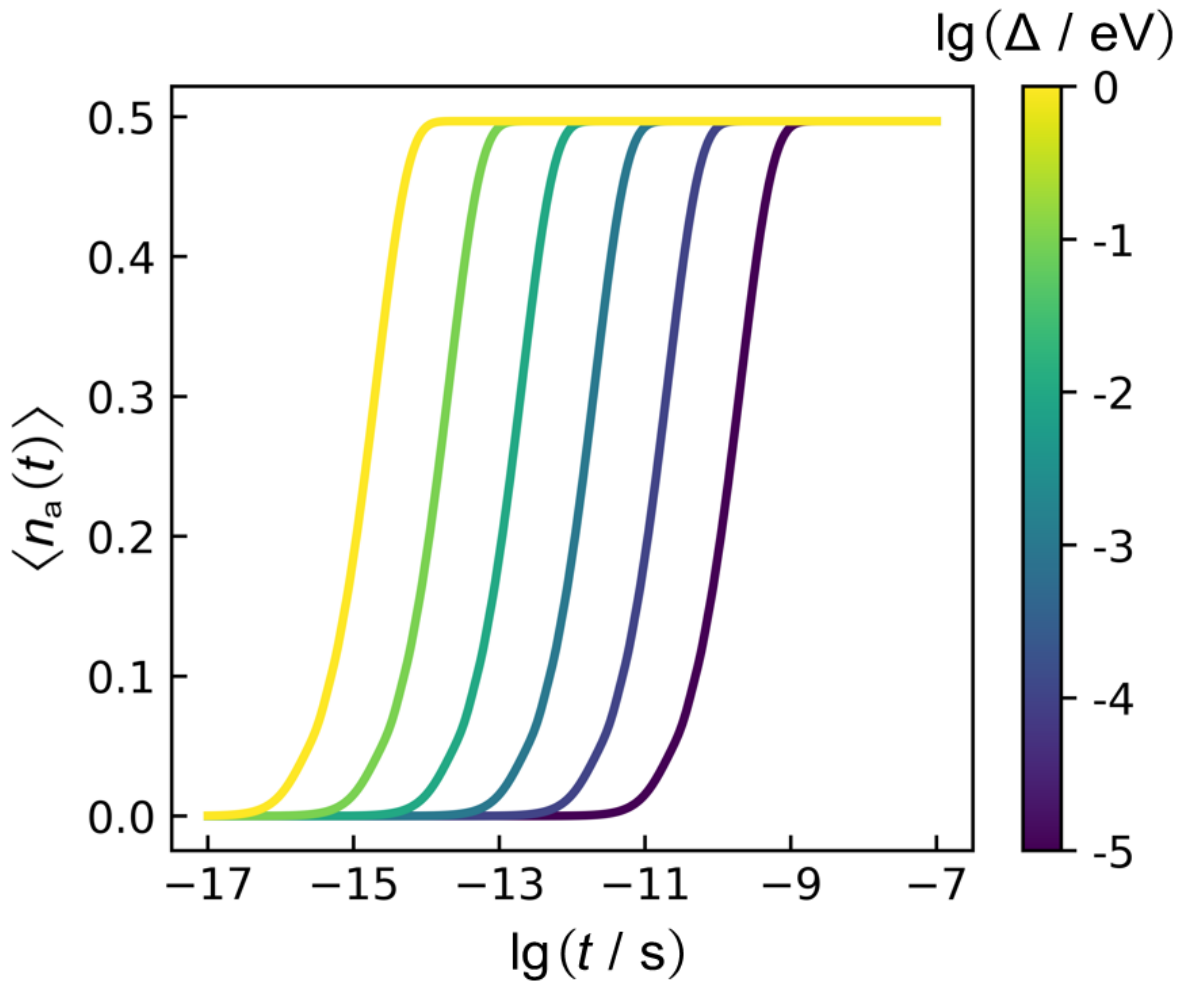

Figure 8. The times evolution of expectation value of electron occupancy number in state $a$ at different electronic coupling strength, $\Delta$. The parameter $\epsilon_a'$ is set to zero.

The relaxation of the electronic state can be clearly observed by plotting $\langle n_a(t)\rangle$ in Eq. 161 as a function of time as shown in Figure 8 for various coupling constants $\Delta$. We can observe that the electron occupation number in state $a$ reaches its equilibrium value faster as the coupling, $\Delta$, is increases. The half time of this relaxation process coincides with the $\tau_{\mathrm{e}}$. In the adiabatic limit, the relaxation to equilibrium electronic state can be always achieved because the coupling is very strong, allowing the electronic state to rapidly relax and reach equilibrium at a given nuclear configuration. By letting time approach infinity, the exponential terms in the numerator of the integrand in Eq. 161 vanish, and the equilibrium expectation value of occupation number in state $a$ is reached. This value, denoted as $\langle n_a\rangle_\xi$, depends on the solvent coordinate $\xi$, which affects the value of $\epsilon_a'$, as defined in Eq. 145. The equilibrium occupation $\langle n_a\rangle_\xi$ is thus obtained by integrating the Lorentzian DOS of the state $a$, centered at $\epsilon_a'$, up to the Fermi level. In the special case where $\epsilon_a'$ is set to zero, the Fermi level aligns with the center of the Lorentzian DOS, resulting in a half-filled state and an equilibrium occupation value of 0.5, as shown in Figure 8. A more general expression of $\langle n_a \rangle_\xi$ can be derived by taking the long-time limit of Eqs. 159 and 160 as

$$\langle n_a(t \to \infty) \rangle = \langle n_a \rangle_\xi = \sum_k \frac{f(\epsilon_k)|H_{ka}|^2}{(\epsilon_a' - \epsilon_k)^2 + \Delta^2}. \tag{164}$$

### 5.3. Adiabatic free energy surface

As $t$ approaches infinity, the electronic occupation number in state $a$ will relax to its equilibrium value. This occupation number can be fractional, ranging between zero and one, depending on the nuclear coordinate $\xi$. The electronic energy of these fractional electrons in state $a$ is referred to as the bonding energy. For system states at different nuclear coordinates, the energy varies due to changes in both the bonding energy and the nuclear free energy of the redox species with respect to the nuclear coordinate. In these states, the electronic states of the system are in equilibrium with the nuclear coordinate and the free energy dependency along the nuclear coordinate defines the adiabatic FES. By approximating the Fermi-Dirac distribution function as the Heaviside step function in Eq. 164, we obtain the adsorbate occupation number as

$$\begin{aligned}
\langle n_a \rangle_\xi &= \sum_{\epsilon_k \leq \epsilon_\mathrm{F}} \frac{f(\epsilon_k)|H_{ka}|^2}{(\epsilon_a' - \epsilon_k)^2 + \Delta^2} \\
&= \int_{-\infty}^{0} \frac{|H_{ka}|^2 \delta(\epsilon - \epsilon_k)}{(\epsilon_a' - \epsilon)^2 + \Delta^2} d\epsilon \\
&= \frac{1}{\pi} \int_{-\infty}^{0} \frac{\Delta}{(\epsilon_a' - \epsilon)^2 + \Delta^2} d\epsilon \\
&= \frac{1}{\pi} \operatorname{arccot}\left(\frac{\epsilon_a'}{\Delta}\right).
\end{aligned} \tag{165}$$

The second identity uses the sifting property of the Dirac delta function. Upon interacting with the metal electronic states, the electronic energy of state $a$ broadens into a continuous band around $\epsilon$ with a Lorentzian distribution for adsorbate DOS:

$$\rho_a(\epsilon) = \frac{1}{\pi} \frac{\Delta}{(\epsilon_a' - \epsilon)^2 + \Delta^2}. \tag{166}$$

The filling $\rho_a$ varies with the nuclear coordinate through $\epsilon_a'$ (see Eq. 145). The covalent binding energy is then calculated by integrating the electronic energy in state $a$ up to the Fermi level,

$$\begin{aligned}
E_\mathrm{cov}(\xi) &= \int_{-\infty}^{0} \epsilon \rho_a(\epsilon) d\epsilon = \frac{\Delta}{\pi} \int_{-\infty}^{0} \frac{\epsilon}{(\epsilon_a' - \epsilon)^2 + \Delta^2} d\epsilon \\
&= \epsilon_a' \langle n_a \rangle_\xi + \frac{\Delta}{2\pi} [\ln((\epsilon - \epsilon_a')^2 + \Delta^2)]_{-\infty}^{0}.
\end{aligned} \tag{167}$$

The divergence of the second term arises as a result of the wide-band approximation[201]. To address this, we can define the renormalized energy scale by taking the free energy of the system at $\xi = 0$ as the energy zero, accounting for both the covalent binding energy and the nuclear free energy. With this the covalent binding energy is

$$E_\mathrm{cov}^\mathrm{ref}(\xi) = \epsilon_a' \langle n_a \rangle_\xi - \epsilon_a'(\xi = 0) \langle n_a \rangle_0 + \frac{\Delta}{2\pi} \ln\left(\frac{(\epsilon_a')^2 + \Delta^2}{(\epsilon_a + \Delta G_\mathrm{n} + \lambda)^2 + \Delta^2}\right). \tag{168}$$

Then the adiabatic FES can be written as a function of the nuclear coordinate,

$$G(\xi) = \epsilon_a' \langle n_a \rangle_\xi - \epsilon_a'(\xi = 0)\langle n_a \rangle_0 + \lambda\xi^2 + \frac{\Delta}{2\pi}\ln\left(\frac{(\epsilon_a')^2 + \Delta^2}{(\epsilon_a + \Delta G_{\mathrm{n}} + \lambda)^2 + \Delta^2}\right). \tag{169}$$

It is important to note that at stronger couplings, the system states at $\xi = 0$ and $\xi = 1$ no longer correspond to the initial and final states of the system, i.e., the minima of the diabatic FES. This is because electronic interactions between the metal surface and redox species can induce partial charges on the redox species, resulting in a shift in their equilibrium nuclear coordinates, as will be seen later in this section. If we denote the nuclear coordinates in the initial and final states as $\xi_i$ and $\xi_f$, respectively, the reaction free energy is

$$\begin{aligned}\Delta G^0 = {}& \left[\epsilon_a + \Delta G_{\mathrm{n}} + \lambda\left(1 - 2\xi_f\right)\right]\langle n_a \rangle_{\xi_f} - \left[\epsilon_a + \Delta G_{\mathrm{n}} + \lambda(1 - 2\xi_i)\right]\langle n_a \rangle_i + \lambda\xi_f^2 \\ & - \lambda\xi_i^2 + \frac{\Delta}{2\pi}\ln\left(\frac{\left(\epsilon_a + \Delta G_{\mathrm{n}} + \lambda\left(1 - 2\xi_f\right)\right)^2 + \Delta^2}{\left(\epsilon_a + \Delta G_{\mathrm{n}} + \lambda(1 - 2\xi_i)\right)^2 + \Delta^2}\right).\end{aligned} \tag{170}$$

Since the Fermi level is taken as the energy zero, changing the electrode potential by an amount of $\Delta\varphi$ will shift $\epsilon_a$ by a corresponding amount of $e_0\Delta\varphi$. In other words, $\epsilon_a = \epsilon_a(\varphi)$ is linearly proportional to the electrode potential by a constant, $e_0$. At the standard equilibrium potential $\varphi^0$, $\Delta G_0 = 0$, which requires that $\epsilon_a(\varphi^0) = -\Delta G_{\mathrm{sol}}$ and $\xi_i + \xi_f = 1$ in Eq. 169. At the electrode potential $\varphi$, we have,

$$\epsilon_a(\varphi) = -\Delta G_{\mathrm{n}} + e_0(\varphi - \varphi^0) = -\Delta G_{\mathrm{n}} + e_0\eta. \tag{171}$$

By substituting Eq. 170 into Eq. 168, the adiabatic FES at the overpotential $\eta$ becomes,

$$\begin{aligned}G(\xi) = {}& [e_0\eta + \lambda(1 - 2\xi)]\langle n_a \rangle_\xi - (e_0\eta + \lambda)\langle n_a \rangle_0 + \lambda\xi^2 \\ & + \frac{\Delta}{2\pi}\ln\left(\frac{(\epsilon_a')^2 + \Delta^2}{(e_0\eta + \lambda)^2 + \Delta^2}\right),\end{aligned} \tag{172}$$

with,

$$\langle n_a \rangle_\xi = \frac{1}{\pi}\operatorname{arccot}\left(\frac{e_0\eta + \lambda(1 - 2\xi)}{\Delta}\right), \tag{173}$$

$$\epsilon_a' = e_0\eta + \lambda(1 - 2\xi). \tag{174}$$

As the coupling constant $\Delta \to 0$, we can recover the diabatic FES for the oxidized state, $G_{\mathrm{ox}} = \lambda\xi^2$ and that for the reduced state, $G_{\mathrm{red}} = e_0\eta + \lambda(\xi - 1)^2$. Figures 9a and 9b present the adiabatic FESs from Eq. 171 and $\langle n_a \rangle_\xi$ from Eq. 172 as a function of the nuclear coordinate at various values of $\Delta$, respectively. When the electronic coupling is very small, such as, $\Delta = 0.01$ eV, a sudden transition of the electronic state occurs at $\xi = 0.5$, as shown in Figure 9b. In this case, the transition region narrows to approximately a single point and the reactant and product regions closely resemble their respective diabatic FESs, as shown in Figure 9a. As $\Delta$ increases, the transition region broadens, and the activation energy decreases. The minima of the adiabatic PES move closer together, and the reactant or product at their respective minima acquires partial charges. The distance dependence of $\Delta$ is not considered here. During the electrosorption process, the redox species approach the metal surface. Since the electronic coupling strength decays rapidly to zero as the redox species moves approximately two angstroms away from the metal surface[199], the reactants typically do not carry any partial charge. However, partial charges may still be present on the chemisorbates.

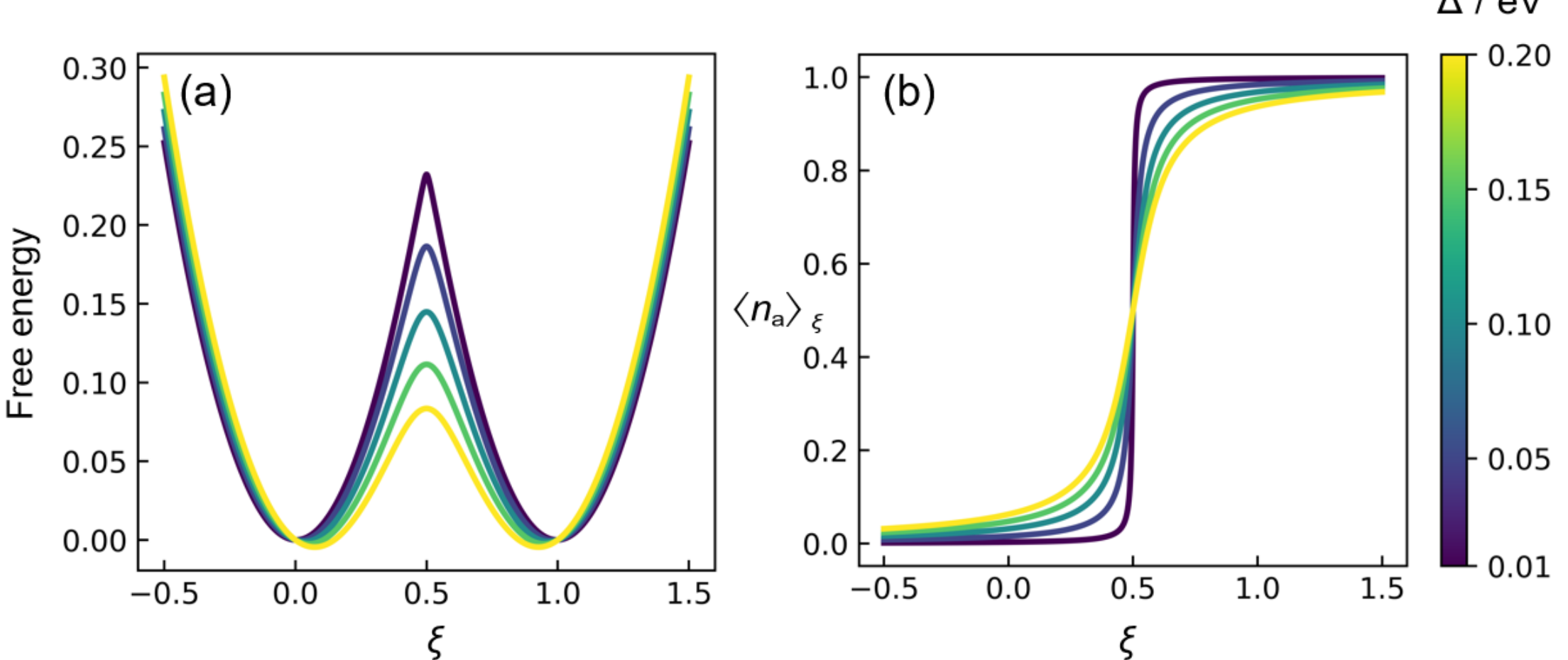

Figure 9. The (a) adiabatic FESs and (b) electron occupancy number in state $a$ as a function of nuclear coordinate $\xi$ at various electronic coupling strength, $\Delta$. The parameters used are $\lambda = 1$ eV, $\eta = 0$ V, $\Delta G_{\mathrm{n}} = 0$ eV.

At small overpotentials, we can find that $\xi = \frac{1}{2}\left(1 + \frac{e_0\eta}{\lambda}\right)$ serves as an approximate solution to $\partial G(\xi)/\partial\xi = 0$, representing an extremum point of the adiabatic FES. Substituting this value of $\xi$ into Eq. 171 yields the activation energy for the reduction reaction,

$$\Delta G_{\mathrm{red}}^{\neq} = \frac{(\lambda + e_0\eta)^2}{4\lambda} - \frac{\Delta}{2\pi}\ln\left(1 + \frac{(\lambda + e_0\eta)^2}{\Delta^2}\right). \tag{175}$$

The first term in the above equation corresponds to the Marcus barrier at the Fermi level, while the second term represents the reduction in the Marcus barrier due to covalent electronic interactions between the metal surface and the redox species.

In the preceding theoretical treatment, all nuclear coordinates are assumed to reorganize collectively during the ET process, and this reorganization is described by a single effective nuclear coordinate $\xi$. In this case, the adiabatic FES is one-dimensional along this reaction coordinate, telling us how the ET process is driven by collective nuclear motions, as discussed in Section 2.4. However, the most frequently encountered ET steps in electrocatalytic reactions are inner-sphere ET involving the bond formation or breaking. We may wish to understand how such ET processes are particularly driven by nuclear modes associated with bond formation or breaking beyond the collective reorganization of the remaining nuclear modes. In this regard, it is informative to construct a multi-dimensional FES by explicitly including the nuclear coordinates associated with bond formation or breaking. For chemisorption of a redox species, e.g., ion adsorption onto an electrode surface, a covalent bond is formed between the redox species and electrode surface. The corresponding nuclear coordinate is the distance between the redox species and electrode surface, denoted as $d$. The adiabatic FES for this process is therefore expected to be two-dimensional, described by the distance $d$ and an effective nuclear coordinate $\xi$, i.e., $G(\xi, d)$. The adiabatic FES in Eq. 171 can be extended to this two-dimensional representation by accounting for the dependence of relevant parameters on the distance $d$. First, the electronic coupling strength increases as the redox species approaches the electrode surface, as discussed above. This dependence can be reasonably approximated by an exponential function of the distance $d$, with coefficients fitted to the $\Delta - d$ relationship obtained from DFT calculations. Second, the electronic level of the redox species in vacuum, $\epsilon_a^0$ in Eq. 37, exhibits an approximately linear dependence on the distance $d$[199,203]. Third, the solvent reorganization energy varies with the distance $d$, as the solvent dielectric polarization becomes distance dependent in the presence of a spatially inhomogeneous interfacial field. Fourth, again due to the inhomogeneous interfacial field, both the electrostatic potential energy and the equilibrium solvation energy of the redox species vary with the distance. This effect can be incorporated by introducing additional work terms into Eq. 171 and will be discussed in Section 7.1. The third and fourth considerations indicate that it is necessary to incorporate EDL modeling to properly describe the local reaction conditions. A promising framework that incorporates both chemisorption and EDL effects is presented in Ref.[125].

Another important inner-sphere ET is bond-breaking ET involving the dissociation of a chemical bond within the redox species. The length of this chemical bond, denoted as $r$, is usually taken as the geometric RC. The reorganization along this nuclear coordinate is assumed to be purely classical and to respond linearly to ET, i.e., to the occupation number in the valence state $a$ of the redox species, consistent with the assumption made for all other nuclear modes in Eq. 142. The key distinction is that the motion along the coordinate $r$ is intrinsically anharmonic. As in Saveant's model[88], the bond energy in the initial state can be described by the Morse potential, whereas in the final state the interaction energy between the two fragments resulting from bond dissociation is represented by the repulsive branch of the Morse potential. A bond-breaking Hamiltonian, $H_{\mathrm{bb}}$, can therefore be introduced in the form of a switching function that interpolates between these two energies[90]:

$$H_{\mathrm{bb}} = D_{\mathrm{e}}(1 - n_a)\left(1 - e^{-\alpha_{\mathrm{M}}(r - r_{\mathrm{e}})}\right)^2 + D_{\mathrm{e}} n_a e^{-2\alpha_{\mathrm{M}}(r - r_{\mathrm{e}})}, \tag{176}$$

where $r_{\mathrm{e}}$ is the equilibrium bond length and $D_{\mathrm{e}}$ the bond dissociation energy. The parameter $\alpha_{\mathrm{M}}$ is related to the bond vibration frequency $\omega_{\mathrm{b}}$ by $\alpha_{\mathrm{M}} = \omega_{\mathrm{b}}(\mu_{\mathrm{b}}/2D_{\mathrm{e}})^{1/2}$, with $\mu_{\mathrm{b}}$ denoting the reduced mass of the atoms involved in the bond dissociation. By incorporating this Hamiltonian into Eq. 143, the total Hamiltonian becomes

$$H = H_{\mathrm{el}} + H_{\mathrm{n}} + H_{\mathrm{bb}} = H'_{\mathrm{el}} + G_{\mathrm{ox}}^{\mathrm{eq}} + \lambda\xi^2 + D_{\mathrm{e}}\left(1 - e^{-\alpha_{\mathrm{M}}(r - r_{\mathrm{e}})}\right)^2. \tag{177}$$

The modified electronic Hamiltonian $H'_{\mathrm{el}}$ retains the same formal structure as in Eq. 144, except that the modified electronic energy $\epsilon'_a$ now depends on both $\xi$ and $r$:

$$\epsilon'_a(\xi, r) = \epsilon_a + \Delta G_{\mathrm{n}} + \lambda(1 - 2\xi) - D_{\mathrm{e}}\left(1 - 2e^{-\alpha_{\mathrm{M}}(r - r_{\mathrm{e}})}\right). \tag{178}$$

The corresponding two-dimensional adiabatic FES can be directly obtained by modifying Eq. 171, adding the Morse potential contribution and replacing $\epsilon'_a$ with Eq. 177, yielding,

$$\begin{aligned} G(\xi, r) = {} & [e_0\eta + \lambda(1 - 2\xi)]\langle n_a\rangle_{\xi,r} - (e_0\eta + \lambda + D_{\mathrm{e}})\langle n_a\rangle_{0,r_{\mathrm{e}}} + \lambda\xi^2 \\ & + \frac{\Delta}{2\pi}\ln\left(\frac{(\epsilon'_a)^2 + \Delta^2}{(e_0\eta + \lambda)^2 + \Delta^2}\right) + D_{\mathrm{e}}\left(1 - e^{-\alpha_{\mathrm{M}}(r - r_{\mathrm{e}})}\right)^2, \end{aligned} \tag{179}$$

where the occupation number of the valence state $a$ is

$$\langle n_a\rangle_{\xi,r} = \frac{1}{\pi}\operatorname{arccot}\left(\frac{e_0\eta + \lambda(1 - 2\xi) - D_{\mathrm{e}}\left(1 - 2e^{-\alpha_{\mathrm{M}}(r - r_{\mathrm{e}})}\right)}{\Delta}\right). \tag{180}$$

Here, $\langle n_a\rangle_{0,r_{\mathrm{e}}}$ denotes the occupation number at the nuclear configuration $\xi = 0$ and $r = r_{\mathrm{e}}$, and the corresponding free energy is taken as the reference. Similarly, at small overpotentials, one finds that $\xi = \frac{1}{2}\left(1 + \frac{e_0\eta}{\lambda + D_{\mathrm{e}}}\right)$ serves as an approximate solution to $\partial G(\xi, r)/\partial\xi = \partial G(\xi, r)/\partial r = 0$, corresponding to an extremum point of the adiabatic FES. The corresponding activation free energy is then given[90] by

$$\Delta G^{\neq}_{\text{red}} = \frac{(\lambda + D_{\text{e}} + e_0\eta)^2}{4(\lambda + D_{\text{e}})} - \frac{\Delta}{2\pi}\ln\left(1 + \frac{(\lambda + D_{\text{e}} + e_0\eta)^2}{\Delta^2}\right). \tag{181}$$

The quantity $(\lambda + D_{\text{e}})$ represents the total reorganization energy, arising from both bond dissociation and the collective adjustment of the remaining nuclear modes. In the present treatment, bond dissociation is treated entirely classically; consequently, it modifies only the activation free energy. For a more rigorous quantum-mechanical description of bond breaking, interested readers are referred to Ref.[59].

## 5.4. Simulating the adiabatic FES

DFT is a ground-state theory where the energy is a functional of the electron density and that external potential defined by the nuclei. Hence, the obtained FES along the nuclear coordinates is by definition the adiabatic one. When combined with enhanced sampling methods and MD simulations, DFT can be readily used compute the adiabatic FES along *geometric* reaction coordinates. However, in ET studies the reaction coordinate is not directly a geometric one but the energy gap coordinate, which depends only indirectly on the nuclear positions, i.e., the system geometry.

As discussed in Section 3.6, sampling the energy gap and constructing the diabatic FES requires the use of diabatic states and a linear mapping Hamiltonian that interpolated between the two diabatic Hamiltonians (Eq. 82). In the adiabatic situation, a very similar approach can be used to compute the adiabatic FES along the energy gap coordinate using Eqs. 83-85 with an adiabatic Hamiltonian instead of the diabatic Hamiltonians ($H_{11}$ or $H_{22}$) used in the construction of the diabatic FESs. The only difference in simulating the diabatic or adiabatic FES is then to find or define the adiabatic Hamiltonian to replace $H_{11}$ and $H_{22}$. In DFT-based methods the most straightforward choice is to just use the normal ground state DFT Hamiltonian. An alternative choice is to approximate the adiabatic energies by diagonalizing the $2 \times 2$ diabatic Hamiltonian (Eq. 75) which yields the adiabatic ground (g) and excited (e) states

$$H_{\text{g/e}}(\boldsymbol{R}) = \frac{H_{11}(\boldsymbol{R}) + H_{22}(\boldsymbol{R})}{2} \mp \frac{1}{2}\sqrt{\Delta E(\boldsymbol{R}) + 4|H_{12}(\boldsymbol{R})|^2}, \tag{182}$$

where $H_{12}(\boldsymbol{R})$ is the electronic coupling matrix element, i.e., the off-diagonal matrix element in Eq. 75. $H_{12}$ may be computed in various ways:

- If both the ground and excited state adiabatic energies or both the diabatic energies and lower adiabatic energy have been evaluated, the coupling matrix element is available from Eq. 181. This is often done in EVB studies employing classical potentials for both diabatic states and the adiabatic ground states. This is also useful in GCE-DFT studies as one avoids computing the coupling matrix explicitly for multiple states with a different number of electrons[81].
- In constrained DFT studies, $H_{12}$ can be readily computed from the diabatic cDFT wave functions using either the cDFT-specific formula by Wu and Van Voorhis[151] or by the more general formalism by Migliore[204]. For GCE-DFT studies, the coupling constant is computed as a grand canonical expectation value over the coupling constant values constant charge as discussed in Section 3.6.
- In EVB with classical potentials, the coupling constant is often "calibrated" by computing it in the gas-phase or for a closely related reference reaction. This is permitted as $H_{12}$ does not appear to be sensitive to the phase[205]. This has not, however, been confirmed for electrochemical interfaces.

- In the Anderson-Newns theory, the effective $H_{12}$ is often approximated with an effective coupling constant $\Delta$, which is related to the density of states of the reacting orbital, Eq. 131. In this case, the coupling constant is computed by fitting Eq. 131 to the DOS from quantum mechanical calculations, most often DFT[199]. Another way of computing the coupling matrix elements is to parametrize the Anderson-Newns Hamiltonian (Eq. 141) by combing DFT with an explicit diabatization scheme developed in Ref.[206].

### 5.5. Rate constant

To compute the rate constant, we need to find the number of electrons in state $a$ at the equilibrium value over a characteristic relaxation time, $\tau_{\mathrm{e}}$. The electron transition rate at a given nuclear coordinate can be taken as the change in the number of electrons in state $a$ over $\tau_{\mathrm{e}}$[11,201]. Therefore, for the reduction and oxidation reactions, we have the corresponding electron transition rates[11,201,202,207] at a certain nuclear coordinate $\xi$,

$$W_{\mathrm{red}}(\xi) = \frac{\langle n_a \rangle_\xi}{\tau_{\mathrm{e}}}, \tag{183}$$

$$W_{\mathrm{ox}}(\xi) = \frac{1 - \langle n_a \rangle_\xi}{\tau_{\mathrm{e}}}. \tag{184}$$

The probability of ET for the reduction or oxidation reaction at a given nuclear coordinate $\xi$ is then the product of the transition rate, $W_{\mathrm{red}}(\xi)$ or $W_{\mathrm{ox}}(\xi)$, and the probabilities of finding the oxidized state, $\rho_{\mathrm{ox}}(\xi)$, or the reduced state, $\rho_{\mathrm{red}}(\xi)$, at that nuclear coordinate, respectively. The overall rate constants can be obtained by integrating the probabilities of electron transfer over the nuclear coordinates[201] $\xi$, i.e.,

$$k_{\mathrm{red}} = \frac{1}{\tau_{\mathrm{e}}} \int \langle n_a \rangle_\xi \rho_{\mathrm{ox}}(\xi) d\xi, \tag{185}$$

$$k_{\mathrm{ox}} = \frac{1}{\tau_{\mathrm{e}}} \int \left(1 - \langle n_a \rangle_\xi\right) \rho_{\mathrm{red}}(\xi) d\xi. \tag{186}$$

We first evaluate $k_{\mathrm{red}}$. By substituting Eqs. 118 and 163 into Eq. 184, we have,

$$k_{\mathrm{red}} = \frac{2\Delta}{\hbar Q_{\mathrm{c}}} \sum_k f(\epsilon_k) |H_{ka}|^2 I_k, \tag{187}$$

with the integral,

$$I_k = \int \frac{e^{-\beta\lambda\xi^2}}{(\epsilon_{ak} - 2\lambda\xi)^2 + \Delta^2} d\xi, \tag{188}$$

$$\epsilon_{ak} = \epsilon_a - \epsilon_k + \Delta G_{\mathrm{n}} + \lambda. \tag{189}$$

Let $u = \epsilon_{ak} - 2\lambda\xi$, then the integral $I_k$ can be rewritten as,

$$I_k = \frac{1}{2\lambda}\int \frac{1}{\epsilon_{ak}^2 + \Delta^2}\cdot e^{-\frac{\beta(\epsilon_{ak}-u)^2}{4\lambda}}du$$
$$= \frac{1}{2\lambda} f_1(\epsilon_{ak}) * f_2(\epsilon_{ak}), \tag{190}$$

with the functions,

$$f_1(\epsilon_{ak}) = \frac{1}{\epsilon_{ak}^2 + \Delta^2}, \tag{191}$$

$$f_2(\epsilon_{ak}) = e^{-\frac{\beta\epsilon_{ak}^2}{4\lambda}}, \tag{192}$$

where $f_1 * f_2$ denotes the convolution of the functions $f_1$ and $f_2$. By performing the Fourier transform on $I_k$ with respect to $\epsilon_{ak}$, we have,

$$\begin{aligned}
\tilde{I}_k(\tau) &= \mathcal{F}[I_k(\epsilon_{ak})] \\
&= \frac{1}{2\lambda}\mathcal{F}[f_1(\epsilon_{ak})] \times \mathcal{F}[f_2(\epsilon_{ak})] \\
&= \frac{1}{2\lambda}\left(\frac{\pi}{\Delta}e^{-\Delta|\tau|}\right) \times \left(\sqrt{\frac{4\pi\lambda}{\beta}}e^{-\frac{\lambda\tau^2}{\beta}}\right) \\
&= \frac{\pi}{\Delta}\sqrt{\frac{\pi}{\beta\lambda}}e^{-\frac{\lambda\tau^2}{\beta}-\Delta|\tau|}.
\end{aligned} \tag{193}$$

In the second equality, the convolution theorem of the Fourier transform is applied, while the third equality utilizes the Fourier transforms of the Lorentzian and Gaussian functions. The integral $I_k$ is obtained by performing the inverse Fourier transform on $\tilde{I}_k(\tau)$,

$$\begin{aligned}
I_k &= \frac{1}{2\pi}\int_{-\infty}^{+\infty}\tilde{I}(\tau)e^{i\epsilon_{ak}\tau}d\tau \\
&= \frac{1}{2\Delta}\sqrt{\frac{\pi}{\beta\lambda}}\int_{-\infty}^{+\infty} e^{-\frac{\lambda\tau^2}{\beta}-\Delta|\tau|}\cdot e^{i\epsilon_{ak}\tau}d\tau \\
&= \frac{1}{\Delta}\sqrt{\frac{\pi}{\beta\lambda}}\int_{0}^{+\infty} e^{-\frac{\lambda\tau^2}{\beta}-\Delta\tau}\cdot \cos(\epsilon_{ak}\tau)\,d\tau \\
&= \frac{1}{\Delta}\sqrt{\frac{\pi}{\beta\lambda}}\mathrm{Re}\left\{\int_{0}^{+\infty} e^{-\frac{\lambda\tau^2}{\beta}-\Delta\tau}\cdot e^{i\epsilon_{ak}\tau}d\tau\right\} \\
&= \frac{1}{\Delta}\sqrt{\frac{\pi}{\beta\lambda}}\mathrm{Re}\left\{\int_{0}^{+\infty} e^{-\frac{\lambda\tau^2}{\beta}+(i\epsilon_{ak}-\Delta)\tau}d\tau\right\}.
\end{aligned} \tag{194}$$

The Gaussian-type integral involved in the above equation can be calculated by the following identity,

$$\int_0^{\infty} e^{-ax^2+bx}dx = \frac{1}{2}\sqrt{\frac{\pi}{a}}e^{\frac{b^2}{4a}}\,\mathrm{erfc}\left(-\frac{b}{2\sqrt{a}}\right), \tag{195}$$

with the complementary error function,

$$\operatorname{erfc}(x) = \frac{2}{\sqrt{\pi}} \int_x^{\infty} e^{-y^2} dy. \tag{196}$$

For the Gaussian-type integral in Eq. 193, we have $a = \lambda/\beta$ and $b = i\epsilon_{ak} - \Delta$. Then we obtain,

$$\begin{aligned} I_k &= \frac{1}{\Delta}\sqrt{\frac{\pi}{\beta\lambda}} \mathrm{Re}\left\{ \frac{1}{2}\sqrt{\frac{\pi\beta}{\lambda}} e^{\frac{\beta(i\epsilon_{ak}-\Delta)^2}{4\lambda}} \operatorname{erfc}\left( -\frac{1}{2}\sqrt{\frac{\beta}{\lambda}}(i\epsilon_{ak} - \Delta) \right) \right\} \\ &= \frac{\pi}{2\lambda\Delta} \mathrm{Re}\left\{ w\left( \frac{1}{2}\sqrt{\frac{\beta}{\lambda}}(\epsilon_{ak} + i\Delta) \right) \right\}, \end{aligned} \tag{197}$$

where $w(z) = e^{-z^2}\operatorname{erfc}(-iz)$ is the complex error function. Substituting $I_k$ into Eq. 186, we obtain the rate constant for the reduction reaction,

$$\begin{aligned} k_{\mathrm{red}} &= \frac{1}{\hbar}\sqrt{\frac{\pi\beta}{\lambda}} \sum_k f(\epsilon_k) |H_{ka}|^2 \mathrm{Re}\{w(z)\} \\ &= \frac{1}{\hbar}\sqrt{\frac{\pi\beta}{\lambda}} \int f(\epsilon) \mathrm{Re}\{w(z)\} \sum_k |H_{ka}|^2 \delta(\epsilon - \epsilon_k)\, d\epsilon, \end{aligned} \tag{198}$$

with,

$$z = \frac{1}{2}\sqrt{\frac{\beta}{\lambda}}(\lambda + e_0\eta - \epsilon + i\Delta), \tag{199}$$

where we use the sifting property of the Dirac delta function in the second identity. By applying the wide-band approximation, we have,

$$k_{\mathrm{red}} = \frac{\Delta}{\hbar}\sqrt{\frac{\beta}{\pi\lambda}} \int f(\epsilon) \mathrm{Re}\{w(z)\}\, d\epsilon. \tag{200}$$

When the electronic coupling strength between the metal surface and redox species is very weak, i.e., the coupling constant $\Delta$ is very small, we have $z \approx \frac{1}{2}\sqrt{\frac{\beta}{\lambda}}(\lambda + e_0\eta - \epsilon)$, which is real-valued. And we get,

$$\begin{aligned} \mathrm{Re}\{w(z)\} &= \mathrm{Re}\left\{ e^{-z^2}\operatorname{erfc}(-iz) \right\} \\ &= \mathrm{Re}\left\{ e^{-z^2}\left( 1 - \frac{2i}{\sqrt{\pi}} \int_0^z e^{-y^2} dy \right) \right\} \\ &= e^{-z^2} = e^{-\frac{\beta(\lambda + e_0\eta - \epsilon)^2}{4\lambda}}. \end{aligned} \tag{201}$$

By inserting this result into Eq. 199, we obtain the rate constant in the non-adiabatic limit, which is the same as the expression derived from time-dependent perturbation theory, as shown in Eq. 132. Herein, the Fermi level $\epsilon_{\mathrm{F}}$ is chosen as the energy reference, which is why it does not appear in the final rate constant expression. If we only consider the ET at the Fermi level, i.e., approximating the Fermi-Dirac distribution function $f(\epsilon)$ as a Dirac delta function at the Fermi level, we have,

$$k_{\mathrm{red}} = \frac{\Delta}{\hbar}\sqrt{\frac{\beta}{\pi\lambda}}\mathrm{Re}\left\{ e^{-\frac{\beta(\lambda+e_0\eta+i\Delta)^2}{4\lambda}} \mathrm{erfc}\left( -\frac{i}{2}\sqrt{\frac{\beta}{\lambda}}(\lambda + e_0\eta + i\Delta) \right) \right\}. \quad (202)$$

A similar derivation can be performed for the oxidation rate constant. The result is presented below without detailed proof. Under the wide-band approximation, we have,

$$k_{\mathrm{ox}} = \frac{\Delta}{\hbar}\sqrt{\frac{\beta}{\pi\lambda}} \int \big(1 - f(\epsilon)\big) \mathrm{Re}\{w(z')\}\, d\epsilon, \quad (203)$$

with,

$$z' = \frac{1}{2}\sqrt{\frac{\beta}{\lambda}}(\lambda - e_0\eta + \epsilon + i\Delta). \quad (204)$$

## 6. Solvent dynamics and interpolating across regions

In the preceding sections we have focused on the theory and simulation of ET kinetics within the adiabatic transition state theory and its non-adiabatic extension. While these allow addressing a wide range of phenomena in ET kinetics, they are based on TST-like theories (Section 2) and therefore do not incorporate the influence of nuclear dynamics on reaction kinetics. Specifically, the nuclear modes are fully equilibrated throughout the ET process. In this section we treat the theory and simulation of nuclear dynamics in ET kinetics, focusing on the environmental (i.e., solvent) dynamics, in two ways: first as a prefactor correction on the TST rate constant, followed by an account for non-ergodic effects in ET kinetics. The inclusion of dynamic effects in the prefactor permits also interpolating across different regimes from non-adiabatic, adiabatic, and nuclear dynamic-controlled limits. This approach may help to answer, e.g., questions whether outer-sphere ET on graphene-like electrodes is adiabatic or non-adiabatic[67,208], whether the nuclear dynamics control the ET rates in deep eutectic solvents[70], or whether ET in redox flow batteries take place through inner-sphere or outer-sphere mechanisms and how this depends on the redox couple and the electrodes[209]. In all these examples, understanding the ET mechanism and phenomena controlling it may aid in the design of electrolytes, electrodes, and the redox couples. Besides using the models with the prefactor correction, the solvent dynamics only affect the rate through the prefactor and do not change the barrier: it is assumed that the sampling of the reaction coordinate, the free energy, and the dynamics are ergodic. The assumption of ergodicity may break for reactions with very small barriers or in slowly relaxing solvents such as ionic liquids[71]; these effects are discussed in Section 6.2.

### 6.1. Prefactor for solvent dynamics

As discussed in Section 2, the TST rate constant corresponds to the zero-time limit in the reactive flux formalism and as such does not depend on the system dynamics. Within this formalism the solvent dynamics can be incorporated through a prefactor $\kappa$ as defined in Eq. 12, which is developed systematically below within a general system-bath model of solvent dynamics.

### 6.1.1. Generalized Langevin equation and friction

A general separation between the system (S) and the bath (B) is achieved using the Zwanzig-Caldeira-Leggett (ZCL) Hamiltonian[210,211]

$$H_{\mathrm{ZCL}}(s,\boldsymbol{x}) = H_{\mathrm{S}}(s) + H_{\mathrm{B}}(\boldsymbol{x}) + H_{\mathrm{SB}}(s,\boldsymbol{x}), \tag{205}$$

where $H_{\mathrm{S}}$ describes the system moving along the RC, $H_{\mathrm{B}}$ the bath, and $H_{\mathrm{SB}}$ their coupling while $s$ is the reaction coordinate and $\boldsymbol{x}$ represents all other coordinates. The bath comprises of uncoupled harmonic oscillators while $H_{\mathrm{SB}}$ is treated as a bilinear coupling between the RC and bath oscillators (note that $H_{\mathrm{SB}}$ is another example of linear coupling between the RC and the solvent environment). Explicitly, the ZCL Hamiltonian for a one-dimensional RC is

$$\begin{aligned} H_{\mathrm{ZCL}}(s,\boldsymbol{x}) &= \underbrace{\frac{p_s^2}{2\mu} + E(s)}_{H_{\mathrm{S}}} + \underbrace{\sum_j \left(\frac{p_j^2}{2m_j} + \frac{1}{2} m_j \omega_j^2 x_j^2\right)}_{H_{\mathrm{B}}} \underbrace{- \sum_j c_j x_j s}_{H_{\mathrm{SB}}} + \sum_j \frac{c_j^2 s^2}{2 m_j \omega_j} \\ &= \frac{p_s^2}{2\mu} + E(s) + \sum_j \left[\frac{p_j^2}{2m_j} + \frac{1}{2} m_j \omega_j^2 \left(x_j - \frac{c_j s}{m_j \omega_j^2}\right)^2\right], \end{aligned} \tag{206}$$

where $p_s$ and $p_j$ denote the system moving along the RC and of the bath DOF $j$, respectively, while $\mu$ and $m_j$ denote the corresponding effective masses, $E(s)$ is the PES along the RC, $m_j$ and $\omega_j$ are the mass and normal mode frequency associated with bath DOF $j$ while $c_j$ is the coupling constant between the RC and the bath. The last term the on the first line renormalizes the free energy and can be included also in the $E(s)$ term[211,212].

It has been shown that the ZCL Hamiltonian corresponds *exactly* to the generalized Langevin equation[210,211]. Here we follow the proof in Ref.[210] and start by writing the equations of motion (EOMs) according to the Hamilton mechanics for the ZCL Hamiltonian

$$\mu \ddot{s} = -\frac{\partial H_{\mathrm{ZCL}}}{\partial s} = -E'(s) + \sum_j c_j \left(x_j - \frac{c_j s}{m_j \omega_j^2}\right), \tag{207}$$

$$m_j \ddot{x}_j = -\frac{\partial H_{\mathrm{ZCL}}}{\partial x_j} = -m_j \omega_j^2 \left(x_j - \frac{c_j s}{m_j \omega_j^2}\right). \tag{208}$$

If a specific trajectory along the RC, $s(t)$, is known, the trajectory $x_j(t)$ corresponds to that of a driven harmonic oscillator with a time-dependent external force due to coupling with RC: $f_{\mathrm{ext}}(t) = c_j s(t)$. The resulting equation can be solved through a Laplace transform ($\hat{f}(\Lambda) \equiv \mathcal{L}[f] = \int_0^\infty f(t) \exp(-\Lambda t)\ dt$. The Laplace transform of both sides of Eq. 207 gives

$$\hat{x}_j(\Lambda) = \frac{\hat{f}_{\mathrm{ext}}(\Lambda)}{m_j \Lambda^2 + m_j \omega_j^2} = \frac{\hat{f}_{\mathrm{ext}}(\Lambda) \hat{g}_{\mathrm{ext}}(\Lambda)}{m_j} = \frac{\mathcal{L}\left[\int_0^t f_{\mathrm{ext}}(t') g(t-t') dt'\right]}{m_j}, \tag{209}$$

where the convolution of two function in Laplace space has been used in the third identity. To go back into the time space, the Laplace transform yielding the function $\hat{g}_{\mathrm{ext}}(\Lambda) = 1/\left(\Lambda^2 + \omega_j^2\right)$ needs to be identified: this is $g(t) = \sin\left(\omega_j t\right)/\omega_j$. Inserting $g(t)$ in the above equation, carrying out the integration, and transforming back to time space one obtains

$$x_j(t) - \frac{c_j s(t)}{m_j \omega_j^2} = \left[x_j(0) - \frac{c_j s(0)}{m_j \omega_j^2}\right] \cos(\omega_j t) + \frac{p_j(0)}{m_j \omega_j} \sin(\omega_j t)$$
$$- \frac{c_j}{m_j \omega_j^2} \int_0^t \cos[\omega_j (t - t')]\, \dot{s}(t')\, dt'. \tag{210}$$

When this is equation is introduced in the EOM for the system along the RC, i.e., Eq. 206, one obtain

$$M\ddot{s} = -E'(s) + \sum_j c_j \left[x_j(0) - \frac{c_j s(0)}{m_j \omega_j^2}\right] \cos(\omega_j t) + \sum_j c_j \frac{p_j(0)}{m_j \omega_j} \sin(\omega_j t)$$
$$- \int_0^t \sum_j \frac{c_j^2}{m_j \omega_j^2} \cos[\omega_j (t - t')]\, \dot{s}(t')\, dt' \tag{211}$$
$$= -E'(s) + F(t) - \int_0^t \Gamma(t - t') \dot{s}(t')\, dt',$$

with the new time-dependent force $F(t)$ and friction kernel $\Gamma(t - t')$

$$F(t) = \sum_j c_j \left[x_j(0) - \frac{c_j s(0)}{m_j \omega_j^2}\right] \cos(\omega_j t) + \sum_j c_j \frac{p_j(0)}{m_j \omega_j} \sin(\omega_j t), \tag{212}$$

$$\Gamma(t - t') = \sum_j \frac{c_j^2}{m_j \omega_j^2} \cos[\omega_j (t - t')]. \tag{213}$$

The force $F(t)$ accounts for the instantaneous impact the bath DOFs on the force along the reaction coordinate while the friction kernel depends on the history and velocity along the reaction coordinate. The last line expresses the motion along the reaction coordinate through a generalized Langevin equation (GLE).

The time-dependent force $F(t)$ depends on the initial conditions of the RC and the solvent bath, which is described as an ensemble consisting of many harmonic oscillators oscillating at different frequencies. As the resulting $F(t)$ will appear random in time and to depend on the unknown initial conditions, it is more convenient to treat $F(t)$ statistically. If both the initial velocity and position of the bath harmonic oscillators have a Gaussian distribution, also the force will have Gaussian distribution with a mean value of zero ($\langle F(t) \rangle = 0$). If the initial positions of the harmonic oscillators are independent, the force autocorrelation function is

$$\langle F(0)F(t) \rangle = k_\mathrm{B} T\, \Gamma(t). \tag{214}$$

This equation is a *fluctuation-dissipation relation* connecting the fluctuation force $F(t)$ with dissipation due to friction. The time-dependent friction can assume various analytical forms but for simulations it's necessary to describe it in terms of solvent dynamics or relaxation at the atomic level. Besides the force-correlation function, this is can be achieved by relating the friction kernel with the bath solvent spectral density. The friction kernel in Eq. 212 can be rewritten in integral form as

$$\Gamma(t) = \int_0^\infty \sum_j \frac{c_j^2}{m_j\omega_j} \frac{\cos(\omega t)}{\omega} \delta(\omega - \omega_j)\, d\omega \\ = \frac{2}{\pi}\int_0^\infty \frac{J(\omega)}{\omega} \cos(\omega t)\; d\omega, \tag{215}$$

with the solvent spectral density

$$J(\omega) = \frac{\pi}{2}\sum_j \frac{c_j^2}{m_j\omega_j} \delta(\omega - \omega_j), \tag{216}$$

which describes the distribution of coupling strengths across different bath mode frequencies. When $J(\omega)$ is proportional to $\omega$, $\Gamma(t)$ is proportional to the $\delta$-function. In this case, the resulting Langevin equation is Markovian, with the friction at time $t$ depending only on the velocity at the same time.

The solvent or outer-sphere reorganization energy can be computed from the solvent spectral density. When the electronic subsystem is located at the initial position $s_{\mathrm{L}}$ or final position $s_{\mathrm{R}}$, the corresponding heat bath Hamiltonians (including $H_{\mathrm{B}}$ and $H_{\mathrm{SB}}$) are given by

$$H_{\mathrm{L/R}} = \sum_j \left[\frac{p_j^2}{2m_j} + \frac{1}{2} m_j \omega_j^2 \left(x_j - \frac{c_j s_{\mathrm{L/R}}}{m_j \omega_j^2}\right)^2\right]. \tag{217}$$

The reorganization energy is then obtained as the ensemble average of the difference between the two heat bath Hamiltonians over the initial state

$$\lambda_{\mathrm{out}} = \langle H_{\mathrm{R}} - H_{\mathrm{L}} \rangle_{\mathrm{L}} = \sum_j \frac{1}{2} m_j \omega_j^2 \left[\left(2\langle x_j \rangle_{\mathrm{L}} - \frac{c_j(s_{\mathrm{L}} + s_{\mathrm{R}})}{m_j\omega_j^2}\right) \frac{c_j(s_{\mathrm{L}} - s_{\mathrm{R}})}{m_j\omega_j^2}\right], \tag{218}$$

where $\langle \ldots \rangle_{\mathrm{L}}$ denotes the ensemble average over the initial state. In classical mechanics, each $x_j$ has a Gaussian distribution with the mean value $\langle x_j \rangle_{\mathrm{L}} = \frac{c_j s_{\mathrm{L}}}{m_j \omega_j^2}$. Substituting this into above equation, we have

$$\lambda_{\mathrm{out}} = \sum_j \frac{1}{2} \frac{c_j^2}{m_j \omega_j^2} (s_{\mathrm{R}} - s_{\mathrm{L}})^2 = \sum_j \frac{1}{2} \frac{{c'_j}^2}{m_j \omega_j^2} = \frac{1}{2\pi} \int_0^\infty \frac{J(\omega)}{\omega}\; d\omega, \tag{219}$$

where in the second step the displacement $(s_{\mathrm{R}} - s_{\mathrm{L}})$ is included in the rescaled coupling constant $c'_j$[213].

### 6.1.2. GLE for electron transfer dynamics

Both the GLE and ET rate theories require specifying the reaction coordinate. As we have discussed in throughout the review, the energy gap in explicit atomistic simulations is typically used as the RC for electron transfer. When the inner-sphere nuclear modes are not included, the non-equilibrium solvent polarization can also serve as the RC, as discussed in Section 3.2. However, it should be noted that these two reaction coordinates are closely related as they describe the same ET process. In particular, continuum models often use the energy gap as the reaction coordinate but dress or write it in terms of e.g. non-equilibrium solvent polarization[214–218]. For this reason, here we will focus on GLE with energy gap as the reaction coordinate while in the following subsections we will present results for both implicit and explicit solvent models. The GLE for dynamics along the energy gap coordinate takes the form

$$\mu\delta\Delta\ddot{E}(t) = -\delta\Delta E(t) - \int_0^t \Gamma(t-t')\delta\Delta\dot{E}(t')\,dt' + F(t), \tag{220}$$

where $\mu$ is the effective mass along the reaction coordinate and $\delta\Delta E(t)$ is the gap fluctuation with respect to its average value. Explicitly these variables have the form

$$\mu = \frac{k_{\mathrm{B}}T}{\left\langle\left(\frac{\partial\Delta E(t)}{\partial t}\right)^2\right\rangle} = \frac{k_{\mathrm{B}}T}{\langle\Delta\dot{E}^2\rangle}, \tag{221}$$

$$\delta\Delta E(t) = \Delta E(t) - \langle\Delta E\rangle. \tag{222}$$

The friction kernel and the random force are still connected by the fluctuation-dissipation relation (Eq. 213) and describe how the solvent influences the energy gap dynamics. A useful measure for dynamics is given by the normalized energy gap time-correlation function

$$\Delta(t) = \frac{\langle\delta\Delta E(0)\delta\Delta E(t)\rangle}{\langle\delta\Delta E(0)\delta\Delta E(0)\rangle}, \tag{223}$$

which can be used to define an effective relaxation timescale for the RC.

To obtain closed form equations for the dynamic effects on ET rates, it is furthermore useful to consider the GLE around a local parabolic minimum or maxima along the RC that correspond to the initial and transition states, respectively. In this case, the GLE for the energy gap coordinate becomes

$$\delta\Delta\ddot{E}(t) = -\omega_i^2\Delta E(t) - \int_0^t \zeta_i(t-t')\delta\Delta\dot{E}(t')\,dt' + f_i(t), \tag{224}$$

where $\omega_i$ is the solvent frequency at the minimum or maximum $i$ while $\zeta_i$ and $f_i$ are the effective mass ($\mu_i$) -weighted friction and random force around $i$, respectively. These variables have the form

$$\mu_i = \frac{k_{\mathrm{B}}T}{\langle\Delta\dot{E}^2\rangle_i}, \tag{225}$$

$$\zeta_i(t) = \frac{k_{\mathrm{B}}T}{\mu_i}\langle f_i(0)f_i(t)\rangle = \frac{\Gamma(t)}{\mu_i}, \tag{226}$$

$$f_i = \mu_i\left(\delta\Delta\ddot{E}(t) + \omega_i^2\delta\Delta E\right). \tag{227}$$

The effective solvent frequency is further related to the force constant, i.e., curvature, of the parabola

$$\omega_i^2 = \frac{k_i}{\mu_i}, k_i = \frac{k_{\mathrm{B}}T}{\langle\delta E^2\rangle_i}. \tag{228}$$

### 6.1.3. Dynamics at the parabolic barrier region: adiabatic ET

For adiabatic reactions, the potential energy surface around the transition state region can be approximated as an inverted parabola along the reaction coordinate. In this case the ZCL Hamiltonian is

$$H_{\mathrm{ZCL}}^{\neq} = \frac{p_s^2}{2\mu_{\neq}} - \frac{1}{2}\mu_{\neq}\omega_{\neq}^2(s - s_{\neq})^2 + \sum_j \left[\frac{p_j^2}{2m_j} + \frac{1}{2}m_j\omega_j^2\left(x_j - \frac{c_j s}{m_j\omega_j^2}\right)^2\right], \tag{229}$$

where $\neq$ denotes the barrier, and $s_{\neq}$ and $\omega_{\neq}$ are the barrier coordinate and frequency, respectively. From this equation it is possible to derive a simple result[219–221] for the dynamic correction in terms of an effective barrier frequency ($w^{\neq}$) that is modified by the friction exerted on the reaction coordinate by solvent bath: this is the celebrated Grote-Hynes model for the prefactor of dynamic solvent effects

$$\kappa_{\mathrm{GH}} = \frac{w^{\neq}}{\omega_{\neq}}, \tag{230}$$

where the effective barrier frequency needs to be self-consistently computed from the equation

$$w^{\neq} = \frac{\omega_{\neq}^2}{w^{\neq} + (1/\mu_{\neq})\hat{\Gamma}(w^{\neq})}, \tag{231}$$

which depends on the Laplace transform of the time-dependent friction kernel

$$\hat{\Gamma}(w^{\neq}) = \int_0^{\infty} \Gamma(t)\exp(-w^{\neq}t)\ dt. \tag{232}$$

The needed parameters, $\omega_{\neq}$, $\mu_{\neq}$ and $\Gamma(t)$ depend on the used reaction coordinate—when the energy gap coordinate is used, they are given by Eqs. 224-227. A simpler expression is obtained by assuming that the friction does not depend on time and is therefore a memoryless constant

$$\Gamma(t) \approx \delta(t)\int_0^{\infty} \Gamma(t)\ dt = \bar{\Gamma} = \mu_{\neq}\zeta, \tag{233}$$

where we have also defined the mass-weighted friction coefficient $\zeta$. This choice simplifies the GLE to an ordinary Langevin equation for the dynamics along the RC. This simplification is justified when memory effects on the friction are modest, which is satisfied when this solvent relaxation is much faster than the time scale of the barrier crossing. When the previous equation is inserted in Eq. 230, the Kramers' result for the effective barrier frequency is obtained

$$w_{\mathrm{KR}}^{\neq} = \sqrt{\omega_{\neq}^2 + \frac{\zeta^2}{4}} - \frac{\zeta}{2}. \tag{234}$$

Inserting this in Eq. 229 gives the famous Kramers' result for the prefactor

$$\kappa_{\mathrm{KR}} = \sqrt{1 + \frac{\zeta^2}{4\omega_{\neq}^2}} - \frac{\zeta}{2\omega_{\neq}}. \tag{235}$$

When the solvent relaxation is extremely fast compared to the barrier crossing, the motion over the barrier is effectively damped and the over-damped, high-friction assumption, $\zeta \gg \omega_{\neq}$, can be used to approximate the solvent effects on the reactive frequency and the transmission coefficient

$$\kappa_{\text{overdamped}} \approx \frac{\omega_{\neq}}{\zeta} \ll 1. \tag{236}$$

The above results are amenable to direct parametrization through MD simulations. However, in many cases it is useful to express the dynamic solvent effects in terms of experimentally measurable relaxation times. This can be achieved by writing the energy gap correlation functions in terms of an analytic functions describing dielectric relaxation dynamics, which are macroscopic measures for the solvent polarization or reorganization dynamics. For this purpose, the (generalized) Langevin equation for the energy gap is written using an implicit dielectric model and solved for the energy gap or polarization dynamics[215,216]. In this case, time-dependent friction describes the solvent polarization dynamics, $\zeta(t) \propto \langle \ddot{\boldsymbol{P}}(0)\ddot{\boldsymbol{P}}(t) \rangle$ and is related to dielectric relaxation timescale $\tau_{\text{rel}}(t)$ through

$$\tau_{\text{rel}}(t) = \Delta(t) = \frac{\zeta(t)}{\omega_L^2}, \tag{237}$$

where $\tau_{\text{rel}}(t)$ depends on time because the dielectric relaxation may have multiple characteristic timescales for different processes such as orientational and translation relaxation. $\Delta(t)$ is the energy gap relaxation timescale introduced in Eq. 222 and $\omega_L$ is the longitudinal solvent frequency[214] for the equilibrium solvent fluctuations that can be computed through Eqs. 224-227 for the initial state. The well and barrier frequencies are related by[214]

$$\omega_{\neq}^2 = \omega_L^2 \left( \frac{\lambda_{\text{out}}}{8V} - 1 \right), \tag{238}$$

where $\lambda_{\text{out}}$ is the solvent reorganization energy and $V$ the electronic coupling constant. The impact of the dielectric relaxation on the barrier crossing dynamics can then be computed from the Grote-Hynes equation of the effective barrier crossing frequency (Eq. 230) using the time-dependent friction

$$w_{\text{rel}}^{\neq} = \frac{\omega_{\neq}^2}{w_{\text{rel}}^{\neq} + \left( \omega_L^2 \hat{\tau}_{\text{rel}}\left( w_{\text{rel}}^{\neq} \right) \right)}, \tag{239}$$

where $\hat{\tau}_{\text{rel}}(w)$ is the Laplace transform of $\tau_{\text{rel}}(t)$. Given a relation for the dielectric relaxation function, $\tau_{\text{rel}}(t)$, the corresponding dynamic correction can be computed[216]. For instance, the widely used model for a Debye solvent has only a single relaxation time and does not depend on frequency. For the longitudinal relaxation, the Debye solvent with the (rotational) relaxation timescale $\tau_{\text{D}}$ has

$$\hat{\tau}_L(s) = \tau_L = \frac{\varepsilon_\infty}{\varepsilon_0} \tau_{\text{D}} \rightarrow \kappa_{\text{GH}} = \sqrt{\frac{(\omega_L \tau_L)^2}{4} + 1} - \frac{\omega_L \tau_L}{2}, \tag{240}$$

with $\zeta_L = \omega_L \omega_{\neq} \tau_L$. For rapid relaxation, this equation reduces the transition state result, while for high friction and slow relaxation, i.e., at the overdamped limit, a Zusman-like equation[218] is obtained

$$\kappa_{\text{overdamped}} \approx \frac{\omega_{\neq}}{\zeta_L} = \frac{1}{\omega_L \tau_L}, \tag{241}$$

which shows that the prefactor is inversely proportional to the longitudinal solvent relaxation time.

### 6.1.4. Dynamics in the initial well: non-adiabatic ET

For a non-adiabatic reaction, the barrier crossing rate or probability is controlled by the electronic coupling constant and ET kinetics often depend more strongly on the crossing probability between the diabatic state than solvent dynamics at the transition state. However, if the solvent dynamics are sufficiently slow, they may influence the equilibrium within the initial state and thereby break the transition state theory assumption of local equilibrium through the reaction coordinate. As a result, in the case of non-adiabatic ET, the solvent dynamics and relaxation in the initial state may control the reaction rate.

The influence of solvent dynamics in initial state equilibration and reaction kinetics may be treated with the GLE or Langevin equation as discussed above for the barrier crossing. As high friction is required for the initial state relaxation to contribute to the reaction rate, over-damped dynamics for solvent dynamics is usually assumed[216,218,222,223]. For the Debye solvent, this leads to the Zusman expression of the prefactor

$$\kappa_{\mathrm{Zus}} \approx \frac{1}{\tau_{\mathrm{D}}\sqrt{\pi\beta\Delta G^{\neq}}}. \tag{242}$$

### 6.1.5. Interpolating between different regions

The previous two sections consider dynamic solvent effects on the prefactor for the adiabatic and non-adiabatic cases. However, in practice it is often necessary to consider also intermediate cases between these two solvent-controlled limits as well as the influence of electronic non-adiabaticity. This can be achieved by constructing a well-defined interpolation between non-adiabatic ET, solvent-controlled adiabatic ET, normal adiabatic ET, and transition between non-adiabatic and solvent-controlled adiabatic ET. One way to achieve this is to use the stable states picture[221,224] of reaction kinetics where the effective prefactor can be written in terms of the barrier crossing dynamics or probability and relaxation dynamics in the initial well. For instance, the barrier crossing can be treated within the Landau-Zener (LZ) theory (section 6.1.6), which interpolates between the adiabatic and non-adiabatic barrier crossing while relaxation dynamics in the initial well may be accounted for with the Zusman model[225]

$$\frac{1}{\kappa_{\mathrm{interpolation}}} \approx \frac{1}{\kappa_{\mathrm{LZ}}} + \frac{1}{\kappa_{\mathrm{Zus}}}. \tag{243}$$

### 6.1.6. Landau-Zener prefactor: Transition from adiabatic to non-adiabatic ET

The Landau-Zener model interpolates between the adiabatic and non-adiabatic electron transfer limits and considers the effective velocity or frequency of the system along the reaction coordinate when crossing the transition state region. As shown in Figure 1, the nuclei first reorganize into a prerequisite configuration at which electron transition can take place. Subsequently, they cross the transition region with an average velocity $v_{\mathrm{avg}}$, while the electronic state gradually transitions from the initial state to the final state. The process occurring in the transition region can be described by Landau-Zener theory[12,226,227].

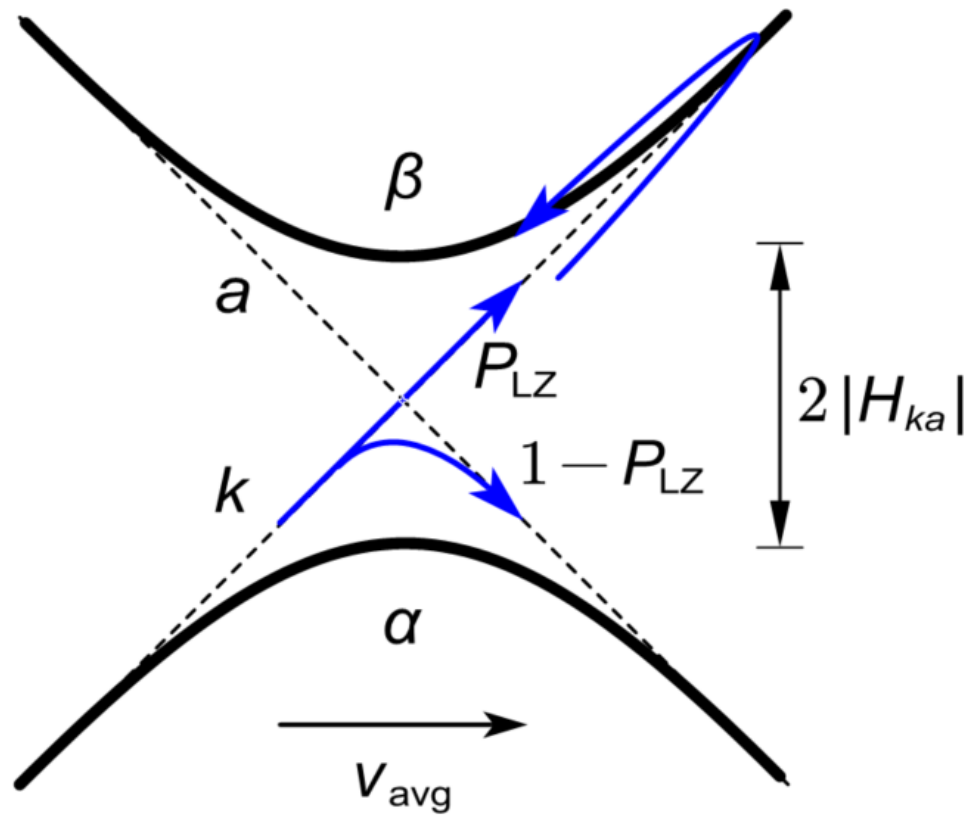

Figure 10. Two-level Landau-Zener model for the description of the system transition from the diabatic state $k$ to the state $a$ in the transition region shown in Figure. 1. The diabatic states are coupled through the matrix element $H_{ka}$, which splits the diabatic states into an upper adiabatic state $\beta$ and a lower adiabatic state $\alpha$. As the nuclei cross the transition region with an average velocity $v_{\mathrm{avg}}$, the probability of system remaining in the initial diabatic state $k$ after the crossing is given by the Landau-Zener transition probability, $P_{\mathrm{LZ}}$. Correspondingly, the probability of the system transitioning to diabatic state $a$ is $1-P_{\mathrm{LZ}}$. In the former case, the nuclei are likely to relax toward the equilibrium nuclear configuration of the diabatic state $k$, leading to a reverse recrossing in the transition region.

To illustrate the Landau-Zener model, we first consider the ET between the diabatic states $k$ and $a$, whose free energy surfaces are described by Eqs. 69 and 70. As shown in Figure 10, the diabatic FESs in the transition region are split into an upper adiabatic state $\beta$ and a lower adiabatic state $\alpha$ by the matrix element $H_{ka}$. As the system crosses the transition region with an average nuclear velocity $v_{\mathrm{avg}}$, the Landau-Zener probability ($P_{\mathrm{LZ}}$) describes the probability with which the system will be excited to the upper state $\beta$, which coincides with the diabatic state $k$ at longer times. Assuming that within the transition region the diabatic PESs vary linearly with the nuclear coordinate $\xi$, with slopes (gradients) $S_k$ and $S_a$ of the corresponding diabatic states $k$ and $a$, respectively, this probability can be described using Landau-Zener theory as:

$$P_{\mathrm{LZ}} = \exp\left(-\frac{2\pi|H_{ak}|^2}{\hbar v_{\mathrm{avg}}|S_k - S_a|}\right). \tag{244}$$

For the parabolas described in Eqs. 69 and 70, it can be shown that the absolute value of the difference between their slopes remains constant and is equal to $2\lambda$ at all nuclear coordinates. $v_{\mathrm{avg}}$ can be considered as the average velocity of the Maxwell-Boltzmann distribution for a single DOF along the reaction coordinate, which for ET is the energy gap, and is given by the equipartition theorem[228], $v_{\mathrm{avg}} = \left(\frac{2k_{\mathrm{B}}T}{\pi\mu}\right)^{\frac{1}{2}}$, where $\mu$ represents the effective mass along the energy gap coordinate in Eq. 220. For the classical and harmonic PESs described in Eqs. 69 and 70, the effective nuclear frequency along the reaction coordinate is given by $\nu_{\mathrm{n}} = \frac{1}{2\pi}\sqrt{\frac{2\lambda}{\mu}}$, from which the reduced mass can be also expressed as $\mu = \frac{\lambda}{2\pi^2\nu_{\mathrm{n}}^2}$. $P_{\mathrm{LZ}}$ can be reformulated by incorporating these considerations as:

$$P_{\mathrm{LZ}} = \exp\left(-\frac{\pi|H_{ak}|^2}{\hbar\nu_{\mathrm{n}}\sqrt{4\pi\lambda k_{\mathrm{B}}T}}\right). \tag{245}$$

In the forward crossing, the probability of transitioning to the diabatic state $a$ is then given by:

$$P_{ka} = 1 - P_{\mathrm{LZ}} = 1 - \exp\left(-\frac{\pi|H_{ak}|^2}{\hbar\nu_{\mathrm{n}}\sqrt{4\pi\lambda k_{\mathrm{B}}T}}\right). \tag{246}$$

When multiple diabatic states $k$ exist, with the electron residing in different electronic states of the metal surface, the system has a probability of $P_{ka}$ of transitioning to the diabatic state $a$ at each transition region between these diabatic states $k$ and $a$. If the coupling between each electronic state $k$ and $a$ is of the same order of magnitude, the crossings at different transition regions can be treated as independent events[78]. In this case, the probability of the system transitioning to the diabatic state $a$ is given by:

$$\begin{aligned} P_a &= 1 - \prod_k (1 - P_{ka}) \\ &= 1 - \exp\left(-\frac{\pi\sum_k |H_{ak}|^2}{\hbar\nu_{\mathrm{n}}\sqrt{4\pi\lambda k_{\mathrm{B}}T}}\right) \\ &= 1 - \exp\left(-\frac{\int \Delta(\epsilon)d\epsilon}{\hbar\nu_{\mathrm{n}}\sqrt{4\pi\lambda k_{\mathrm{B}}T}}\right). \end{aligned} \tag{247}$$

The probability $1 - P_a$ then indicates the probability for the system to remain in the original diabatic states $k$, which gives the system a chance to reverse and move backwards towards another recrossing of the transition region. Considering this backward recrossing, there is also a probability $P_a$ for the system to transition to the diabatic state $a$. Therefore, the probability of a successful electron transition when the solvent nuclei cross and recross the transition regions during each fluctuation. This leads to the electron transmission coefficient within the Landau-Zener model

$$\kappa_{\mathrm{LZ}} = \frac{2P_a}{1 + P_a}. \tag{248}$$

If we consider ET at the Fermi level, we have $\Delta(\epsilon) = \Delta\delta(\epsilon - \epsilon_{\mathrm{F}})$. Consequently, $P_a$ can be rewritten as

$$\kappa_{\mathrm{LZ}} = \frac{2\left(1 - \exp\left(-\frac{\nu_{\mathrm{el}}}{2\nu_{\mathrm{n}}}\right)\right)}{2 - \exp\left(-\frac{\nu_{\mathrm{el}}}{2\nu_{\mathrm{n}}}\right)}, \tag{249}$$

with the electronic frequency,

$$\nu_{\mathrm{el}} = \frac{2\Delta}{\hbar} \cdot \frac{1}{\sqrt{4\pi\lambda k_{\mathrm{B}}T}}. \tag{250}$$

When combined with Eq. 1, the rate constant becomes

$$k = \nu_{\mathrm{n}} \frac{2\left(1 - \exp\left(-\frac{\nu_{\mathrm{el}}}{2\nu_{\mathrm{n}}}\right)\right)}{2 - \exp\left(-\frac{\nu_{\mathrm{el}}}{2\nu_{\mathrm{n}}}\right)} e^{-\frac{\Delta G^{\neq}}{k_{\mathrm{B}}T}}, \tag{251}$$

where it should be noted that the effective frequency is related to the corresponding angular frequency though $\nu_{\mathrm{n}} = \omega_{\mathrm{n}}/(2\pi)$. In the weak coupling limit ($\Delta \to 0$), the exponential term can be estimated as $\exp(-x) \approx 1 - x$ and the pre-exponential factor in Eq. 250 is equal to $\nu_{\mathrm{el}}$ and identical to that in Eq. 134 such that the pre-exponential factor is fully determined by the electronic coupling strength between the metal surface and redox species. In the strong coupling limit, where $\exp(-x) \to 0$, the $\kappa_{\mathrm{LZ}}$ simplifies to $\nu_{\mathrm{n}}$, and is entirely determined by the effective nuclear frequency. The activation energy $\Delta G^{\neq}$ can be determined from the adiabatic FES in Eq. 171 across the entire range of coupling strengths. For cases with small overpotentials, substituting Eq. 174 into Eq. 250 yields:

$$k_{\mathrm{red}} = \nu_{\mathrm{n}} \frac{2\left(1 - \exp\left(-\frac{\nu_{\mathrm{el}}}{2\nu_{\mathrm{n}}}\right)\right)}{2 - \exp\left(-\frac{\nu_{\mathrm{el}}}{2\nu_{\mathrm{n}}}\right)} e^{-\frac{\beta(\lambda+e_0\eta)^2}{4\lambda}} e^{\frac{\beta\Delta}{2\pi}\ln\left(1+\frac{(\lambda+e_0\eta)^2}{\Delta^2}\right)}. \tag{252}$$

Figure 11a shows the above reduction rate constant as a function of the coupling strength at zero overpotential. The parameters used are $\nu_{\mathrm{n}} = 10^{11}\ \mathrm{s}^{-1}$ and $\lambda = 1$ eV. As the coupling strength increases, the rate constant initially rises rapidly in the weak coupling regime, then grows very slowly in the intermediate coupling regime, and finally raises rapidly again in the strong coupling regime. As shown in Figure 11b, the behavior of the rate constant at different coupling strengths can be can be understood by dissecting the rate into contributions from the pre-exponential factor and the barrier. In the weak coupling regime, the pre-exponential factor increases rapidly with electronic coupling. However, after reaching a certain value, it becomes limited by solvent dynamics and remains independent of coupling strength. In the strong coupling regime, the rate constant is primarily determined by the activation energy, where stronger coupling reduces the activation energy due to stronger hybridization between the electrode and the reactant—this is sometimes referred to as the electrocatalytic effect. In the intermediate coupling regime, both the pre-exponential factor and activation energy exhibit a weak dependence on coupling strength.

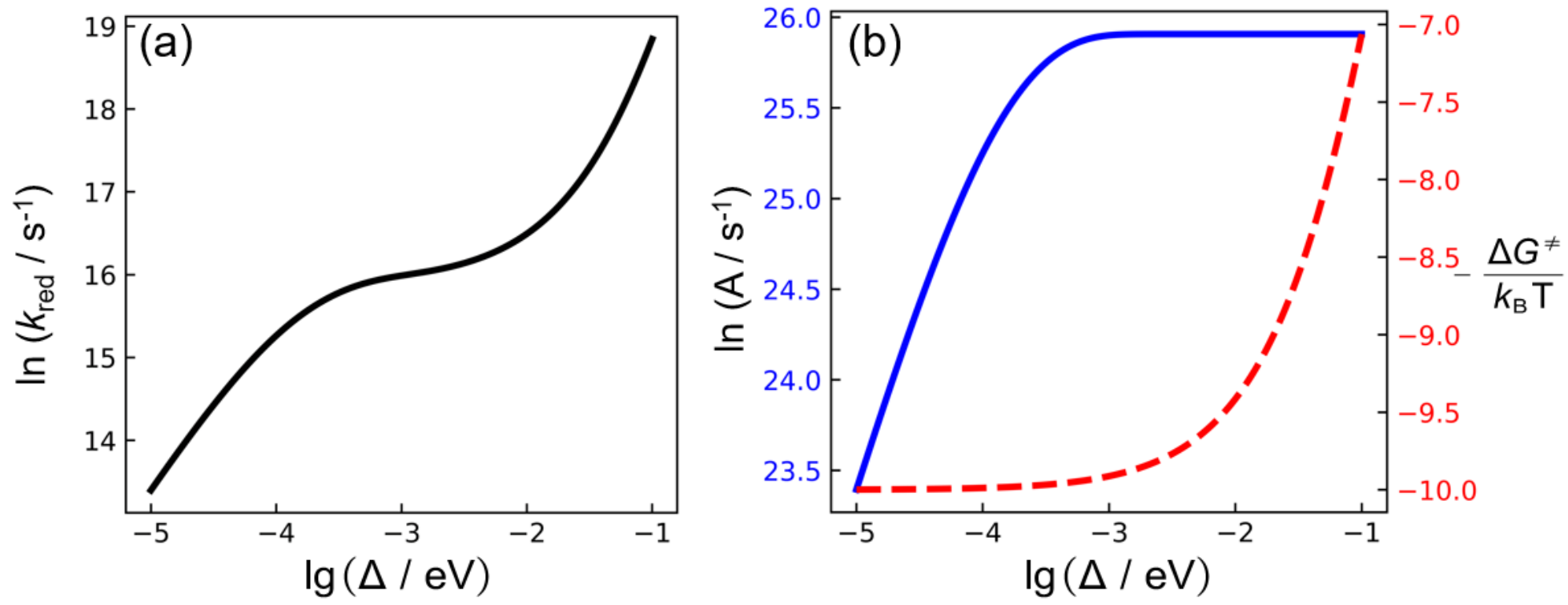

Figure 11. The (a) reduction rate constant, and (b) pre-exponential factor (solid line) as well as activation energy (dashed line) a function of the coupling constant $\Delta$ at zero overpotential. The parameters used are $\nu_{\mathrm{n}} = 10^{11}\ \mathrm{s}^{-1}$ and $\lambda = 1$ eV.

### 6.1.7. Interpolation across regions

The stable states equation, Eq. 242, achieves interpolation between electronically non-adiabatic and adiabatic limits and to the solvent dynamics controlled non-adiabatic limit. It does not, however, interpolate to adiabatic solvent-controlled reactions where the barrier crossing dynamics is described by the Kramers-Grote-Hynes-like equations. This can be tentatively corrected by recognizing that for adiabatic reactions $\kappa_{\mathrm{LZ}} \to 1$ and should be replaced by $\kappa_{\mathrm{GH}}$ in Eq. 239. On the other hand, for non-adiabatic reactions $\kappa_{\mathrm{LZ}} \ll \kappa_{\mathrm{GH}}$ and the solvent dynamics depend on $\kappa_{\mathrm{Zus}}$.

To obtain a uniform interpolation across different regions, we therefore propose the following formula

$$\frac{1}{\kappa_{\mathrm{interpolation}}} = \frac{1}{\kappa_{\mathrm{LZ}}\kappa_{\mathrm{GH}}} + \frac{1}{\kappa_{\mathrm{Zus}}}. \tag{253}$$

When the solvent is described with the Debye model, the interpolation between different regions results in Figure 12.

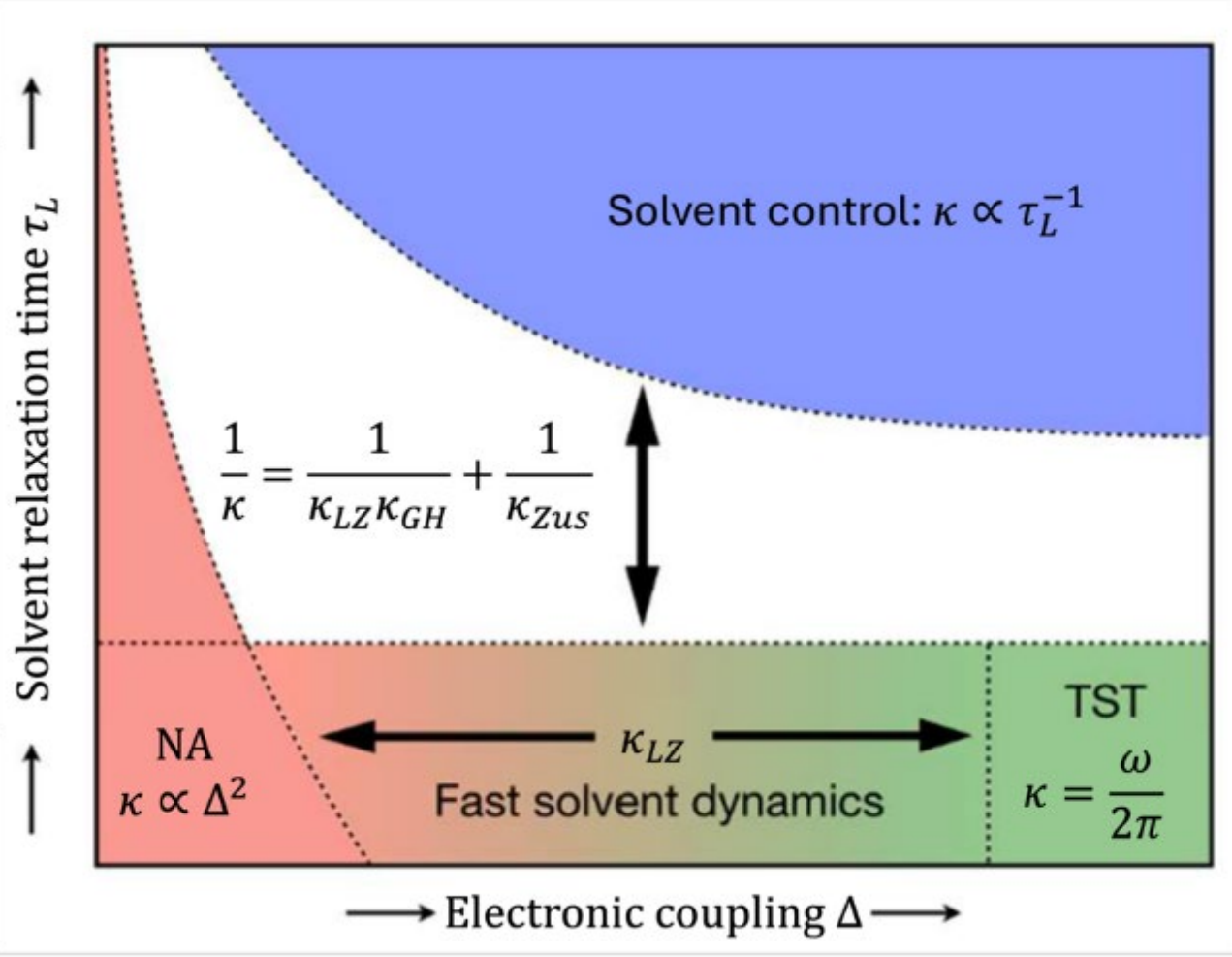

Figure 12. Interpolation of the prefactor for different regions of electron transfer. Adapted from Ref.[225], with permission from the American Chemical Society. Copyright (2019), ACS.

### 6.1.8. Simulating the dynamic prefactor

The dynamic solvent effects can be simulated either through direct MD simulations of the transmission coefficient in Eq. 12 or by using MD simulations to parametrize the semi-analytical models of the transmission coefficient. The former approach is in principle simple but computationally expensive as it requires launching several short MD trajectories at the dividing surface and studying whether they end up in the reactant or product regions. In general, parametrizing the semi-analytical models is computationally less demanding and requires evaluation of the parameters through Eqs. 220-227, which give the needed parameters in terms of the time correlation function for the energy gap or its time derivative. Once an EVB model and the diabatic are constructed and the energy gap is sampled, the needed correlation functions can be obtained through standard techniques[229]. However here it is important to notice that because the prefactor depends explicitly on the time and the system dynamics, results obtained from canonical and grand canonical can no longer be interconverted through a Legendre transform. This can be appreciated by considering e.g. Eqs. 220-227 which show that the quantities entering the definition of the dynamics prefactor depend explicitly on time, fluctuations, and systems dynamics, which are different in the canonical and grand canonical ensembles. Hence, choosing the appropriate ensemble at the start of the simulation is pivotal in the description of solvent dynamics in ET kinetics.

## 6.2. Non-ergodic rate theory of ET kinetics

Non-ergodicity due to slow system dynamics have been reported e.g., for ET in ionic liquids[71] and has led to reconsideration of the basic foundations of rate theories which typically assume that the ergodic hypothesis is valid. The ergodic hypothesis assumes that the ensemble and infinite-time averages of an observable $O$ are equal and time-independent. For the canonical ensemble this is written as

$$\langle O\rangle = \frac{\int d\boldsymbol{p}d\boldsymbol{r}\; O(\boldsymbol{r},\boldsymbol{p})\exp[-\beta H(\boldsymbol{r},\boldsymbol{p})]}{Q} = \int_{\Upsilon}\rho(\Upsilon)O(\Upsilon)d\Upsilon \\ = \lim_{\tau\to\infty}\frac{1}{\tau}\int_0^{\tau} O(t)dt, \tag{254}$$

where the second equality on the first line introduces the sample phase space $\Upsilon$ and the phase space probability distribution $\rho(\Upsilon)$. The ergodic hypothesis can naturally be extended to the grand canonical ensemble[120].

When applied in the simulation of reaction kinetics within the TST and its corrections, the ergodic hypothesis dictates that all needed quantities are time-independent and ergodic; as a result, the free energies, partition functions, and dynamic prefactors are time-independent and should be simulated at the infinite time limit or equivalently through complete phase space sampling of the relevant portion of the phase space.

Most results of statistical thermodynamics and thereby the rate theory as presented in Section 2 build on the assumption of ergodicity, and in most treatments and theories of reaction kinetics, the system is assumed to be fully ergodic and therefore ergodic hypothesis is deeply ingrained in chemical kinetics. In a strict TST perspective the ergodic hypothesis indicates that all the timescales of an electrochemical systems, discussed in Section 2.5., from sub-femtosecond electron motion to bond vibrations on the femtosecond scale, picosecond solvent reorganization, or double layer dynamics on the microsecond scale, should contribute to the partition functions, barrier, and kinetics and should therefore be included in the simulations. While this is not possible in practical simulations, this leads to very difficult questions: which timescales are relevant to the reaction kinetics? Which timescales should be included in the simulation of reaction rates? Should some motions or degrees freedom appear frozen on the timescale of an electrochemical reaction?

These questions can be been analyzed within the non-ergodic ET rate theory developed by Matyushov[230,231] for protein electrochemistry. However, the formalism is applicable more generally to also electrochemical ET[232] and the key insight of this theory is that some motions and degrees of freedom appear dynamically frozen on the timescale of reactive ET events. The apparent freezing of some slow degrees of freedom leads to ergodicity breaking, which influences both the reaction barrier and the prefactor, which become functions of system dynamics and thereby explicitly time-dependent. As non-ergodicity places restrictions on which regions of the phase space contribute to the partition function and reaction barrier, it dictates that the phase space averages and sampling times should be limited to regions where the environment dynamics ($\tau_{\mathrm{env}}$) are faster or equal to the reaction timescale $\tau_{\mathrm{react}} = k^{-1}$, $\tau_{\mathrm{env}} \geq \tau_{\mathrm{react}}$ or $\tau_{\mathrm{env}}k \geq 1$. For free energy this is enforced by constraining the phase space sampling as

$$G(\tau_{\mathrm{env}}) = \lim_{\tau\to\tau_{\mathrm{env}}}\frac{1}{\tau}\int_0^{\tau} G(t)dt = \int_{\tau_{\mathrm{env}}k\geq 1}\rho(\Upsilon)G(\Upsilon)d\Upsilon. \tag{255}$$

In practice, the phase space sampling needs to be limited to the degrees of freedom that are faster than the ET kinetics of the studied reaction. For this it is beneficial to cast the phase space in the frequency domain through a Fourier transform such that phase space becomes $d\boldsymbol{p}d\boldsymbol{r} = \prod_{\omega}\prod_i dp_i(\omega)dr_i(\omega)$ for the frequency ($\omega$) for different degrees of freedom $i$. In the non-ergodic sampling of the phase space, only frequencies higher than the reaction rate or frequency are considered and the slower degrees are filtered out. Hence, only the sufficiently high-frequency motions, $\tau \geq \tau_{\mathrm{s}}$, of the environment contribute to the rate and free energy:

$$G(\tau_{\mathrm{s}}) = \int_{\Upsilon} \rho(\Upsilon) G(\Upsilon) \prod_{\omega \geq \tau_{\mathrm{s}}} \prod_{i} dp_i(\omega) dr_i(\omega). \tag{256}$$

Qualitatively, this means that the system does not have enough time to roam the entire phase space on the reaction timescale which places strict restrictions on which regions of the phase space contribute to the partition function and free energy. More quantitatively, the system dynamics ($\tau_{\mathrm{env}}$) that are slower than the reaction timescale ($\tau_{\mathrm{react}}$) are dynamically frozen and do not contribute to the (time-dependent) thermodynamic quantities. Hence, non-ergodicity is relevant when $\tau_{\mathrm{env}} \geq \tau_{\mathrm{react}}$ as the contributions from dynamic processes slower than the reaction do not contribute to the free energy, the partition functions, rate or other quantities. This can be explicitly seen by re-writing the TST rate equations, Eqs. 9 and 12, in terms of the non-ergodic quantities, which leads to the non-ergodic (ne) TST

$$k_{\mathrm{ne-TST}}(\tau_{\mathrm{env}}) = \kappa_{\mathrm{dyn}}(\tau_{\mathrm{env}}) \frac{k_{\mathrm{B}}T}{h} \frac{Q^{\neq}(\tau_{\mathrm{env}})}{Q_i(\tau_{\mathrm{env}})} = \kappa_{\mathrm{dyn}}(\tau_{\mathrm{env}}) \frac{k_{\mathrm{B}}T}{h} \exp[-\beta \Delta G^{\neq}(\tau_{\mathrm{env}})]. \tag{257}$$

This equation shows that for a non-ergodic system the dynamics influence both the prefactor and the barrier. Another way to look at the influence of non-ergodicity on ET kinetics is to consider the reorganization energy. By restricting the sampled frequencies according to $\omega \geq \tau_{\mathrm{env}}$ in the integral of Eq. 218 for the solvent reorganization energy in terms of the spectral density, one obtains

$$\lambda_{\mathrm{out}}(\tau_{\mathrm{env}}) = \frac{1}{2\pi} \int_{\omega \geq \tau_{\mathrm{env}}} \frac{J(\omega)}{\omega} d\omega, \tag{258}$$

which shows that also the solvent reorganization energy and thereby the Marcus kinetics may depend on dynamics in non-ergodic systems.

#### 6.2.1. Computational and practical considerations

The equations appearing in non-ergodic rate theories should be solved self-consistently or iteratively to satisfy the $\tau_{\mathrm{env}} \geq \tau_{\mathrm{react}}$ condition; this iterative process makes computation of the non-ergodic rate constants very difficult and time-consuming as the simulation of the system dynamics and the rate constant become dependent on each other. This requires very thorough sampling of the phase space and the dynamics, which has thus far been achieved through classical MD simulations for small molecules[233], proteins[234], and even a molecule on a metallic electrode[235]. Future studies may extend to choice of methods to QM/MM and machine learning potentials but to our knowledge this has not yet been done. It is also important to notice that both environment relaxation and reactions times depend explicitly on the sampling time, fluctuations, and systems dynamics, which are different in the canonical and grand canonical ensembles, which makes the ensemble choice a critical issue.

In DFT-level studies it is currently not possible to achieve the needed sampling but it is important to estimate in which cases non-ergodicity should be accounted for in ET simulations. Such an analysis was done in Ref.[120], where the ergodicity-breaking was inspected from the perspective of system dynamics. As the system dynamics naturally depend on the system, it is not possible to provide general guidelines but for ET reactions in aqueous electrolytes it was concluded that for reactions with $\Delta G^{\neq} > 0.3$ eV or $\lambda > 1.2$ eV the system is expected to be ergodic for systems that can be studied using DFT. Hence, non-ergodic ET is expected in systems with very fast kinetics or very slow system dynamics, which may be observed in e.g., ionic liquids.

## 7. Electric double layer effects

As ET occurs within a local region of the EDL, the local properties and reactant concentration in the reaction environment can differ significantly from those in the bulk solution. Incorporating these differences and EDL effects into the ET rate is essential for a comprehensive understanding of experimental results, particularly in studies of electrolyte effects. The EDL effects greatly complicate the understanding of ET reaction even “simple” outer-sphere couples as the electrode charge, potential of zero charge, and the electrostatic potential within the EDL may or may not influence ET rate depending on the electrolyte concentration. More generally, as electrolyte- and potential-dependent electrostatic or electric field effects are often used to rationalize electrochemical interfaces[236], understanding of the EDL is paramount for understanding how and why ET kinetics depend on the local reaction conditions under varying operational conditions[237,238]. A particular example of topical interest is understanding cation effects on $CO_2$ reduction[238]; are the cations effects on ET kinetics due to global, long-range electric field effects[239] or local, direct, and short-range Coulombic interactions[114,240]. Answering such questions may help to design optimal reaction conditions and control ET but this requires e.g., atomic-scale insight on electrostatic effects within the EDL[241] and the development and application of microkinetic models that account for the variety of EDL-related phenomena[125,242]. In this section, we examine how the EDL effects on the ET rate can be described and affected through three key factors: the work terms, local reactant concentration, and reorganization energy.

### 7.1. Work terms

As discussed in Section 1.2.1, work $w_{\mathrm{red}}$ and $w_{\mathrm{ox}}$ is required for the reduced and oxidized species, respectively, to move from the bulk solution to the reaction plane for ET to occur. In this case, the equilibrium free energies of the redox species, as expressed in Eqs. 41 and 42, at the reaction plane differ from those in the bulk solution. Assuming the equilibrium free energies of the redox species at the reaction site are $G^0_{\mathrm{red},a}$ and $G^0_{\mathrm{ox},a}$, respectively, we have:

$$G^0_{\mathrm{red},a} = G^0_{\mathrm{red}} + w_{\mathrm{red}}, \qquad G^0_{\mathrm{ox},a} = G^0_{\mathrm{ox}} + w_{\mathrm{ox}}. \tag{259}$$

Here, $G^0_{\mathrm{red}}$ and $G^0_{\mathrm{ox}}$ are the equilibrium free energies of the reduced and oxidized species in the solution bulk. By combining Eqs. 47, 63 and 258, the reaction free energy for the reaction in Eq. 63 at the reaction plane is corrected to,

$$\begin{aligned}\Delta G^0(\epsilon_k) &= G^0_{\mathrm{red},a} - G^0_{\mathrm{ox},a} - \epsilon_k \\ &= e_0\eta + \epsilon_{\mathrm{F}} - \epsilon_k + w_{\mathrm{red}} - w_{\mathrm{ox}}.\end{aligned} \tag{260}$$

The work terms $w_{\mathrm{red}}$ and $w_{\mathrm{ox}}$ primarily describe changes in the electrostatic potential energy and equilibrium solvation energy of the redox species and can be decomposed into $w_i = w_i^{\mathrm{elec}} + w_i^{\mathrm{solv}}$. For chemisorption reactions at electrochemical interfaces, the process is accompanied by changes in the covalent interaction energy and the desolvation of the redox species and electrode surface. These effects and their corrections to the reaction free energy can be self-consistently accounted for by combining the EDL theory and chemisorption theory to construct a two-dimensional FES, as introduced in Section 5.3 and in Ref.[125]. The reaction free energy is then given by the difference between the minimum of the chemisorbed product and that of the reactant. The electrostatic potential contribution, $w_i^{\mathrm{elec}}$, can be approximated by using a simple model where the redox species are treated as point charges. The corresponding work terms, associated with the in electrostatic potential energy of the redox species, is then given by:

$$\Delta w^{\mathrm{elec}} = w_{\mathrm{red}}^{\mathrm{elec}} - w_{\mathrm{ox}}^{\mathrm{elec}} = (z_{\mathrm{red}} - z_{\mathrm{ox}}) e_0 (\phi_a - \phi_{\mathrm{S}}) = -e_0 \Delta\phi_{\mathrm{S}}^{a}, \quad (261)$$

where $\Delta\phi_{\mathrm{S}}^{a} = \phi_a - \phi_{\mathrm{S}}$ represents the difference between the electrostatic potential at the reaction plane within the EDL and the inner potential of the bulk solution. $z_{\mathrm{red}}$ and $z_{\mathrm{ox}}$ denote the valencies of the reduced and oxidized species, respectively. If the reaction plane is located at the outer Helmholtz plane (OHP), $\Delta\phi_{\mathrm{S}}^{a}$ corresponds to the potential drop across the diffuse layer of the EDL. In this case, $\Delta\phi_{\mathrm{S}}^{a}$ increases monotonically with the surface free charge of the electrode, $\sigma_{\mathrm{free}}$, which is equal in magnitude but opposite in sign to the excess ionic charge in the diffuse layer, ensuring the electroneutrality of the EDL. $\sigma_{\mathrm{free}}$ typically increases monotonically with the electrode potential in the range near the potential of zero free charge (PZFC). However, the accumulation of partially charged chemisorbates, which contribute to the dipole potential at the electrode surface, may lead to a non-monotonic relationship between surface free charge and electrode potential[243–246].When $\sigma_{\mathrm{free}} > 0$ , $\Delta\phi_{\mathrm{S}}^{a}$ is positive, the overpotential is effectively decreased by $\Delta\phi_{\mathrm{S}}^{a}$ , which facilitates reduction reactions. Conversely, when $\sigma_{\mathrm{free}} < 0$ , the overpotential is effectively increased, which facilitates oxidation reactions.

The difference between the local solvation free energy and its bulk counterpart arises from solvent screening of the interfacial electric field, which makes the local dielectric response differ from that of the bulk. In a qualitative description of local field effects on ET, the local dielectric permittivity can be assumed to be a constant, $\varepsilon_{\mathrm{s}}^{\mathrm{loc}}$, for the redox species at a given electrode potential. Considering the reaction of a spherical ion with a negligible change in size during ET, the outer-sphere contribution to $w_i^{\mathrm{solv,o}}$ can be expressed though the Born solvation model as,

$$w_i^{\mathrm{solv,o}} = \frac{z_i^2 e_0^2}{8\pi R}\left(\frac{1}{\varepsilon_{\mathrm{s}}^{\mathrm{loc}}} - \frac{1}{\varepsilon_{\mathrm{s}}^{\mathrm{b}}}\right), \quad (262)$$

where $R$ is the radius of spherical ion, $\varepsilon_{s}^{\mathrm{b}}$ is the static dielectric permittivity of the bulk solution, and $z_i$ is the ion valence. The corresponding work terms, associated with the change in the solvation energy of the redox species, is then given by:

$$\Delta w^{\mathrm{solv,o}} = w_{\mathrm{red}}^{\mathrm{solv,o}} - w_{\mathrm{ox}}^{\mathrm{solv,o}} = \frac{\left(z_{\mathrm{red}}^2 - z_{\mathrm{ox}}^2\right) e_0^2}{8\pi R}\left(\frac{1}{\varepsilon_{\mathrm{s}}^{\mathrm{loc}}} - \frac{1}{\varepsilon_{\mathrm{s}}^{\mathrm{b}}}\right), \quad (263)$$

where we assume that the reaction plane for both the reduced and oxidized species are identical. When the electrode potential deviates from the PZFC, $\varepsilon_{s}^{\mathrm{loc}} < \varepsilon_{s}^{\mathrm{b}}$. For reactions where $z_{\mathrm{red}}^2 > z_{\mathrm{ox}}^2$, $\Delta\omega^{\mathrm{solv,o}} > 0$. In this case, $\Delta\omega^{\mathrm{solv,o}}$ effectively increases the overpotential, facilitating oxidation reactions while hindering reduction reactions. Conversely, for reactions where $z_{\mathrm{red}}^2 < z_{\mathrm{ox}}^2$ , $\Delta\omega^{\mathrm{solv,o}} < 0$ and the overpotential felt by the redox couple is effectively decreased: this favors reduction reactions and suppresses oxidation reactions. As the metal surface accumulates more excess charge, $\varepsilon_{s}^{\mathrm{loc}}$ decreases further, which amplifies the effect of $\Delta w^{\mathrm{solv,o}}$. Apparently, the correction of the work terms to the ET rate depends on the chosen position of the reaction plane.

However, we should not expect more than a qualitative understanding of the work terms related to the solvation from Eq. 262, which has some defects that limit its application. First, a continuum description of electrostatic or electric field effects cannot fully mimic the microscopic electrostatic interactions arising from direct Coulombic interactions between molecules themselves and with the electrode[241]. Second, the redox species are not simply spherical and the metal surface may not be perfectly flat and structureless. Third, the local dielectric constant $\varepsilon_{\mathrm{s}}^{\mathrm{loc}}$ is not constant and it varies spatially near the metal surface and the redox species due to the dielectric screening of the interfacial field by the solvent. To address the latter two issues, we can model both the redox species and the metal surface with arbitrary shapes and specific charge density distributions in 3D space. For such 3D modeling it is necessary to reformulate the outer-sphere components in Eqs. 41 and 42 in terms of scalar quantities, such as electric potential and charge density, rather than using vector fields. This can be achieved by transforming the integration of electric displacement fields into those of charge density and electrostatic potential:

$$\begin{aligned}&\int \frac{\boldsymbol{D}_i}{\varepsilon_{\mathrm{s}}}\boldsymbol{D}_j dV = \int \boldsymbol{E}_i\boldsymbol{D}_j dV = \int (-\nabla\phi_i)\boldsymbol{D}_j dV \\ &= \int \left(-\nabla\left(\phi_i\boldsymbol{D}_j\right) + \phi_i\nabla\cdot\boldsymbol{D}_j\right)dV \\ &= \int \phi_i\varrho_j dV,\end{aligned} \tag{264}$$

where $\boldsymbol{D}_i$ and $\boldsymbol{D}_j$ are the electric displacement fields of two charging state $i$ and $j$ at their equilibrium polarizations, $\phi_i$ and $\boldsymbol{E}_i$ the electric potential and field in charging state $i$, $\varrho_j$ the charge distribution in charging state $j$. The first three identities are due to Eq. 24, the relation between the electric field and potential, and the product rule for the divergence operator. The fourth identity assumes a finite system, where the first term in the second line of Eq. 263 vanishes by applying the divergence theorem. According to Eq. 263, the equilibrium solvation free energy in Eqs. 41 and 42, can be reformulated as,

$$G_i^{\mathrm{eq,o}} = \frac{1}{2}\int \phi_i\varrho_i \, dV \ \ (i = \mathrm{ox, red}). \tag{265}$$

As mentioned, for the system where the redox species are embedded in the dielectric medium near the metal surface, we can account only for the charge distribution on the redox species in the integral of Eq. 264. If the charge of the redox species is assumed to be distributed on its surface, Eq. 264 can be reduced to a surface integral as,

$$G_i^{\mathrm{eq,o}} = \frac{1}{2}\int \phi_i^{\mathrm{s}}\sigma_i \, dS \ \ (i = \mathrm{ox, red}), \tag{266}$$

where $\sigma_i$ represents the charge density distribution on the surface of species $i$, and $\phi_i^{\mathrm{s}}$ denotes the corresponding electrostatic potential distribution on this surface. To calculate the electrostatic potential distribution, an EDL model is needed to account for the solvent polarization, as the response of solvent molecules to the interfacial electric field and the charge of the redox species significantly affects the local dielectric permittivity and, consequently, the solvation free energy. Such simulations can be achieved by using the comprehensive continuum EDL theory that accounts for both the electron response of the metal electrons and structured solvent[43]. This theory also incorporates the effect of short-range correlations between solvent molecules and between ions and solvent molecules into solvent polarization, and provides a computationally efficient approach for describing solvation at electrified interface with a more realistic description, as demonstrated in Figure. 14.

### 7.2. Solvent reorganization free energy

The inner-sphere reorganization energy is generally independent of local conditions, as it is primarily determined by the intrinsic structural properties of the redox species. In this subsection, we therefore focus on how the outer-sphere (i.e., solvent) reorganization energy depends on local conditions. The solvent reorganization energy, $\lambda_{\mathrm{out}}$, is given in Eq. 58 within the continuum description of the solvent. This expression shows that $\lambda_{\text{out}}$ depends on the local dielectric permittivity and the charge distribution of the redox species, the latter being related to their shape. We begin by developing a qualitative understanding of $\lambda_{\text{out}}$ by examining the reduction of a spherical ion near a metal surface, as depicted in Figure 13. In its oxidized state the spherical ion is located at a distance $d$ from the metal surface and has the charge $z_{\text{ox}}e_0$ and the radius $R$. If the ion is very close to the metal surface, i.e., $d$ is very small, the metal electrons may sense the electric field due to the ionic charge and are either repelled or attracted, resulting in the formation of induced (surface) charge. The electrostatic effects of this induced charge can be equivalently represented by an image charge, which mirrors the ionic charge distribution across the metal surface but with the opposite sign. Assuming the ionic charge generated by both the ionic and image charge is uniformly distributed on the surface of the ionic sphere, $\boldsymbol{D}_{\text{ox}}$ is given by,

$$
\begin{aligned}
&\boldsymbol{D}_{\text{ox}} = \frac{z_{\text{ox}}e_0}{4\pi r_a^2}\left(\frac{\boldsymbol{r}_a}{r_a}\right) - \frac{z_{\text{ox}}e_0}{4\pi r_b^2}\left(\frac{\boldsymbol{r}_b}{r_b}\right)\ (|\boldsymbol{r}_a| \geq R, |\boldsymbol{r}_b| \geq R), \\
&\boldsymbol{D}_{\text{ox}} = 0\ (|\boldsymbol{r}_a| < R, |\boldsymbol{r}_b| < R),
\end{aligned}
\tag{267}
$$

where $\boldsymbol{r}_a$ and $\boldsymbol{r}_b$ denote the radial vectors from the centers of the ionic sphere and image charge sphere, respectively, with their magnitudes $r_a$ and $r_b$. If $R$ changes negligibly during the ET process, $\boldsymbol{D}_{\text{red}}$ can be formulated by simply replacing $z_{\text{ox}}$ with $z_{\text{red}}$ in Eq. 266. Then we have,

$$
\lambda_{\text{out}} = \frac{e_0^2}{32\pi^2}\int\left(\frac{1}{\varepsilon_\infty} - \frac{1}{\varepsilon_{\text{s}}}\right)\left(\nabla\frac{1}{r_a} - \nabla\frac{1}{r_b}\right)^2 dV_0. \tag{268}
$$

Here, $V_0$ excludes the volume space inside the spheres. On the solution side, we assume a constant local dielectric permittivity, $\varepsilon_{\text{s}}^{\text{loc}}$, while on the metal side the dielectric constant equals $\varepsilon_\infty$. The above integral can then be explicitly evaluated[1], yielding $\lambda_{\text{out}}$ as,

$$
\lambda_{\text{out}} = \frac{e_0^2}{8\pi}\left(\frac{1}{\varepsilon_\infty} - \frac{1}{\varepsilon_{\text{s}}^{\text{loc}}}\right)\left(\frac{1}{R} - \frac{1}{2d}\right). \tag{269}
$$

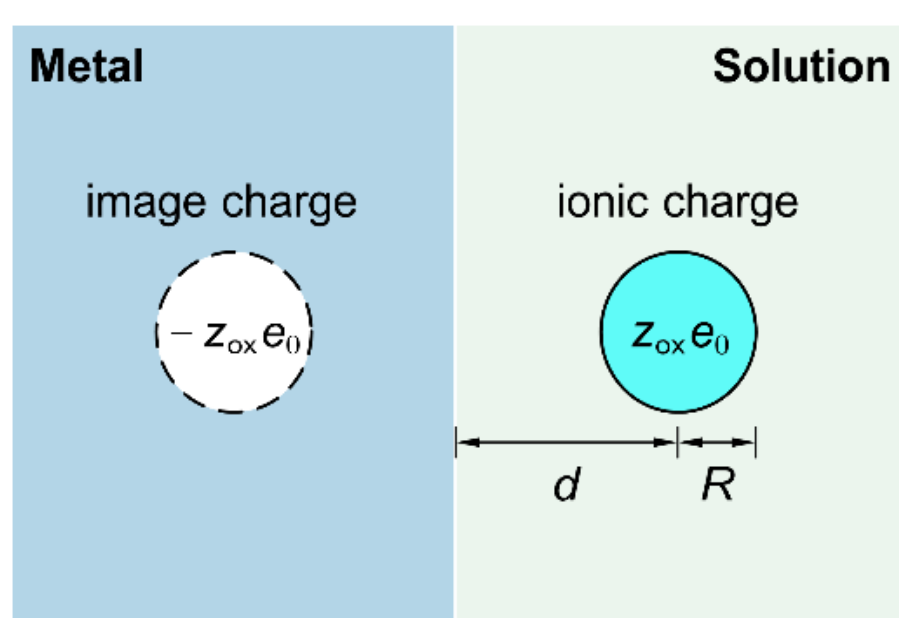

Figure 13. The spherical ion in its oxidized state, with the charge $z_{\text{ox}}e_0$ and the radius $R$, is positioned at a distance $d$ from the metal surface. Its image charge, having the same magnitude but an opposite charge distribution, is mirrored by the metal surface.

Eq. 268 can provide some fundamental insights into the factors influencing the solvent reorganization energy. An ion with a smaller radius $R$ exerts a stronger electric force on the nearby solvent molecules, which makes it more difficult for the solvent molecules to reorient and which leads to a larger reorganization energy. A larger reorganization energy is also expected for a shorter distance $d$ of the ion from the metal surface, where the solvent molecules experience a stronger electric field from the image charge. A larger $\varepsilon_{\text{s}}^{\text{loc}}$ indicates that the solvent molecules have a greater ability to screen the external electric field, which causes them to sense a weaker electric field from the charge of the redox species and its image charge. The weaker electric field experienced by the solvent molecules from both the ion and image charges allows them to reorient more freely, implying a smaller reorganization energy. The value of $\varepsilon_{\mathrm{s}}^{\mathrm{loc}}$ depends on the nature of the solvent as well as the local electric field in the EDL, which in turn depends on the surface charge and electrode potential. The solvent properties of a dipolar liquid are primarily determined by the solvent dipole moment such that a larger dipole moment implies a stronger screening ability and thus a higher $\varepsilon_{\mathrm{s}}^{\mathrm{loc}}$. The field effects reflect the importance of local reaction conditions, namely EDL effects. At a charged metal surface, solvent molecules feel a stronger electric field and become more ordered as they approach the surface, resulting in a lower $\varepsilon_{\mathrm{s}}^{\mathrm{loc}}$ at shorter distance from the metal surface. The solvent reorganization energy at the interface is thus generally smaller than that in the bulk solution, as confirmed by computations and experiments[247,248].

For more accurate and convenient computation for $\lambda_{\mathrm{out}}$ at the continuum level, we first express $\lambda_{\mathrm{out}}$ in Eq. 58 in terms of scalar quantities. Similar to Eq. 264, we obtain,

$$\int \frac{\boldsymbol{D}_i}{\varepsilon_\infty}\boldsymbol{D}_j dV = \int \boldsymbol{E}_i^\infty \boldsymbol{D}_j dV = \int \phi_i^\infty \boldsymbol{D}_j dV, \tag{270}$$

with,

$$\boldsymbol{E}_i^\infty = \frac{\boldsymbol{D}_i}{\varepsilon_\infty}, \tag{271}$$

where $\boldsymbol{E}_i^\infty$ represents the electric field in the charging state $i$ with only fast polarization present, while $\phi_i^\infty$ denotes the corresponding electric potential. Combined with Eqs. 264 and 269, we have,

$$\int c\boldsymbol{D}_i\boldsymbol{D}_j dV = \int \left(\phi_i^\infty \varrho_j - \phi_i \varrho_j\right) dV = -\int \phi_i^{\mathrm{n}} \varrho_j dV, \tag{272}$$

with,

$$\phi_i^{\mathrm{n}} = \phi_i - \phi_i^\infty, \tag{273}$$

where $\phi_i^{\mathrm{n}}$ represents the electric potential change due to the introduction of slow polarization in charging state $i$. $\lambda_{\mathrm{out}}$ in Eq. 58 is then reformulated to

$$\begin{aligned}\lambda_{\mathrm{out}} &= \frac{1}{2}\int (-\phi_{\mathrm{red}}^{\mathrm{n}}\rho_{\mathrm{red}} - \phi_{\mathrm{ox}}^{\mathrm{n}}\rho_{\mathrm{ox}} + \phi_{\mathrm{red}}^{\mathrm{n}}\rho_{\mathrm{ox}} + \phi_{\mathrm{ox}}^{\mathrm{n}}\rho_{\mathrm{red}}) dV \\ &= -\frac{1}{2}\int (\rho_{\mathrm{red}} - \rho_{\mathrm{ox}})(\phi_{\mathrm{red}}^{\mathrm{n}} - \phi_{\mathrm{ox}}^{\mathrm{n}}) dV,\end{aligned} \tag{274}$$

which is the same as in Ref.[125]. If the charge density is assumed to be distributed on the surfaces, $\lambda_{\mathrm{out}}$ is replaced with a surface integral,

$$\lambda_{\mathrm{out}} = -\frac{1}{2}\int (\sigma_{\mathrm{red}} - \sigma_{\mathrm{ox}})(\phi_{\mathrm{red}}^{\mathrm{n}} - \phi_{\mathrm{ox}}^{\mathrm{n}}) dS, \tag{275}$$

where $\sigma_{\mathrm{red}}$ and $\sigma_{\mathrm{ox}}$ are the surface charge distributions in the reduced and oxidized states of the system. As mentioned, when considering a specific ET near the metal surface, $\sigma_{\mathrm{red}}$ and $\sigma_{\mathrm{ox}}$ can be taken as the charge distributions of the reduced and oxidized species, respectively.

### 7.3. Local concentration

The discussion above focused on the formulation of the rate constant and the influence of local reaction conditions on it but the mass action law, which is valid for elementary reactions, tells that the overall reaction rate is proportional to both the rate constant and the local reactant concentration at the reaction site. The local concentration of electroactive species involved in ET reactions is determined by both the reaction kinetics and mass transfer effects. While the reaction kinetics can be described by ET theory, mass transfer is often modeled using the Poisson-Nernst-Planck equation or its variants[25,249], which account for diffusion, electromigration, and convection. As the interfacial electric field is confined within the EDL, typically a nanoscale region near the metal surface, electromigration influences the EDL region whereas mass transfer beyond the EDL side naturally occurs through diffusion and convection. As a result, a concentration gradient extending to the microscale or beyond, may form across the EDL: this is known as the diffusion layer. Experimental techniques such as rotating disk electrodes (RDEs), microelectrodes, and pulse methods have been developed to mitigate the concentration gradient or to decouple mass transfer effects from reaction kinetics within the diffusion layer. Therefore, this issue will not be the focus of the following discussion. However, caution should be exercised in the case of very fast reactions, where even strong convection methods may not completely eliminate the concentration gradient, potentially leading to controversial observations on electrocatalytic activity[250–252] or ET kinetics[67]. When the reaction kinetics are slow, mass transfer effects within the EDL can be neglected and the local concentrations can be approximately described by the equilibrium state theory of the EDL, wherein the species concentrations follow an equilibrium distribution, such as the Poisson-Boltzmann distribution in the simplest case. In this subsection, we restrict our attention to the concentration distribution within an equilibrium EDL.

In most electrochemical reactions, the reactants do not specifically adsorb on the electrode surface. The concentration distributions of such non-specifically adsorbing reactants are controlled by two main factors. Firstly, for ionic reactants the electrostatic interactions between the free charge on the metal surface and ionic charge lead to an excess concentration of ions in the diffuse layer, the extent of which decreases toward the bulk solution over a distance characterized by the Debye length. The response of the free charge to the electrode potential may be affected by electron spillover and surface dipoles formed at the metal surface. Assuming a Poisson-Boltzmann distribution of ions, the local reactant concentration can be expressed as,

$$c_i^{\mathrm{loc}} = c_i^{\mathrm{b}} \exp\left(-\frac{z_i e_0 \Delta\phi_{\mathrm{S}}^{a}}{k_{\mathrm{B}} T}\right), \tag{276}$$

where $c_i^{\mathrm{b}}$ is the bulk concentration of species $i$. The modifications to reaction kinetics introduced by accounting for the local reactant concentration in Eq. 275 and the work term in Eq. 260 are commonly known as the Frumkin correction[253]. The Frumkin correction is often based on the notation of a reaction plane, typically designated at the OHP in the simulations[17,32,249,252,254].

Secondly, the interactions between reactants and the solvent medium, structured e.g., by hydrogen bonds, lead to a layered reactant concentration profile, as observed in molecular dynamic (MD) simulations[255]. The separation of these solvent layers is characterized by the periodic length of the spatial correlation function of the longitudinal solvent polarization[256]. While such a layered reactant profile is absent in classical EDL models and requires high computational cost in DFT-based simulations, the recent density-potential-polarization functional theoretical approach (DPPFTA)[43,257] offers a semi-classical and computationally efficient approach to model it under constant-potential conditions. As shown in Figure 14a and 14b, the DPPFTA approach captures two essential features of the EDL: the electronic response on the metal side, which makes making the extent of electron spillover depend on the electrode potential, and the structured solvent on the solution side leading to damped oscillations in solvent polarization extending toward the bulk solution. Oscillations in the solvent polarization further lead to an oscillatory electric potential on the solution side, as shown in Figure 14c. Anions and cations tend to form layered structures as they accumulate near the peaks and throughs of the electric potential, which corresponds to their energetically preferred distributions within the solvent layers, as shown in Figure 14d and 14e.

The importance of the layered ion concentration profile on local reaction conditions is reflected in several aspects. First, the layered structure of ions allows co-ions to have appreciable densities near the metal surface. As shown in the inset of Figure 14d, even at an electrode potential as negative as -0.2 V vs potential of zero charge (PZC), the anion concentration in the first layer remains approximately one-third of that in the bulk solution. In fact, at a potential of -0.1 V vs PZC, the anion concentration in the first layer even slightly exceeds the bulk concentration. The situation is opposite for cations, as shown in Figure 14e. The anomalous accumulation of co-ions near the metal surface is attributed to the structured solvent stabilizing the co-ions. The non-negligible concentration of anions near the negatively charged metal surface may be responsible for the apparent anion effect observed in the electrochemical $CO_2$ reduction reaction[258]. Second, the assumption of a single OHP as the position of the closest non-specifically adsorbed ions in classical EDL models may not be reasonable. As shown in Figure 14d and 14e, anions and cations approach the metal surface to different extents, each adjusting to better fit between solvent layers, suggesting that the use of a single, predefined OHP may be an oversimplification. Third, the layered structure of ions can help identify the position of the reaction plane, which would be located at the position of the peak concentration of ions in the first layer. In classical EDL models, the reaction plane is chosen based on the sizes of ions and solvent molecules. However, how closely non-specifically adsorbed ions can approach the metal surface also depends on their compatibility within solvent layers. Fourth, the position of the reaction plane, chosen based on the first ion layer, is dependent on the electrode potential. Instead of considering the ET rate only at the reaction plane, the calculation can be improved by integrating the ET rate over a reaction volume, as discussed in Section 1.2.1. Such a reaction volume may be chosen as the first layer of reaction adjacent to the electrode surface. For subsequent layers of reactant, the distance from the electrode surface is typically too large for a non-negligible electron tunnelling probability. The distance-dependent parameters include the reorganization energy, the coupling matrix elements, and the local concentration of the reactant. It should also be noted that, when considering ET rate over the broad electronic spectrum of the electrode surface, the integration should first be carried out over the distance from the electrode surface and then over the electronic energy spectrum[56].

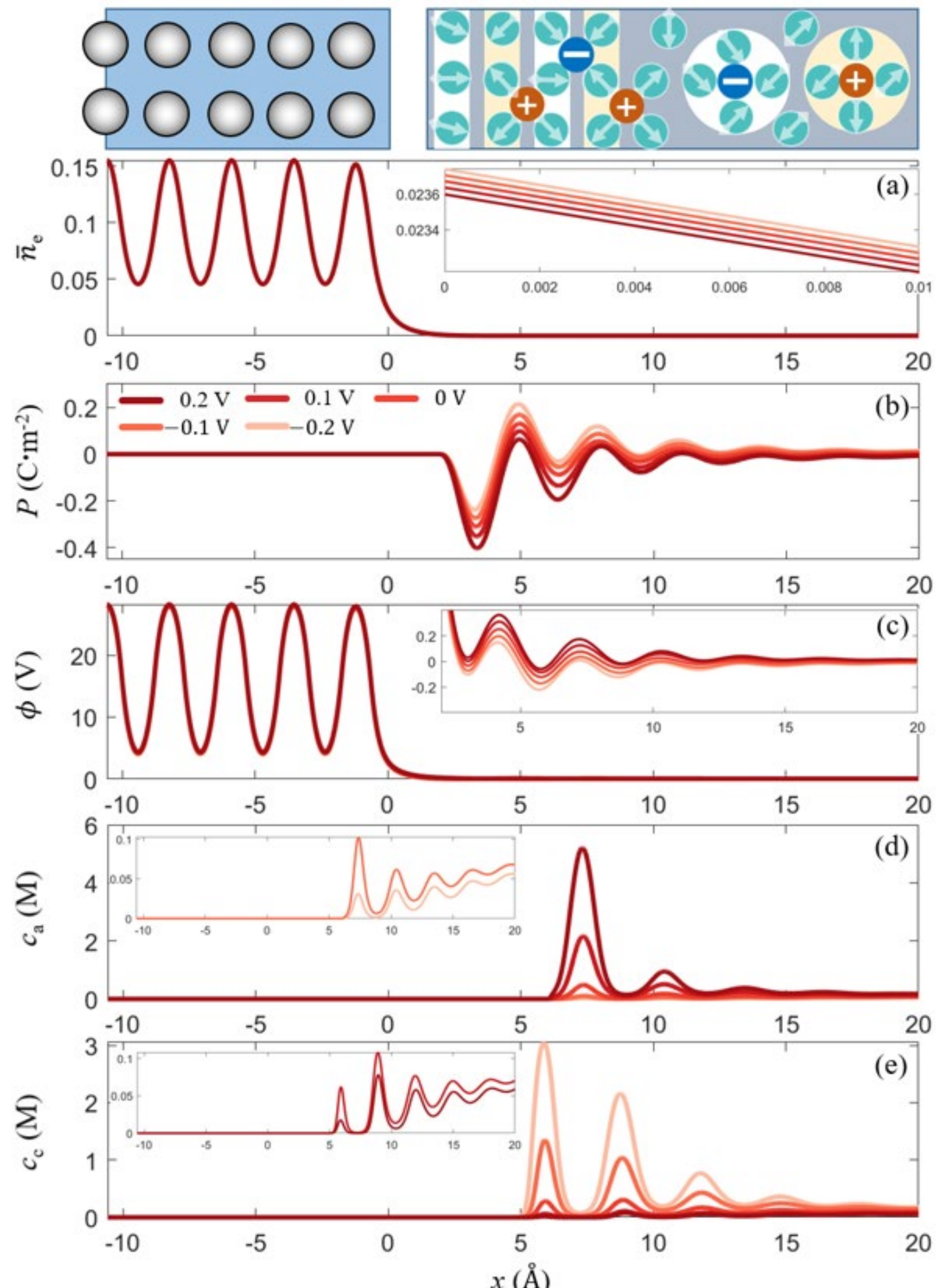

Figure 14. Model results[43] of density-potential-polarization functional theory for the Ag(110)-0.1 M $KPF_6$ aqueous solution interface at five electrode potentials, referenced to the potential of zero charge, as indicated in the legend of Figure 14b: (a) distribution of the dimensionless electron density, with the inset presenting an enlarged vies near the metal surface, (b) distribution of the solvent polarization, (c) distribution of the electric potential, with the inset showing an enlarger view on the solution side, (d) distribution of the anion concentration, with the inset presenting an enlarged view at negative electrode potentials, (e) distribution of the cation concentration, with the inset presenting an enlarged view at positive electrode potentials. In all of these plots, the metal edge is located at $x = 0$ Å. The top panel provides a schematic illustration of the electric double layer structure, featuring periodically arranged metal cationic cores, and layered structures of ions and solvent molecules near the metals surface. Adapted from Ref.[43], licensed under Creative Commons CC BY 4.0.

## 8. Conclusion and outlook

We have reviewed the theory and simulation of electron transfer (ET) kinetics, aiming to provide a unified and pedagogical description of the key concepts, derivations of central equations, and their parametrization through atomistic simulations. On the theoretical side, we first derived the Marcus activation free energy, in which the inner sphere (i.e., the redox species) is treated as a set of harmonic vibrational modes and the outer sphere (i.e., the solvent) as a dielectric continuum. This was then complemented by a treatment of quantum transitions between the electronic states of the metal and redox species using time-dependent perturbation theory in the non-adiabatic limit. We subsequently explored the adiabatic ET regime through the Anderson-Newns-Schmickler model Hamiltonian, which allows analytic construction of the free energy surfaces (FESs) from the non-adiabatic limit to the adiabatic limit. Since these derivations do not explicitly include solvent dynamics, we further discussed corrections to the pre-exponential factor based on the generalized Langevin equation. Through the above systematic review of nuclear reorganization, electronic coupling, and solvent dynamics, we established a comprehensive conceptual framework capable of addressing a broad spectrum of electrochemical ET phenomena and reactions. We have also shown how the central parameters entering the key equations, i.e., the reorganization energy, electronic coupling constant, and solvent nuclear frequency, can be extracted from atomistic simulations. In addition, we devoted special attention to the role of the electrical double layer (EDL), which influences ET rates through local changes in work terms, reorganization energy, and concentrations. Taken together, the considered topics provide a comprehensive understanding of how electrochemical ET can be achieved through the integration of ET theory with the theory of the EDL and atomistic simulations. Throughout the review, we have highlighted exemplary questions that may be answered by the methods discussed within the section.

In general, our atomistic understanding of electrochemical ET can be significantly advanced when experiments are combined with analytical theories and accurate numerical simulations. In general, the computational treatment of electrochemical ET requires a reliable quantum mechanical treatment of the electrode, a thermodynamics treatment of the solvent/electrolyte as well as sampling of the reaction coordinate to compute the FESs. Currently, this can be achieved using diabatic DFT methods, such as constrained DFT, coupled with molecular dynamics (MD) simulations and the mapping Hamiltonian approach to construct the FES along the energy gap coordinate. While such DFTEVB-MD simulations can provide an atomistic and electronic description of ET, it should be noted that even small errors (~0.1 eV) due to the exchange-correlation functionals, basis set, k-point sampling or finite-size or -time effects translate to orders of magnitude errors in rate constant predictions. Hence, computing accurate absolute rate constants is challenging. As measuring absolute rate constants is also difficult, we consider that the best outcome from experiment-theory-simulation works is best achieved by focusing on differences, trends, and phenomena rather than absolute values.

Throughout the review we have emphasized the central assumption of linear response, or equivalently, linear coupling between the electron transfer and the response or reorganization of nuclear modes. Specifically, we have shown how the linear response approximation influences the continuum electrostatic description in Marcus theory, the EVB simulations and computation of the reorganization energy, the Anderson-Newns-Schmickler model, and even the formulation of dynamic solvent effects starting from the generalized Langevin equation. In many instances, the linear response theory leads to significant simplifications, most importantly the possibility to obtain closed form equations, or more facile simulations of the key parameters. In general, the linear response theory is valid whenever the coupling between ET and nuclear reorganization of the reaction medium rather weak and when the solvation properties of the redox couple do not drastically change during the ET reaction[111]; such cases correspond to (bulky) outer-sphere redox couples, weakly solvated redox couples, and whenever the solvation of the reactant and products is "sufficiently" similar. The precise conditions of "sufficiently similar" is difficult to know a *priori,* for instance, the case of $Cu^{2+}$ reduction can be treated using a linear coupling model but the $Zn^{2+}$ reduction cannot[259].

It is not, however, generally understood when the linear response approximation is valid and stronger and non-linear coupling between the ET and nuclear reorganization is expected because the validity of the linear response assumption is rarely tested. There might be several cases where it breaks down –most obvious being electrocatalytic or inner-sphere ET reactions where to solvent goes through significant structural changes during reactions. Understanding the limits and cases when the linear response theory remain valid is not only of theoretical importance but is also of high experimental relevance as the assumption of linear response is *implicitly* included in the vast majority of ET theories used to analyze experiments. For instance, whenever a Marcus-like model is used to explain the ET kinetics or to obtain the reorganization energy, it is implicitly assumed that the linear response theory holds. For instance, recently[190,260–262] the Marcus-Hush-Chidsey equation (Eq. 132) has started to gain traction in analyzing solvent effects of electrocatalytic reactions and the reorganization energy is used as a key metric; the Marcus-Hush-Chidsey equation is strictly valid only for non-adiabatic ET within the linear response region, and it is not guaranteed that electrocatalytic reactions fulfill either criterion. While such analysis can be highly useful, it is pertinent to realize the underlying limitations of the model used.

To understand reactions with significant changes in the solvation and chemical structure of the redox species, it is necessary to go beyond the linear response approximation[111], and here simulations and theory can be highly complementary[263]. The most direct way to study the ET kinetics without relying on linear response is to use the mapping Hamiltonian approach for direct enhanced sampling EVB-MD simulations of the free energy surfaces along the energy gap coordinate. Such simulations can yield both the diabatic and adiabatic FESs, and in particular the diabatic FESs can inform whether the linear response assumption is valid; if the diabatic surfaces are parabolic, the energy gap distribution is Gaussian, and linear response holds. Any departures from parabolic diabatic FESs are an indication that the linear response approximation is violated. While such calculations at the force field[144,264] and DFT levels[157,176,265] indicate that the linear response is a good approximation for outer-sphere reactions, much less is known about electrochemical ET or inner-sphere reactions at the electrode surface. While force field simulations[172] of outer-sphere electrochemical ET and DFT-level studies[186] of inner-sphere ET at an oxide surface show that electrochemical ET reactions may be within the linear response regime, the validity of linear response should not be taken for granted. For instance, recent constrained DFT-MD studies showed[115] that the initial outer-sphere electron transfer step in electrocatalytic oxygen reduction reaction at Pt surface, $\mathrm{O_2(sol) + e^- \rightarrow O_2^-(sol)}$, is not well described by the linear response approximation. On the other hand, using the same computational approach to study a very similar reaction, the outer-sphere ET in the $CO_2$ reduction reaction on an Au surface, $\mathrm{CO_2(sol) + e^- \rightarrow CO_2^-(sol)}$, shows that this reaction falls within the linear response region[114]. As it is not currently known how well the linear response assumption holds for electrochemical ET reactions in general, EVB-MD studies, especially at the DFT level, are urgently needed to understand the ET at the atomic scale on realistic electrode models; to our knowledge, only few such studies have been carried out.

The possibility and role of non-linear coupling between ET and the solvent should also be investigated within analytical theoretical models. One way to achieve this is to use non-Gaussian energy gap distribution functions[111], which allows the formulation of more complex but still closed form equations resembling the Marcus theory. Such models are also amenable to parametrization through EVB simulations which enables the integration between theory and simulations. Another way to account for the non-linear solvent response in analytical theories is to include multiple solvent reorganization energies e.g. in the Anderson-Newns-Schmickler model[259]. Also, in this case it is possible to combine theory and atomistic simulations as multiple reorganization energies can be used to fit the Anderson-Newns-Schmickler to reproduce the EVB simulations carried out with the mapping Hamiltonian. Such a combined approach could be further extended to systematically study e.g. electrolyte and electrode potential effects on ET kinetics within a mixed quantum-classical models[42,43,125,266] of ET and the electrochemical double layer.

Besides the linear coupling or response approximation, the ET models discussed in section 4 are made for non-adiabatic ET where the electronic coupling between the electrode and the redox species is weak. The adiabaticity emerges in two ways: in the prefactor and the barrier. In non-adiabatic ET the prefactor contains the coupling matrix elements describing electron tunnelling probability. In the barrier, the adiabaticity is more implicitly present as the barrier is obtained as the crossing point of two *diabatic* states, which is valid for weakly coupled states only. Both conditions underlying non-adiabatic TST are expected to hold e.g., long-distance outer-sphere ET but to break down at short distances and strong electronic interactions encountered in e.g. electrocatalysis where the ET is expected to be adiabatic. Theoretically, the transition from non-adiabatic to adiabatic is well-understood as discussed in Sections 5 and 6.1.6-6.1.7. From the computational side the questions of (non-)adiabaticity are most directly addressed through the EVB-MD simulations of both diabatic and adiabatic barriers but for electrochemical systems such as simulations are rare and only just emerging[267]. More well-established ways include the computation of Landau-Zener factors to interpolate between adiabatic and non-adiabatic ET[107,268]. We expect that wider utilization and further development of both the EVB and interpolation methods will enable more detailed understanding of factors controlling both the reaction prefactor and barrier, and thereby yield improved and more general understanding of electrochemical ET. This is not relevant only for simulations and theory but also for experiments where ET kinetics are often understood through Marcus-Hush-Chidsey-type non-adiabatic models even in cases with presumably strong electronic interactions.

Furthermore, one should also account for the possibility that nuclear dynamics control the ET kinetics; while it is again possible to simulate the (solvent) nuclear dynamics either directly (Eq. 12) by computing the direct friction (kernel) for the Kramers-Grote-Hynes models (Section 6.1), or using the non-ergodic rate theory (Section 6.2). All of these are computationally intensive approaches and we are not aware of any DFT level studies where these would have been computed for electrochemical interfaces. To go beyond the time and length scales of DFT methods and to achieve more comprehensive sampling while retaining the needed accuracy, developing diabatic tight-binding DFT approaches[269], improved classical EVB potentials[146], and machine learning methods[181] to construct EVB models might be beneficial. The description and understanding of dynamic nuclear effects is, however, expected to be highly beneficial to understand fast ET reactions or ET in complex, non-polar media such as encountered in e.g., electrosynthesis or batteries.

Finally, the models and methods addressed above in this discussion section assumes that the energy gap is the only relevant reaction coordinate. For a more general description it might be necessary to include also other reaction coordinates, such as the distance of the redox couple from the electrode surface, the bond length coordinate in bond-breaking ET, and the distance between a redox species and ions in ion-coupled ET; this calls for the extension to multiple reaction coordinate which can be done through standard methods[106] but at the expense of significant increase in the computational cost. As the distance between the electrode and redox couple or the breaking bond length is sampled, it will also be necessary to account for the possibility of switching between adiabatic and non-adiabatic ET; this again calls for the evaluation of the electronic coupling matrix element and evaluating the Landau-Zener transmission probability. Even though this is possible even for electrochemical ET, the computational cost is high and more efficient simulation approaches are needed. Despite these difficulties in treating multiple reaction coordinates, we expect that facing and overcoming them will be highly beneficial, as it greatly opens the scope of reactions that can be addressed. Some notable examples of topical interest electrodeposition and dissolution[270], as well as ion-coupled electron transfer[242,271] where the combination of simulations, theory, and experiments is needed to understand the complex chemistry underlying these processes.

Overall, our review shows that the basic theory and simulation methods to address electrochemical electron transfer can now be considered well-established. At the same time, our review highlights the outstanding challenges and areas for further refinement, including the treatment of non-linear coupling between ET and the solvent, various timescales and system dynamics, and a more detailed treatment of the reaction environment, i.e., the electrical double layer. As the challenges are common to both analytical theory and simulation, we consider that a closer integration of the theory, atomistic simulations, and well-controlled experiments to be very beneficial. For instance, the adiabaticity of outer-sphere electron transfer kinetics is a basic question in fundamental and applied electrochemistry but answering this question has required the careful integration of well-defined electrodes, highly detailed and sensitive electrochemical experiments, GCE-DFT simulations, and a model Hamiltonian description of ET kinetics[67]. Such studies will push the boundaries of experiments, theory, and simulations which is beneficial to not only a more detailed understanding of ET and electrochemical interfaces but also in addressing even more complex electrochemical reactions encountered in electrocatalysis.

## 9. Acknowledgements

M.Z. and Y.C. thank the financial support from the National Natural Science Foundation of China under grant number 22472162, National Key Research and Development Program of China (No. 2023YFA1509004) and international partnership program of the Chinese Academy of Science. J.H. is supported by the Initiative and Networking Fund of the Helmholtz Association (No. VH-NG-1709), and European Research Council (ERC) Starting Grant (MESO-CAT, Grant agreement No. 101163405). M.M.M acknowledges funding the by Research Council of Finland (Grant #338228) and support by the Central Finland Mobility Foundation (Cefmof).

## 10. Appendix

### 10.1. Electrostatic energy

The total energy of an electrostatic system is the sum of Coulomb interactions between all charged particles,

$$U = \sum_{i<j} \frac{q_i q_j}{4\pi\varepsilon|\boldsymbol{r}_i - \boldsymbol{r}_j|} = \frac{1}{2}\sum_{i,j} \frac{q_i q_j}{4\pi\varepsilon|\boldsymbol{r}_i - \boldsymbol{r}_j|} = \frac{1}{2}\sum_i q_i \sum_j \frac{q_j}{4\pi\varepsilon|\boldsymbol{r}_i - \boldsymbol{r}_j|}, \tag{277}$$

where $q_i$ and $q_j$ are the charges of particle $i$ and $j$, and $\boldsymbol{r}_i$ and $\boldsymbol{r}_j$ are their positions. Then we have,

$$U = \frac{1}{2}\iint \sum_i q_i \delta(\boldsymbol{r} - \boldsymbol{r}_i) \sum_j \frac{q_j \delta(\boldsymbol{r}' - \boldsymbol{r}_j)}{4\pi\varepsilon|\boldsymbol{r} - \boldsymbol{r}'|} d\boldsymbol{r} d\boldsymbol{r}' = \frac{1}{2}\iint \frac{\varrho(\boldsymbol{r})\varrho(\boldsymbol{r}')}{4\pi\varepsilon|\boldsymbol{r} - \boldsymbol{r}'|} d\boldsymbol{r} d\boldsymbol{r}', \tag{278}$$

with the charge density distribution,

$$\varrho(\boldsymbol{r}) = \sum_i q_i \delta(\boldsymbol{r} - \boldsymbol{r}_i). \tag{279}$$

When we charge the system with an infinitesimal amount of charge $\delta\rho$, the reversible work required corresponds to the difference in electrostatic energy before and after the charging process, i.e.,

$$\delta U = \delta W = \int \left( \int \frac{\varrho(\boldsymbol{r}')}{4\pi\varepsilon|\boldsymbol{r} - \boldsymbol{r}'|} d\boldsymbol{r}' \right) \delta\varrho(\boldsymbol{r}) d\boldsymbol{r} = \int \phi(\boldsymbol{r}) \delta\varrho(\boldsymbol{r}) d\boldsymbol{r}, \tag{280}$$

where the electrostatic potential is defined as $\phi(\boldsymbol{r}) = \int \frac{\varrho(r')}{4\pi\varepsilon|r-r'|} d\boldsymbol{r}'$. The reversible work required to charge the system from $\varrho_i$ to $\varrho_j$ is then given by,

$$W = \int \left( \int_{\varrho_i}^{\varrho_j} \phi \delta\varrho \right) d\boldsymbol{r}. \tag{281}$$

The reversible work can be equivalently reformulated as follows,

$$\begin{aligned} W &= \int \left( \int_{\boldsymbol{D}_i}^{\boldsymbol{D}_j} \phi \delta(\nabla \cdot \boldsymbol{D}) \right) d\boldsymbol{r} \\ &= \int \left( \int_{\boldsymbol{D}_i}^{\boldsymbol{D}_j} \phi \nabla \cdot \delta\boldsymbol{D} \right) d\boldsymbol{r} \\ &= \int \left( \int_{\boldsymbol{D}_i}^{\boldsymbol{D}_j} \nabla \cdot (\phi \delta\boldsymbol{D}) \right) d\boldsymbol{r} - \int \left( \int_{\boldsymbol{D}_i}^{\boldsymbol{D}_j} \nabla\phi \cdot \delta\boldsymbol{D} \right) d\boldsymbol{r} \\ &= \int \left( \int_{\boldsymbol{D}_i}^{\boldsymbol{D}_j} \boldsymbol{\mathcal{E}} \cdot \delta\boldsymbol{D} \right) d\boldsymbol{r}, \end{aligned} \tag{282}$$

where $\boldsymbol{\mathcal{E}} = -\nabla\phi$ is the electric field. The first three identities follow from the electrostatic relation, the linearity of the divergence operator, and the application of the divergence product rule. The fourth identity assumes a finite system, where the first term in the third line of Eq. 281 vanishes by applying *the* divergence theorem.

### 10.2. Equations of motion in the Heisenberg picture

The Heisenberg picture confines the dynamical evolution of a quantum system to the operators rather than the quantum state as done in the Schrödinger picture. A Schrödinger operator $A_\mathrm{S}$ has a corresponding Heisenberg operator $A_\mathrm{H}(t)$ defined as,

$$A_\mathrm{H}(t) = U^\dagger(t,t_0)A_\mathrm{S}U(t,t_0), \tag{283}$$

where $U(t,t_0)$ is a unitary operator for time evolution, which evolves the quantum state at $t_0$ to that at $t$ by,

$$|\Psi,t\rangle = U(t,t_0)|\Psi,t_0\rangle, \tag{284}$$

where $U^\dagger(t,t_0)$ is the adjoint operator of $U(t,t_0)$. The equations of motion (EOMs) for Heisenberg operators, i.e., the differential equation determining the time evolution of the Heisenberg operators, have the form,

$$i\hbar\frac{\partial A_\mathrm{H}(t)}{\partial t} = [A_\mathrm{H}(t), H_\mathrm{H}(t)] + i\hbar\left(\frac{\partial A_\mathrm{S}}{\partial t}\right)_\mathrm{H}, \tag{285}$$

where $H_\mathrm{H}(t)$ is the total Hamiltonian of the system in the Heisenberg picture and where the commutation relation between two operators is,

$$[A,B] = AB - BA. \tag{286}$$

Let us consider a quantum system with the Hamiltonian being $H'_\mathrm{el}$ in Eq. 144. Since the Schrödinger operators $c_i^\dagger$ and $c_i$ are time-independent, the EOMs of the Heisenberg operators $c_a(t)$ and $c_k(t)$, which respectively corresponds to the Schrödinger operators $c_a$ and $c_k$, are directly given by Eq. 284 after neglecting the last term,

$$i\hbar\frac{\partial c_a(t)}{\partial t} = [c_a(t), H'_\mathrm{el}(t)], \tag{287}$$

$$i\hbar\frac{\partial c_k(t)}{\partial t} = [c_k(t), H'_\mathrm{el}(t)]. \tag{288}$$

To explicitly express the right-hand-sides of the above equations, we make use of the fact that the Heisenberg operator $C_\mathrm{H}(t)$ associated with the product of Schrödinger operators $C_\mathrm{S} = A_\mathrm{S}B_\mathrm{S}$ is equal to the product of their corresponding Heisenberg operators $A_\mathrm{H}(t)B_\mathrm{H}(t)$. This can be shown by,

$$\begin{aligned} C_\mathrm{H} &= U^\dagger(t,t_0)A_\mathrm{S}B_\mathrm{S}U(t,t_0) \\ &= U^\dagger(t,t_0)A_\mathrm{S}U(t,t_0)U^\dagger(t,t_0)B_\mathrm{S}U(t,t_0) \\ &= A_\mathrm{H}(t)B_\mathrm{H}(t). \end{aligned} \tag{289}$$

From this it follows that the commutation relation between two Heisenberg operators is the same as that between the corresponding Schrödinger operators. The commutation relations in Eqs. 286 and 287 for the corresponding Schrödinger operators can be respectively calculated as,

$$[c_a, H'_\mathrm{el}] = \epsilon'_a[c_a, n_a] + \sum_k \epsilon_k[c_a, n_k] + \sum_k H_{ka}[c_a, c_k^\dagger c_a] + \sum_k H^*_{ka}[c_a, c_a^\dagger c_k], \tag{290}$$

$$[c_k, H'_\mathrm{el}] = \epsilon'_a[c_k, n_a] + \sum_{k'} \epsilon_{k'}[c_k, n_{k'}] + \sum_{k'} H_{k'a}[c_k, c_{k'}^\dagger c_a] + \sum_k H^*_{k'a}[c_k, c_a^\dagger c_{k'}]. \tag{291}$$

The above calculations are straightforward by using the following relations for fermions,

$$\{c_i, c_j^\dagger\} = \delta_{ij}, \{c_i, c_j\} = 0, \{c_i^\dagger, c_j^\dagger\} = 0, \tag{292}$$

where $\{A, B\} = AB + BA$ denotes the anti-commutation relation between the two operators $A$ and $B$. Then the EOMs can be obtained as,

$$i\hbar\dot{c}_a(t) = \epsilon_a' c_a(t) + \sum_k H_{ka}^* c_k(t), \tag{293}$$

$$i\hbar\dot{c}_k(t) = \epsilon_k c_k(t) + H_{ka} c_a(t). \tag{294}$$

## 10.3.Alternative derivation of the non-adiabatic ET rate constant

This derivation starts from Eq. 115, which is the Fermi Golden rule describing the transition probability between two vibronic states $km$ and $an$. By assuming that the system is in thermal equilibrium, the overall rate between the electronic states $k$ and $a$ is given by transitions between all vibrational states (Eq. 100) and the transition probability is

$$W_{ka} = \frac{2\pi}{\hbar}|H_{ak}|^2 \sum_{km,an} \rho_{km} S_{an,km}^2 \delta(E_{km} - E_{an}), \tag{295}$$

where the summation goes over all nuclear wave functions and $p_{km} = \frac{\exp(-\beta E_{km})}{\sum_m \exp(-\beta E_{km})}$ is the Boltzmann weight of the initial vibronic state $|\Psi_{km}^0\rangle$ in Eq. 96. To proceed, the delta function is replaced by its Fourier transform

$$\begin{aligned}
W_{ka} &= |H_{ak}|^2 \sum_{km,an} p_{km} \int S_{an,km}^2 \exp\left[\frac{it(E_{km} - E_{an})}{\hbar}\right] \\
&= |H_{ak}|^2 \sum_{km,an} p_{km} \int |\langle\chi_{an}|\chi_{km}\rangle|^2 \exp\left[\frac{it(E_{km} - E_{an})}{\hbar}\right] dt \\
&= |H_{ak}|^2 \sum_{km,an} p_{km} \int \langle\chi_{km}| \exp\left[\frac{it(E_{km} - E_{an})}{\hbar}\right] |\chi_{an}\rangle\langle\chi_{an}|\chi_{km}\rangle dt \\
&= |H_{ak}|^2 \sum_{km} p_{km} \int \langle\chi_{km}| \exp\left[\frac{it(H_{km} - H_{an})}{\hbar}\right] |\chi_{km}\rangle dt,
\end{aligned} \tag{296}$$

where on the second line the overlap integral is written using the vibrational wave functions. On the fourth line we make use of the fact the vibrational wave functions are eigenfunctions of the nuclear Schrödinger equation (Eqs.78-79, $H_{km}|\chi_{km}\rangle = E_{km}|\chi_{km}\rangle$), and the summation over the final state vibrational states is eliminated by the using the completeness of vibrational wave functions: $\sum_{an}|\chi_{an}\rangle\langle\chi_{an}| = 1$. While the final form is very general and can be used to compute the transition probability through the energy gap correlation function ($\exp\left[\frac{it(H_{km}-H_{an})}{\hbar}\right]$), it is too complex for most practical applications. To obtain a more manageable expression, the vibrational wave functions are assumed to be those of independent harmonic oscillators, which are displaced during the transition from the initial to the final state. In this case, the time integral can be written in terms of the thermal Franck-Condon factor for the energy gap ($\mathcal{D}(\Delta E/\hbar)$)

$$W_{ka} = |H_{ak}|^2 \sum_{km} p_{km} \int \langle \chi_{km}| \exp\left[\frac{it(H_{km} - H_{an})}{\hbar}\right] |\chi_{km}\rangle dt \\ = \frac{|H_{ak}|^2}{\hbar} \mathcal{D}(\Delta E(\epsilon_k)/\hbar), \tag{297}$$

where the Franck-Condon factor is[272]

$$\mathcal{D}(\Delta E(\epsilon_k)/\hbar) = \frac{1}{2\pi\hbar} \exp[-R(0)] \int \left( \exp\left[\frac{i\Delta E(\epsilon_k)t}{\hbar}\right] + R(t) \right) dt, \tag{298}$$

and $R(t)$ is

$$R(t) = \int_0^\infty J(\omega)\big[\exp(-i\omega t)\left(1 + n(\omega)\right) + \exp(i\omega t)\, n(\omega)\big]\, d\omega, \tag{299}$$

and $J(\omega)$ is the spectral density of the vibrational states and $n(\omega)$ is the Bose-Einstein distribution function of the harmonic oscillator states. The spectral density $J(\omega)$ represents the coupling strength between the reaction coordinate (energy gap) and vibrational modes at frequency $\omega$. In other words, the presents the density of oscillator states weighted by the electron–vibrational coupling constant. The spectral density can be computed in terms of the vibrational normal modes (Eq. 215) and is closely related to the reorganization energy

$$\lambda = \frac{1}{\hbar} \int_0^\infty \omega J(\omega) d\omega. \tag{300}$$

As discussed in Section 6.1, the spectral density can be computed using MD simulations and if it is split different contributions, such as low and high frequencies or outer-sphere and inner-sphere vibrations, the reorganization energy can be split into the corresponding parts. While the introduction of the harmonic oscillators and the Franck-Condon factors simplifies the formulation of the electron transition probability, the expression for $R(t)$ and its time integral are still difficult to treat, and this prevents obtaining a closed form equation. There are several different ways to proceed to derive results for different temperature and frequency limits[13,187,272,273]. Here, we show the results for the high-temperature case where the nuclear vibrations can be considered classical; in this case[187], $k_B T \gg \hbar\omega$ and $1 + 2n(\omega) \approx \frac{2k_B T}{\hbar\omega} \gg 1$. Using this and splitting $R(t)$ is split into it real and imaginary gives:

$$R(t) = \int_0^\infty \cos\left(\omega t\big(1 + 2n(\omega)\big)\right) J(\omega) d\omega - i \int_0^\infty \sin\big(\omega t J(\omega)\big)\, d\omega \\ \approx \int_0^\infty \cos\left(\omega t \frac{2k_B T}{\hbar\omega}\right) J(\omega) d\omega - i \int_0^\infty \sin\big(\omega t J(\omega)\big)\, d\omega. \tag{301}$$

One can expand the cosine and sine terms as a Taylor series and truncate after the leading terms that depend on $\omega t$. This so-called "slow-fluctuation", high-temperature limit gives

$$R(t) \approx \int_0^\infty \frac{(\omega t)^2}{2} \frac{2k_B T}{\hbar\omega} J(\omega) d\omega - i \int_0^\infty \omega t J(\omega) d\omega. \tag{302}$$

Comparison with Eq. 299 shows that both integrals now contain the reorganization energy. Replacing the spectral density with the reorganization and inserting $R(t)$ in Eq. 299 gives

$$\mathcal{D}(\Delta E(\epsilon_k)/\hbar) = \frac{1}{2\pi\hbar} \int_{-\infty}^{+\infty} \exp\left(\frac{it(\Delta E_k - \lambda)}{\hbar}\right)\left(-\frac{k_B T t^2 \lambda}{\hbar^2}\right) dt. \tag{303}$$

This integral can be evaluated analytically to give

$$\mathcal{D}(\Delta E(\epsilon_k)/\hbar) = \frac{1}{\sqrt{4\lambda k_\mathrm{B} T}} \exp\left(-\frac{(\Delta E(\epsilon_k) + \lambda)^2}{4\lambda k_\mathrm{B} T}\right), \tag{304}$$

which now contains the Marcus result for the barrier. Inserting this in Eq. 296 gives the transition probability as

$$W_{ka} = \frac{2\pi |H_{ak}|^2}{\hbar} \sqrt{\frac{\beta}{4\pi\lambda}} \exp\left(-\frac{\beta(\Delta E(\epsilon_k) + \lambda)^2}{4\lambda}\right). \tag{305}$$

Because of the Franck-Condon assumption, energy is conserved during the ET event and the entropy does not change. Hence, the energy $\Delta E(\epsilon_k)$ can be replaced with the correspond reaction free energy $\Delta G^0(\epsilon_k)$.

### 10.4.Separating the total reorganization energy into inner- and outer-sphere contributions

In Section 3.6 we discussed EVB-MD simulations and showed how the total nuclear reorganization energy can be computed. However, elsewhere in the review we have often treated the inner- and outer-sphere reorganization components separately. Here we show how the total reorganization energy can be decomposed into inner- and outer-sphere components. Here we focus on the separation in the slow fluctuation, high-temperature limit valid for classical vibrational modes.

For this, the spectral density of Eq.300 is separated in two contributions, the high frequency inner-sphere part and the low frequency outer-sphere part:

$$\begin{aligned} \lambda = \frac{1}{\hbar}\int_0^\infty \omega\, J(\omega) d\omega &= \frac{1}{\hbar}\left(\int_0^{\omega_\mathrm{outer}} \omega\, J(\omega) d\omega + \int_{\omega_\mathrm{outer}}^\infty \omega\, J(\omega) d\omega\right) \\ &= \lambda_\mathrm{outer} + \lambda_\mathrm{inner} \end{aligned}$$

The same outcome can be realized by splitting the spectral density is separated into its inner- and outer-sphere parts:

$$J(\omega) = J_\mathrm{in}(\omega) + J_\mathrm{out}(\omega)$$

A particular case arises when the inner part can be characterized by a few specific vibrational frequencies ($\omega_i$) as done, for example, in Eq. 57,

$$J_\mathrm{in}(\omega) = \sum_i c_i\ \omega_i\, \delta(\omega - \omega_i)$$

where $c_i$ is the coupling constant between the vibrational mode and the energy gap coordinate. The outer part is then $J_\mathrm{out}(\omega) = J(\omega) - J_\mathrm{in}(\omega)$ and the reorganization energy becomes

$$\begin{aligned} \lambda = \frac{1}{\hbar}\int_0^\infty \omega\, J(\omega) d\omega &= \frac{1}{\hbar}\int_0^\infty \omega\big(J_\mathrm{outer}(\omega) + J_\mathrm{inner}(\omega)\big) d\omega \\ &= \lambda_\mathrm{outer} + \lambda_\mathrm{inner} \end{aligned}$$

## 11. Biographies

Mengke Zhang obtained his Doctorate degree from University of Science and Technology of China under co-supervision of Prof. Yanxia Chen and Prof. Jun Huang in June 2025. He obtained his bachelor's degree in applied chemistry from Hefei University of Technology in June 2019. His research interests currently include theory of electrochemical interfaces and ion-coupled electron transfer processes. He planned to continue postdoctoral research with Prof. Jun Huang in Juelich, Germany, and has been waiting for the Germany Visa for around eight months since July 2025.

Yanxia Chen is a full professor at the University of Science and Technology of China. She graduated from Xiamen University in 1993 and received her doctorate degree from the Department of Chemistry of Xiamen University in 1998. From March 1999 to December 2000, she worked as a Humboldt postdoctoral fellow at the University of Düsseldorf in Germany, with Professor A. Otto as her co-supervisor; from January to December 2001, she worked as a postdoctoral fellow at the Technical University of Munich, with Professor U. Stimming as her co-supervisor; from January to March 2002, she worked as a COE researcher in the group of Professor M. Osawa at Hokkaido University in Japan; from April 2002 to December 2006, she worked as a postdoctoral fellow at the University of Ulm in Germany, with Professor R.J. Behm as her co-supervisor. She joined the Department of Physics of the University of Science and Technology of China in January 2007 and is a recipient of the Hundred Talents Program of the Chinese Academy of Sciences. Her main research directions are surface electrochemistry, spectroelectrochemistry and electrocatalysis, focusing on basic research on electrochemical interface structure, fuel cells and electrocatalytic reaction mechanisms and kinetics related to bioelectrochemistry, as well as the development of related research technologies. She has published more than 160 papers, with an H-index of 31. She has published one translated book, three monograph chapters, and three invention patents. She is currently the co-executive editor of Electrochemistry Communication, an editorial board member of the Journal of Electrochemistry, and an advisory editorial board member of Electrochimica Acta and Current Opinion of Electrochemistry.

Marko M. Melander is currently a University Lecturer and an Academy of Finland Research Fellow at the Department of Chemistry and the Nanoscience Center of University of Jyvaskyla, Finland. He also holds a docenture (Adjunt professorship) in physical electrochemistry. He obtained Doctor of Science under supervision of Prof. Kari Laasonen from Aalto University of Chemical Technology in 2015. His thesis was chosen as the best thesis in catalysis between 2013-2015 by the Finnish Catalysis Society and best thesis at the Aalto University of Chemical Technology in 2015. He did postdoctoral research firstly in Denmark Technical University from 2015 to 2017 and then in University of Jyvaskyla from 2017 to 2020. He was awarded Academy of Finland Fellow in 2021 and started his independent research group. His research focuses on using, developing, and combining a variety of computational and theoretical methods based on quantum mechanics, statistical mechanics, thermodynamics and rate theory to enable multiscale simulations and theory of electrochemical systems and to provide fundamental understanding of their chemical physics from the atomic scale.

Jun Huang is a Helmholtz Young Investigators Group leader in IET-3, FZ Julich, Germany, since May 2022, and a Junior Professor at RWTH Aachen University. He obtained his bachelor's and Ph.D. degrees from the department of automotive engineering of Tsinghua University in 2012 and 2017, respectively, both under the supervision of Prof. Jianbo Zhang. From Oct. 2015 to May 2016, he was a visiting Ph.D. student in Prof. Michael Eikerling's research group at Simon Fraser University. From July 2017 to November 2019, he was an associate professor at Central South University, China. He moved to Germany in 2020. After an initial three-month visit at FZ Julich, he worked as a Humboldt fellow with Prof. Axel Groß at Ulm University until he took his current position. During the Humboldt fellowship, he had a four-month research stay at the Institute of Electrochemistry at Alicante University (Host: Professors Juan Feliu and Victor Climent) in 2021. His team works mainly on developing density-potential functional theoretic methods for understanding thermodynamics and nonequilibrium dynamics of electrocatalytic double layers, especially on the mesoscale. He was awarded Hellmuth Fischer Medal (2025) in recognition of his theoretical studies of electrochemical interfaces, impedance spectroscopy and electrocatalysis.

TOC Graphic

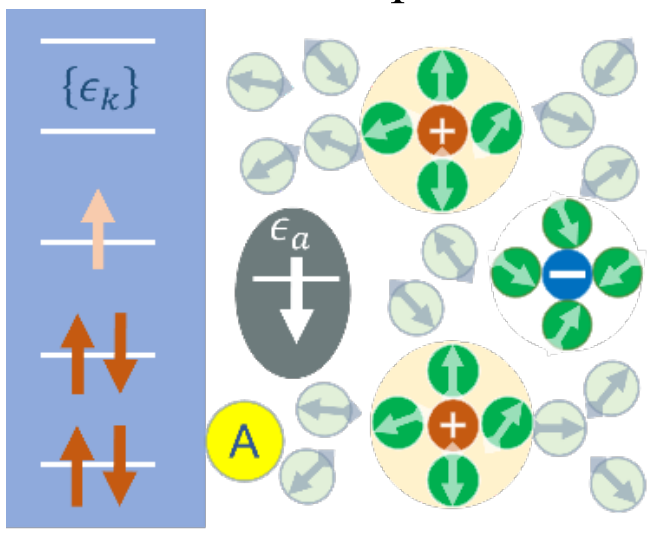

$$\mathrm{ox} + \mathrm{e}^{-}(\epsilon_k) \rightleftharpoons \mathrm{red}$$